\documentclass{aa}  

\usepackage{graphicx}
\usepackage{txfonts}
\usepackage{threeparttablex}
\usepackage{longtable}
\usepackage{hyperref}
\usepackage{multicol}
\usepackage[switch]{lineno}
\usepackage{hyperref}

\newcommand{\orcid}[1]{\href{https://orcid.org/#1}{\includegraphics[scale=0.05]{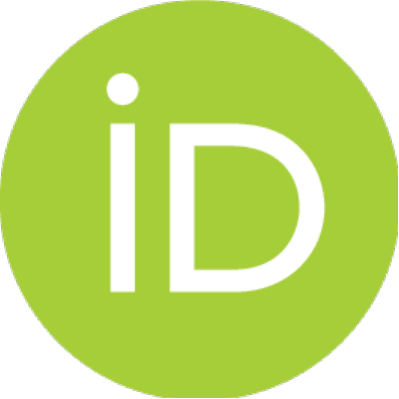}}}

\begin{document}

   \title{Sample of hydrogen-rich superluminous supernovae from the Zwicky Transient Facility}

    \titlerunning{SLSNe~II from ZTF}
    
   \author{P.J.~Pessi \inst{1}\fnmsep\thanks{\email{priscila.pessi@astro.su.se}}\orcid{0000-0002-8041-8559}
    \and R. Lunnan\inst{1}\orcid{0000-0001-9454-4639}
    \and J. Sollerman\inst{1}\orcid{0000-0003-1546-6615}
    \and S. Schulze\inst{2}\orcid{0000-0001-6797-1889}
    \and A. Gkini\inst{1}\orcid{0009-0000-9383-2305}
    \and A. Gangopadhyay\inst{1}\orcid{0000-0002-3884-5637}
    \and L. Yan\inst{3}\orcid{0000-0003-1710-9339}
    \and A. Gal-Yam\inst{4}\orcid{0000-0002-3653-5598}
    \and D.A. Perley\inst{5}\orcid{0000-0001-8472-1996}
    \and T.-W. Chen\inst{6}\orcid{0000-0002-1066-6098}
    \and K.R. Hinds\inst{5}\orcid{0000-0002-0129-806X}
    \and S.J. Brennan\inst{1}\orcid{0000-0003-1325-6235}
    \and Y. Hu\inst{1}
    \and A. Singh\inst{1}
    \and I. Andreoni\inst{7,8,9}\orcid{0000-0002-8977-1498}
    \and D.O. Cook\inst{10}\orcid{0000-0002-6877-7655}
    \and C. Fremling\inst{3,11}\orcid{0000-0002-4223-103X}
    \and A.Y.Q.~Ho\inst{12}\orcid{0000-0002-9017-3567}
    \and Y. Sharma\inst{11}\orcid{0000-0003-4531-1745}
    \and S. van Velzen\inst{13}
    \and T. Kangas\inst{14,15}
    \and A. Wold\inst{10}\orcid{0000-0002-9998-6732}
    \and E.C. Bellm\inst{16}\orcid{0000-0001-8018-5348}
    \and J.S. Bloom\inst{17,18}\orcid{0000-0002-7777-216X}
    \and M.J. Graham\inst{11}\orcid{0000-0002-3168-0139}
    \and M.M. Kasliwal\inst{11}\orcid{0000-0002-5619-4938}
    \and S.R. Kulkarni\inst{11}\orcid{0000-0001-5390-8563}
    \and R. Riddle\inst{19}\orcid{0000-0002-0387-370X}
    \and B. Rusholme\inst{10}\orcid{0000-0001-7648-4142}    
          }

   \institute{The Oskar Klein Centre, Department of Astronomy, Stockholm University, AlbaNova 106 91, Stockholm, Sweden
    \and Center for Interdisciplinary Exploration and Research in Astrophysics (CIERA), Northwestern University, 1800 Sherman Ave., Evanston, IL 60201, USA
    \and Caltech Optical Observatories, California Institute of Technology, Pasadena, CA 91125, USA
    \and Department of Particle Physics and Astrophysics, Weizmann Institute of Science, Rehovot 76100, Israel
    \and Astrophysics Research Institute, Liverpool John Moores University, 146 Brownlow Hill, Liverpool L3 5RF, UK
    \and Graduate Institute of Astronomy, National Central University, 300 Jhongda Road, 32001 Jhongli, Taiwan
    \and Joint Space-Science Institute, University of Maryland, College Park, MD 20742, USA
    \and Department of Astronomy, University of Maryland, College Park, MD 20742, USA
    \and Astrophysics Science Division, NASA Goddard Space Flight Center, Mail Code 661, Greenbelt, MD 20771, USA
    \and IPAC, California Institute of Technology, 1200 E. California Blvd, Pasadena, CA 91125, USA
    \and Division of Physics, Mathematics and Astronomy, California Institute of Technology, Pasadena, CA, 91125, USA
    \and Department of Astronomy, Cornell University, Ithaca, NY 14853, USA
    \and Leiden Observatory, Leiden University, PO Box 9513, 2300 RA Leiden, The Netherlands
    \and Finnish Centre for Astronomy with ESO (FINCA), FI-20014 University of Turku, Finland
    \and Department of Physics and Astronomy, FI-20014 University of Turku, Finland
    \and DIRAC Institute, Department of Astronomy, University of Washington, 3910 15th Avenue NE, Seattle, WA 98195, USA    
    \and Department of Astronomy, University of California, Berkeley, CA 94720
    \and Physics Division, Lawrence Berkeley National Laboratory, 1 Cyclotron Road, MS 50B-4206, Berkeley, CA 94720, USA
    \and Department of Astronomy, California Institute of Technology, 1200 E. California Blvd, Pasadena, CA, 91125, USA
             }

   \date{Received August 27, 2024; Accepted January 20, 2025}

  \abstract
{Hydrogen-rich superluminous supernovae (SLSNe II) are rare. The exact mechanism producing their extreme light curve peaks is not understood. Analysis of single events and small samples suggest that circumstellar material (CSM) interaction is the main mechanism responsible for the observed features. However, other mechanisms can not be discarded. Large sample analysis can provide clarification.
}
{We aim to characterize the light curves of a sample of 107 SLSNe~II to provide valuable information that can be used to validate theoretical models.
}
{We analyze the $gri$ light curves of SLSNe~II obtained through ZTF. We study peak absolute magnitudes and characteristic timescales. When possible we compute $g-r$ colors and pseudo-bolometric light curves, and estimate lower limits for their total radiated energy. We also study the luminosity distribution of our sample and estimate the fraction that would be observable by the LSST. Finally, we compare our sample to other H-rich SNe and to H-poor SLSNe~I.
}
{SLSNe~II are heterogeneous. Their median peak absolute magnitude is $\sim -20.3$ mag in optical bands. Their rise can take from $\sim$ two weeks to over three months, and their decline times range from $\sim$ twenty days to over a year. We found no significant correlations between peak magnitude and timescales. SLSNe~II tend to show fainter peaks, longer declines and redder colors than SLSNe~I. 
}
{We present the largest sample of SLSN~II light curves to date, comprising 107 events. Their diversity could be explained by 
different CSM morphologies, although theoretical analysis is needed to explore alternative scenarios. Other luminous transients, such as Active Galactic Nuclei, Tidal Disruption Events or SNe~Ia-CSM, can easily become contaminants. Thus, good multi-wavelength light curve coverage becomes paramount. LSST could miss $\sim$ 30\% of the ZTF events in its footprint in the $gri$ bands. 
}


   \maketitle
%

\section{Introduction}
\label{sec:intro}

Core-collapse supernovae (CCSNe) result from the explosive death of massive stars (M > 8~M$_{\odot}$). These have historically been classified based on their observed spectral features \citep{1997ARA&A..35..309F}. The lack or presence of hydrogen (H) in the spectra will result in a Type I or Type II classification respectively. Type II supernovae (SNe~II) represent the highest observed fraction of CCSNe \citep[e.g.][]{2020ApJ...904...35P}. SNe~II can be further divided into different subclasses depending on particular spectroscopic or photometric properties. Among the spectroscopic subclasses we can find Type IIb SNe (SNe~IIb), whose spectral sequence progressively shifts from being dominated by H lines to being dominated by helium (He) lines \citep{1993ApJ...415L.103F}; and Type IIn SNe (SNe~IIn), whose spectra show narrow H emission features \citep{1990MNRAS.244..269S}. Whereas among the photometric subclasses we can find Type IIP and IIL SNe (SNe~IIP and SNe~IIL respectively, hereafter collectively referred to as regular SNe~II), the former showing a light-curve ``plateu'' after peak and the latter declining linearly after peak \citep{1979A&A....72..287B}.
Furthermore, Luminous SNe (LSNe~II) show light curve peaks that are more luminous than those of regular SNe~II \citep{2023MNRAS.523.5315P}; and superluminous SNe (SLSNe~II) present extremely luminous light curves \citep[peaking at magnitudes $\lesssim -20$~mag in optical bands, although this limit is somewhat arbitrary, e.g.][and references therein]{2019ARA&A..57..305G}, that can not be explained with typical CCSN powering mechanisms. This work focuses on this last subclass of SLSNe~II events and aims at characterizing their light curves in order to better understand the physical mechanisms that power them.  

The fraction of SLSNe~II is among the 
lowest observed fractions of SNe \citep[e.g.][]{2020ApJ...904...35P} 
and so, the H-deficient SLSNe~I have historically been given more attention as they seem to be more numerous \citep{2019ARA&A..57..305G}, with the current number of classified SLSNe~I exceeding few hundred events \citep[e.g.][]{2023ApJ...943...41C,2024MNRAS.535..471G}. Studies of SLSNe~I allowed to constrain the possible powering mechanisms that may be driving their extreme luminosities. Four main mechanisms have been proposed as the most likely powering sources of SLSNe~I. Three of these consist of the thermalization of the energy produced by a process that can be either the spin-down of a magnetar; the accretion of fallback material into a black hole; or the interaction of the SN ejecta with surrounding circumstellar material (CSM). The fourth mechanism considers extremely massive progenitors (M $\sim$ 140--260~M$_{\odot}$) that undergo a thermonuclear explosion triggered by electron--positron pair production, producing events known as Pair Instability SNe (PISNe) that are powered by the radioactive decay of the large amounts of $^{56}$Ni synthesized by the explosion. These four mechanisms can be considered in stand alone models or can be combined to explain the unusual behaviour of SLSNe~I \citep[see][and references therein]{2017hsn..book..939K,2019ARA&A..57..305G}. 

In principle, the same powering mechanisms can be invoked to explain SLSNe~II. 
Studies of a few SLSNe~II (e.g: SN~2010jl, \citealt{2011ApJ...730...34S};  SN~2016aps, \citealt{2021ApJ...908...99S}; SN~2021adxl, \citealt{2023arXiv231213280B}), as well as small samples analysis (e.g. \citealt{2018MNRAS.475.1046I}, two SLSNe~II; and \citealt{2022MNRAS.516.1193K}, ten SLSNe~II) conclude that CSM interaction is probably the main driving mechanism for these events. The presence of CSM interaction becomes obvious when the spectra show narrow H lines \citep[e.g.][and references therein]{2017hsn..book..403S}, although the absence of such lines does not discard the presence of CSM interaction \citep[e.g.][]{2022MNRAS.516.1193K,2023MNRAS.523.5315P}. A SN with narrow H lines in the spectra will be typically classified as a SN~IIn. 
SNe~IIn tend to be luminous, with an average peak magnitude of $\sim -19$~mag \citep{2020A&A...637A..73N}, placing them at the edge of the SLSN class. It has long been debated whether SNe~IIn should also be considered as SLSNe~II (or SLSNe~IIn) when their light curve peak luminosities exceed that of classical events \citep[e.g.][]{2017hsn..book..431H}. One of the arguments of such a debate is whether the classification is connected to the physical processes that power the light curves or not. 

All massive stars experience mass loss either via steady winds, outbursts before the death of the star, or mass transfer in multiple systems \citep[e.g.][]{2008A&ARv..16..209P}. Thus, all SNe will show signs of interaction at some point in their evolution. The exact observational evidence of interaction will depend on the CSM morphology, density and extension \citep[e.g.][]{2011ApJ...729L...6C,2017hsn..book..843B,2017ApJ...838...28M,2023ApJ...952..119B,2024arXiv240504259D}, if the CSM is optically thin to electron scattering, narrow lines will be absent \citep[e.g.][and references therein]{2022A&A...660L...9D}. It has been argued that considering steady winds as the prevalent mass-loss mechanism would result in overestimated mass-loss rates, and it has been proposed that the mass loss should occur through eruptions shortly before explosion instead \citep[][]{2020MNRAS.492.5994B,2022MNRAS.517.1483D,2023A&A...677A.105D}. This scenario is supported by observed evidence of pre-explosion activity \citep[e.g.][]{2021ApJ...907...99S,2023ApJ...945..104T}. Eruptive mass loss is most commonly observed in Luminous Blue Variables \citep[LBVs, see][for a review on the characteristics of this broad class of stars]{2020Galax...8...20W}. Eruptions could also occur in very massive stars (M $\sim$ 70--260~M$_{\odot}$) due to pulsations driven by pair production instabilities, which will be energetic, but not enough to disrupt the whole star. The pulsations will continue until the mass of the star has been reduced enough to avoid producing pulsations, and the star will continue to evolve until it finishes its life as a CCSNe. These events are known as Pulsational Pair Instability SNe \citep[PPISNe, e.g.][]{2017ApJ...836..244W}. 

The uncertainties in the CSM distribution and degree of interaction contribution to the energy budget of H-rich SNe makes it difficult to create a full picture of the progenitor systems and explosion energetics of some of these events, particularly when pre-explosion images are not available. Among SLSNe~II there are two events considered to be prototypical: SN~2006gy \citep{2007ApJ...659L..13O,2007ApJ...666.1116S}, that shows persistent narrow lines in its spectral evolution and thus would be classified as a SLSN~IIn; and SN~2008es \citep{2009ApJ...690.1303M,2009ApJ...690.1313G}, that does not show persistent narrow spectroscopic emission lines. Regardless of the spectroscopic differences, the light curves of both events have been explained invoking CSM interaction. Although SN~2008es has also been suggested to be magnetar-powered \citep{2018MNRAS.475.1046I}, \cite{2019MNRAS.488.3783B} disfavored the magnetar model and favored CSM interaction based on the late-time light curve. It is still unclear whether interaction is able to account for the observed characteristics of all SLSNe~II when larger samples are considered.

In this work we analyze the light curves of a large sample of H-rich SLSNe~II, regardless of their particular spectral features, in order to see if there are distinct photometric characteristics that can point towards a common progenitor configuration and explosion mechanism.
This paper is organized as follows in Sect.~\ref{sec:sample} we describe the sample. Sect.~\ref{sec:analysis} presents the light curves analysis. In Sect.~\ref{sec:extev} we highlight events in the extremes of the analyzed parameter distributions and in Sect.~\ref{sec:comparisons} we present comparisons to other events. Sect.~\ref{sec:contaminants} describes possible contaminants of the sample. We discuss our findings in Sect.~\ref{sec:discussion} and conclude in Sect.~\ref{sec:conclusions}.

\section{SLSN~II sample description}
\label{sec:sample}

The sample presented in this paper was collected by the Zwicky Transient Facility \citep[ZTF,][]{2019PASP..131a8002B,2019PASP..131g8001G,2019PASP..131a8003M,2020PASP..132c8001D}. ZTF is a high-cadence (from minutes to days depending on the science case, with an average of three days for the public survey in $gr$ bands, \citealt{2019PASP..131f8003B}), wide-field (47-square-degree field of view) survey that covers the whole northern sky using the 48 inch aperture Samuel Oschin Telescope at the Palomar Observatory. In this work we include all spectroscopically classified Type II SNe whose peak brightness surpass the SLSN threshold, without making any spectroscopic distinction based on the presence or absence of narrow emission lines. We consider that if events with and without narrow lines are sufficiently photometrically distinct, we will observe multi-modality in the distributions of the considered features (see Sect.~\ref{sec:analysis}). To test this hypothesis, we decided to exclude events previously published by \cite{2022MNRAS.516.1193K} from our full sample and present comparisons of the photometric parameters of both samples in Sect.~\ref{sec:kangas} instead. Since the events published by \cite{2022MNRAS.516.1193K} were specifically selected due to the absence of narrow lines in their spectra, if a multi-modality existed between events with and without narrow lines, the light curve parameters of these events should map any possible multi-modality. To remain consistent with our classification scheme, we exclude the events in the sample of \cite{2022MNRAS.516.1193K} classified as SLSNe~I.5. We also exclude SN~2020yue as this event is a contaminant in their sample (see Sect.~\ref{sec:contaminants} for further discussion on contaminants) that has been re-classified as a Tidal Disruption Event (TDE) by \cite{2023ApJ...955L...6Y}.
We consider only three simple selection criteria to build our sample:
\begin{itemize}
    \item[\textbf{\Huge{.}}] The source must have at least one observed spectrum from which a classification as H-rich can be inferred. This means that we require the presence of Balmer lines in the spectra, but we do not discriminate between narrow and broad lines;
    \item[\textbf{\Huge{.}}] It should be possible to perform baseline correction to the forced photometry (see Sect.~\ref{sec:forced_phot_ztf}), in the time interval between March 17th 2018 (MJD $58194.0$) and December 12th 2022 (MJD $59925.0$)\footnote{The considered observation time range was arbitrarily selected to include events observed from the beginning of the survey up to the beginning of this work.};
    \item[\textbf{\Huge{.}}] At some point of the evolution, the rest frame absolute magnitudes (see Sect.~\ref{sec:abs_mag}) of each source must be $\leq -19.9$~mag in any of the ZTF $gri$ bands (see Sect.~\ref{sec:red}).
\end{itemize}
 All sources were selected from the \texttt{GROWTH} Marshal \citep{2019PASP..131c8003K} and \texttt{Fritz} platform \citep{2019JOSS....4.1247V,2023ApJS..267...31C}. Sources with no classification on these databases are not considered in this work. Transient classification depends both on the interest of the community for a given source and on availability of observing resources. Many of the sources included in this sample were deemed interesting by the ZTF SLSN working group based on properties such as long-lived light curves and/or small/faint hosts, etc. Subsequent efforts were invested in obtaining further classification for these events. Some sources were classified by other groups interested in potential SLSNe, and some sources by dedicated classification surveys. All the classification reports are presented in Table~\ref{tab:sample}. Eleven ambiguous events were found that show evidence that indicate they could be classified as either Active Galactic Nuclei \citep[AGN,][]{1963RvMP...35..947B} or Tidal Disruption Events \citep[TDEs,][]{1988Natur.333..523R} thus, they are excluded from the sample and discussed in Sect.~\ref{sec:contaminants}. The final sample includes a total of 107 SLSNe~II. The general characteristics of the sample are presented in Table~\ref{tab:sample}. 

\subsection{Classification as hydrogen rich}
\label{sec:class}

All events presented in this paper have at least one observed spectrum used to secure the Type II classification, this means that a H feature\footnote{Most classifications rely on the presence of H$\alpha$ in the spectra, except for SN~2020uaq that only shows H$\beta$ because the H$\alpha$ region falls outside of the wavelength range of the data in hand.} can be found in the available spectra. In each case, line identification is supported by spectral matching to the spectral template library of the Supernova Identification (SNID; \citealt{2007ApJ...666.1024B}) software. All classification spectra are publicly available on the Transient Name Server (TNS\footnote{\url{https://www.wis-tns.org/}.}).

While the majority of objects have spectra consistent with a classification as Type~IIn by SNID, a significant fraction ($\sim$ 26\%) have only (or mostly) low-resolution spectra from the Spectral Energy Distribution Machine \citep[SEDM,][]{2018PASP..130c5003B,2019A&A...627A.115R} or the Spectrograph for the Rapid Acquisition of Transients \citep[SPRAT,][]{2014SPIE.9147E..8HP} available, precluding any line profile analysis and further sub-classification beyond Type II.
Beyond classification, any further spectroscopic analysis is deferred to future work. Ambiguous cases and possible contaminants are discussed in Sect.~\ref{sec:contaminants}.

\subsection{Redshift determination and classification as SLSNe}
\label{sec:red}

The considered heliocentric redshifts ($z$) were obtained from spectral lines as $z = (\lambda - \lambda_{0}) / \lambda_{0}$, where $\lambda_{0}$ is H$\alpha$ rest wavelength and $\lambda$ the H$\alpha$ observed wavelength, obtained by fitting a Gaussian close to the centre of the emission line using the \texttt{lmfit} package \citep{2014zndo.....11813N}. If we have multiple spectra ($\sim$ 24\% of the sample has more than three observed spectra) we calculate $z$ for each spectrum and use each of these independently obtained $z$ to calculate the absolute magnitude (Sect.~\ref{sec:abs_mag}) of the corresponding object at peak (Sect.~\ref{sec:lcpeak}). The largest standard deviation obtained when doing this exercise corresponds to a variation of $0.1$~mag. Thus, we consider as SLSNe~II any hydrogen rich event that reaches magnitudes $\leq -19.9$~mag in any of the ZTF $gri$ bands. Any spectral analysis beyond classification and redshift determination is beyond the scope of this work.

\begin{figure}
   \includegraphics[width=8.5cm]{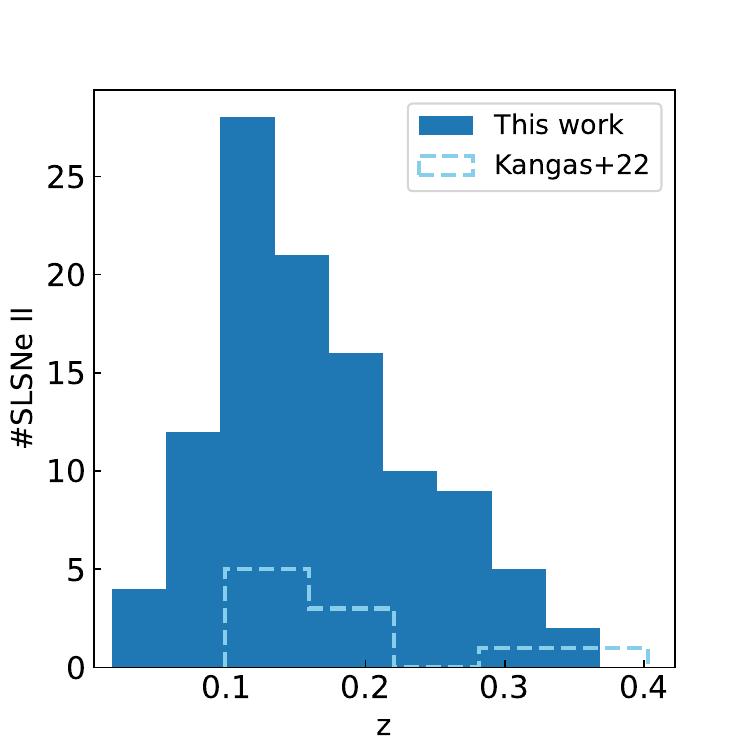}
      \caption{Redshift distribution of the 107 events in our ZTF SLSN~II sample in blue, filled regions. In skyblue dashed, empty regions, the redshift distribution of the 11 events in the ZTF SLSN~II sample presented by \cite{2022MNRAS.516.1193K}, see Sect.~\ref{sec:kangas} for a discussion.
              }
         \label{fig:z}
   \end{figure}

Fig.~\ref{fig:z} shows in blue the distribution of $z$ for our sample of 107 SLSNe~II and for the sample of 11 SLSNe~II presented by \cite{2022MNRAS.516.1193K} in skyblue (see Sect.~\ref{sec:kangas}). Except for SN~2021adxl ($z = 0.018$), all SLSNe~II are in the Hubble flow ($z > 0.02$), with the next closest event in our sample being SN~2022mma ($z = 0.038$).

\subsection{Photometry}
\label{sec:forced_phot}

The bulk analysis of this sample is performed using ZTF data. When possible and for comparison purposes, we also include public photometry from the Asteroid Terrestrial-impact Last Alert System \citep[ATLAS,][]{2018PASP..130f4505T,2020PASP..132h5002S}. 
The sample's average S/N among bands is similar with a mean value of 10. The average cadence of observation (calculated as the rate of observations in the observed rest-frame time range) is $\sim$ 3.6 days in $g$ and $r$ bands, $\sim$ 6 days in $i$ band, $\sim$ 7 days in $o$ band, and $\sim$ 16 days in $c$ band. Light curves with less than ten observed points are not considered for the analysis. . 
The light curves of each event are presented in Figs.~\ref{fig:alllcs1} to ~\ref{fig:alllcs5}. 

\subsubsection{ZTF}
\label{sec:forced_phot_ztf}

ZTF images can be found in the Public Data Releases\footnote{\url{https://irsa.ipac.caltech.edu/data/ZTF/docs/releases/ztf_release_notes_latest}}. Photometry is processed by the Science Data System at IPAC\footnote{Formerly referred to as the Infrared Processing \& Analysis Center (\url{https://www.ipac.caltech.edu}).} \citep{2019PASP..131a8003M}. The photometry for every transient is obtained by subtracting the observed science image from a reference image using \texttt{ZOGY} \citep{2016ApJ...830...27Z}. The reference image is generated by co-adding 15 to 40 high-quality historical images obtained with the same CCD quadrant and filter as the science. The main problem with obtaining photometry in this way is that the reference image can be contaminated with transient flux, this issue is particularly problematic for transients observed at the beginning of the survey when pre-transient images were scarce. In order to improve the quality of the light curves, we requested IPAC forced point-spread function (PSF) photometry, following the steps outlined in the ZTF forced photometry guideline\footnote{\url{http://web.ipac.caltech.edu/staff/fmasci/ztf/forcedphot.pdf}}. The forced photometry is delivered together with several quality flags. We considered them and removed photometric points associated to bad pixels, difference image cutouts off image or too close to the edge, and catastrophic errors. We also removed points observed with seeing $> 4\arcsec$. Following the ZTF forced photometry guideline, we also assess the associated \texttt{scisigpix} value to each photometric point. The \texttt{scisigpix} parameter is defined by ZTF as the robust sigma per pixel. To estimate the threshold of this parameter, we consider the median \texttt{scisigpix} of all the photometric observations taken with the same filter for all the retrieved sources, which resulted to be $\sim$ 23. Thus, we removed all phtometric points with \texttt{scisigpix} $>$ 23.

After applying all the quality cuts, we reprocess the forced photometry to correct for possible offsets produced either by transient contamination in the reference images, or by observations of the same source utilizing different quadrants or chips of the camera. 
If present, these offsets will be constant, and can be corrected by subtracting the median stationary signal of the transient with respect to the reference image, also referred to as the ``baseline''. To find the baseline we first combine the flux observations in one day bins. We then get a rough estimation of the light curve's peak epoch by applying a Savitzky-Golay filter \citep{1964AnaCh..36.1627S} using the \texttt{savgol\_filter} module on the \texttt{SciPy} package \citep{2020SciPy-NMeth}, and calculating the maximum of the resulting filtered light curve. SLSNe can have very long rise times and longer declines from peak than fainter events, we consider that images taken prior to six months (180~d) or later than three years (1095~d) from the estimated peak should not contain transient light.
These time ranges are loosely based on previous SLSN studies \citep[e.g.][]{2017hsn..book..431H,2019ARA&A..57..305G}. If there are no observations in the considered range, the time range for the baseline is chosen arbitrarily by hand, making sure that the observed flux in the selected region is consistent with an straight line, indicating that the transient is no yet present in the observations. There are a few cases for which a baseline can be defined both before and after peak. Nonetheless, we typically consider the baseline after peak for events that occurred before 2019, and the baseline before peak for events that occurred after 2019. We only consider baselines with at least 20 observed photometric points in the selected time range. If a baseline has less that 20 photometric points, we deemed the correction impossible and discard the observations associated to the corresponding field in the corresponding CCD chip. Baseline correction is performed following \cite{2021ApJ...907...99S}. This is, we iterate over the photometric points removing those that are further away from the median baseline flux until 20\% of the points are left. The computed baseline median is then subtracted from the overall flux of the transient. The baseline corrected flux is then converted to AB magnitudes. The associated error bars were calculated through the Computer Calculation of Uncertainties method \citep{2003drea.book.....B}, this method provides asymmetric error bars, we adopt as error bar the absolute magnitude of the larger associated uncertainty. If a random photometric point appeared to deviate from the general shape on the light curve, we inspected the IPAC images visually to check if the transient is present or if an artifact is introducing a spurious detection. In case of the latter, the point was removed from the light curve. 

\subsubsection{ATLAS}
\label{sec:ATLAS}

The ATLAS survey scans the sky with a two day cadence \citep{2020PASP..132h5002S,2021TNSAN...7....1S}. ATLAS observes in two wide filters $c$ (or ``cyan'' band, that covers the wavelength range 4200–6500~\AA, roughly corresponding to the $g + r$ range) and $o$ (or ``orange'' band, that covers the wavelength range of 5600–8200~\AA, roughly corresponding to the $r + i$ range). We retrieved the ATLAS photometry from the forced-photometry server\footnote{\url{https://fallingstar-data.com/forcedphot/}} and processed the output utilizing the pipeline developed by \cite{Young_plot_atlas_fp}. ATLAS photometry has less associated quality flags than ZTF, and ATLAS filters are wider than the ZTF ones. Therefore we mostly use the ATLAS photometry as check of the taxonomic description of the light curves and do not perform any analysis on it.

\subsubsection{Swift}
\label{sec:swift}

Fourteen objects in our sample were observed with the UV/optical Telescope \citep[UVOT,][]{2005SSRv..120...95R} aboard the Neil Gehrels Swift Observatory \citep{2004ApJ...611.1005G}. We retrieve the level-2 data from UK Swift Data Archive\footnote{\url{https://www.swift.ac.uk/swift_portal/}}. For each object, we co-added all sky exposures for a given epoch and filter to boost the S/N using \texttt{uvotimsum} in HEAsoft\footnote{\url{https://heasarc.gsfc.nasa.gov/docs/software/heasoft} version 6.32.2.}. Afterwards, we measured the brightness of the event with the Swift tool \texttt{uvotsource}. The source aperture had a radius of $5''$, while the background region had a radius of $30''$. All measurements were calibrated with the calibration files from November 2021 and converted to the AB system following \cite{2011AIPC.1358..373B}.

\subsection{Absolute magnitudes}
\label{sec:abs_mag}

\begin{figure}
   \includegraphics[width=8.5cm]{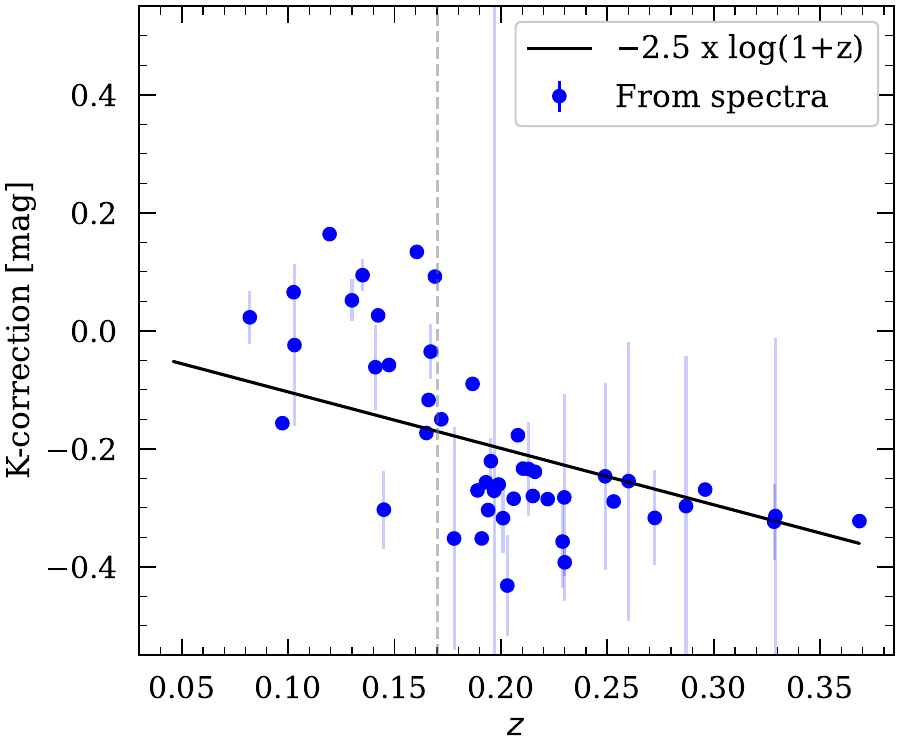}
      \caption{K-correction approximation obtained as $-2.5\times\log(1+z)$ (black line), compared to full K-corrections obtained using \texttt{SNooPy} (blue dots). The inverse of the S/N of the spectrum considered to calculate full K-corrections is presented as associated error bars. The grey vertical dashed line indicates $z = 0.17$, at which we switch from $g$ to $r$ band to calculate the correction.
              }
         \label{fig:Kcorr}
   \end{figure}

To obtain absolute magnitudes we first use the corresponding heliocentric $z$ of each SLSN~II (see Sect.~\ref{sec:red}) to compute their distance modulus ($\mu$). We employ the \texttt{astropy.cosmology} software and adopt the NASA/IPAC Extragalactic Database's (NED\footnote{The NASA/IPAC Extragalactic Database (NED) is funded by the National Aeronautics and Space Administration and operated by the California Institute of Technology.}) canonical cosmological parameters (H$_{0} = 73$ km s$^{-1}$ Mpc$^{-1}$, $\Omega _{\mathrm{Matter}} = 0.27$, $\Omega _{\mathrm{Lambda}} = 0.73$). We do not consider uncertainties due to the host galaxy peculiar velocities as all our SLSNe~II except SN~2021adxl are in the Hubble flow ($z > 0.02$, see Sect.~\ref{sec:red}) and such uncertainties will be small. In the case of SN~2021adxl we use the distance modulus presented by \cite{2023arXiv231213280B}. Milky Way extinction is calculated using NED's Galactic Extinction Calculator\footnote{NED's Extinction Calculator considers the recalibration presented by \citep{2011ApJ...737..103S} to the extinction map presented by \citep{1998ApJ...500..525S}, assuming a \citep{1999PASP..111...63F} reddening law with R$_{\mathrm{v}} = 3.1$.}, accessed through the \texttt{ned\_extinction\_calc} script\footnote{\url{https://github.com/mmechtley/ned_extinction_calc}}. We do not consider host extinction. For all the SLSNe~II in the sample, we adopt the cosmological term for the K-correction \citep{2002astro.ph.10394H} as $-2.5\times\log(1+z)$, as presented in \cite{2023ApJ...943...41C}. This is because the spectral coverage of our sample is rather poor. To investigate the uncertainties introduced by adopting this approximation, we compute full K-corrections for those events with available spectra within 30~days of the $r$-band peak (as this is the best observed band). To do this, we use the SuperNovae in Object Oriented Python \citep[\texttt{SNooPy,}][]{2011AJ....141...19B} software. Each considered spectrum was corrected for Milky Way extinction. In addition, following \cite{2023ApJ...943...41C}, the full K-corrections consider $g$ band for events at $z \leq 0.17$ and $r$ band for events at $z > 0.17$. After obtaining the full K-corrections, we compared them to the adopted approximation. The comparison can be seen in Fig.~\ref{fig:Kcorr}. In this figure, the inverse of the S/N of each considered spectrum is indicated as associated error bars so, larger error bars indicate lower S/N. This is not the error associated to the K-correction but serves as an indication of the quality of the considered spectrum. We note that the adopted approximation for K-corrections follows the behavior of the full K-correction, with a dispersion of 0.2~mag. We conclude that the adopted K-correction approximations is a good approximation and consider it when computing absolute magnitudes. Absolute magnitudes are then calculated in the AB system as $\mathrm{M}_{\mathrm{\lambda}} = \mathrm{m}_{\mathrm{\lambda}} - \mathrm{\mu} - \mathrm{A}_{\mathrm{\lambda}} - \mathrm{K_{corr}}$, where $\mathrm{\mu}$ is the calculated distance modulus, $\mathrm{A}_{\mathrm{\lambda}}$ is the Milky Way extinction in the corresponding wavelength and $\mathrm{K_{corr}}$ is the K-correction approximation. 

\section{Analysis}
\label{sec:analysis}

In this section we describe the analysis methods and present a general description of the studied sample. Most of the light curve parameters were estimated using light curve interpolation in either flux or magnitude space. We use two main methods of interpolation, Gaussian Process \citep[GP, e.g.][]{2006gpml.book.....R}; and locally estimated scatterplot smoothing \citep[LOESS,][]{1992sms..book.....C}, considering a second order polynomial. The former was implemented using the \texttt{GPy} package\footnote{\url{https://gpy.readthedocs.io/en/deploy/}.} and the latter using the Automated Loess Regression (\texttt{ALR}) pipeline presented by \cite{2019MNRAS.483.5459R}. Similarly to GP, the method implemented by \texttt{ALR} is non-parametric and allows to compute confidence regions. If \texttt{ALR} can not fit a second order polynomial, it will default to do the interpolation considering a first order polynomial. \texttt{ALR} is less sensitive to big data gaps than GP, allowing for a better representation of our light curves at later times. Still, artificial wiggles can be seen in the interpolation when the data is too sparse, but the effect is not as strong as that observed when using GP\footnote{It has become popular to interpolate SN light curves using multiband light curve information \citep[e.g.][]{2019AJ....158..257B}, here we avoid such techniques as they assume the color evolution of the event, which we do not know a priori for our sample of SLSNe~II.}.
   
\subsection{Light curve peak estimation}
\label{sec:lcpeak}

\begin{figure}
   \includegraphics[width=8.5cm]{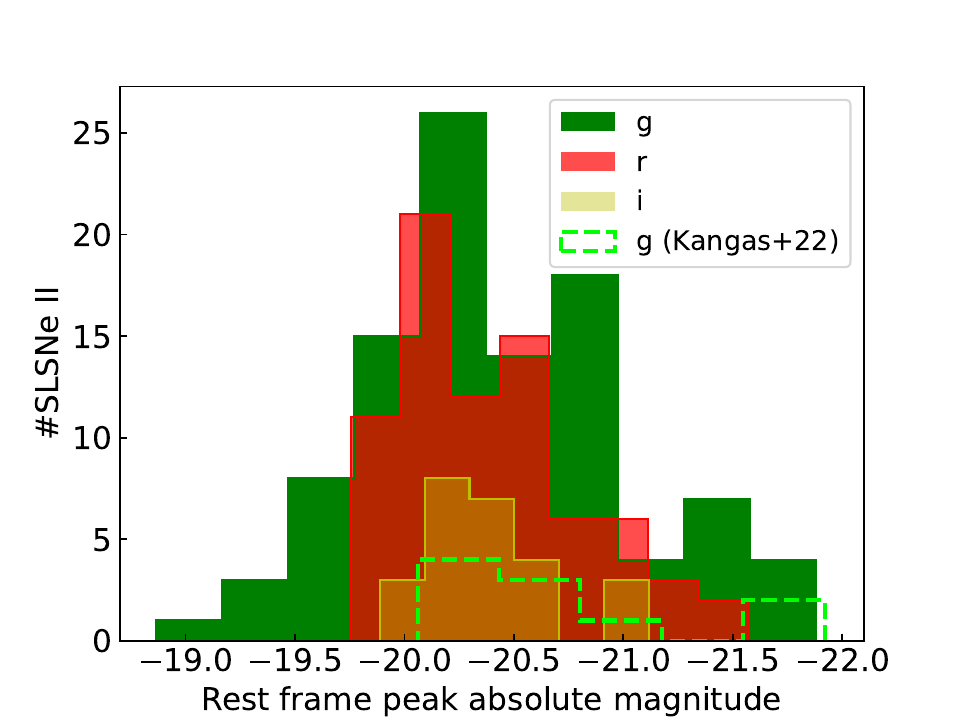}
      \caption{Rest frame peak absolute magnitude distribution of SLSNe~II in $gri$ bands in filled steps. Empty, dashed lime steps show the distribution of rest frame $g$ band peak absolute magnitude of the ZTF SLSN~II sample presented by \cite{2022MNRAS.516.1193K}, see Sect.~\ref{sec:kangas} for a discussion.
              }
         \label{fig:peakabsmag}
   \end{figure}

SLSNe are classified based on their peak absolute magnitude, thus this is their most important characterization parameter. To estimate the peak epoch and absolute magnitude in each photometric band, we interpolate the light curves in flux space through GP (see Sect.~\ref{sec:analysis}). We use a Monte Carlo approach, drawing 800 samples from the posterior distribution of each GP interpolation and calculating the peak epoch for every sample. The median of the resulting distribution is considered to be the associated peak epoch of an event, the 15.9\%ile and 84.1\%ile of the distribution is taken as the associated lower and upper asymmetric error bars respectively. The peak absolute magnitude is obtained from the interpolation at peak epoch, the associated asymmetric error bars are obtained from the interpolation's confidence interval at the 15.9\%ile and 84.1\%ile. The distribution of peak absolute magnitude  in $gri$ bands is presented in Fig.~\ref{fig:peakabsmag}. All the three distributions present a median absolute magnitude of $\sim$ 20.3~mag, consistent with our selection criteria. Given that we consider superluminous any event that presents a magnitude $\leq -19.9$~mag at some point of the evolution in any of the ZTF $gri$ bands, we can conclude that events in the left hand side of $g$ band peak absolute magnitude distribution are either intrinsically redder or suffering from non-negligible host extinction. Further spectral analysis is needed in order to properly quantify host extinction; such analysis is out of the scope of this paper. The $g$ band distribution also extends to the higher values of peak absolute magnitude. Events in the right hand side of $g$ band peak absolute magnitude distribution are mainly located at larger $z$. The median value and dispersion of the rest frame peak absolute magnitude in $gri$ bands are presented in Table~\ref{table:peakAbsmagANDrisedecay}. 

\subsection{Rise and decline times}
\label{sec:riseanddecline}

\begin{figure*}
   \centering
   \includegraphics[width=19cm]{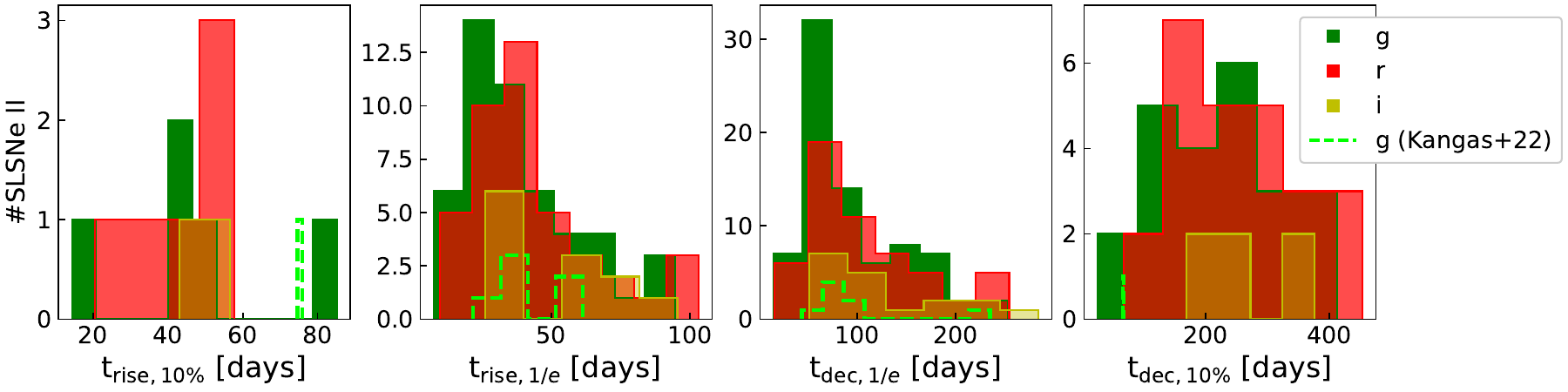}
      \caption{$gri$ band (green, red and yellow respectively) distribution of t$_{\mathrm{rise,10\%}}$ (first panel), t$_{\mathrm{rise,1/e}}$ (second panel), t$_{\mathrm{dec,1/e}}$ (third panel) and t$_{\mathrm{dec,10\%}}$ (last panel). Empty, dashed lime steps show the corresponding parameter distribution of the ZTF SLSN~II sample presented by \cite{2022MNRAS.516.1193K}, see Sect.~\ref{sec:kangas} for a discussion.
              }
         \label{fig:risedecaydist}
   \end{figure*}

Although GP interpolation is useful to calculate the peak epoch and associated error bar, it is not ideal to use as a representation of the full light curves. This is because GP is sensitive to gaps and we do not put any constraints on the observed cadence of the light curves. In general, our SLSN~II light curves have densely sampled peaks and sparsely sampled declines that can present big gaps. ALR interpolation (see Sect.~\ref{sec:analysis}) is used to characterize the timescales of the sample. We follow the work of \cite{2023ApJ...943...41C} and consider the t$_{\mathrm{rise/dec,x}}$ parameters to describe the rise and decline times of our SLSN~II light curves. These represent the time intervals between peak and an $x$ fraction of the flux. In particular, we consider the same fractions as \cite{2023ApJ...943...41C}, which are 10\% ($\Delta$mag $=$ 2.5) and 1/e ($\Delta$mag $=$ 1.09). In each case, the associated error bar corresponds to the peak epoch error bar.
In Fig.~\ref{fig:risedecaydist} we show the distribution of t$_{\mathrm{rise/dec,x}}$ for the $gri$ bands of our full sample. Note that we can only measure t$_{\mathrm{rise,10\%}}$ for a handful of events, mainly due to the lack of data at very early times. We do not see any multi-modality in the rise and decline time distributions. To confirm this, we use the \texttt{gaussian\_kde} module available in the \texttt{SciPy} package \citep{2020NatMe..17..261V} to calculate the probability density function of the distributions via kernel density estimation (KDE), using a Silverman bandwidth. The resulting KDE for each time parameter 
supports the lack of multiplicity in the sample. 
In Table~\ref{table:peakAbsmagANDrisedecay} we include the median value and dispersion for each calculated time parameter in each band. 

We note that the values of t$_{\mathrm{rise,1/e}}$ are shorter for bluer bands, with the median value in $g$ band being $\sim$ 3 rest frame days shorter than that in $r$ band, and the median $r$ band rise time being $\sim$ 10 rest frame days shorter than that in $i$ band. 
Bluer light curves not only rise faster but they also decline faster. t$_{\mathrm{dec,10\%}}$ and t$_{\mathrm{dec,1/e}}$ are $\sim$ 13 rest frame days shorter for $g$ band than for $r$ band. The faster evolution of the $g$ band light curves can also be seen in the timescale distributions presented in Fig.~\ref{fig:risedecaydist}. This is followed by the evolution in the $r$ band and the $i$ band seems to be the slowest evolving one. 
This hints towards the potential of redder bands to be used in the search and follow up of SLSNe~II.

\begin{figure*}
   \centering
   \includegraphics[width=19cm]{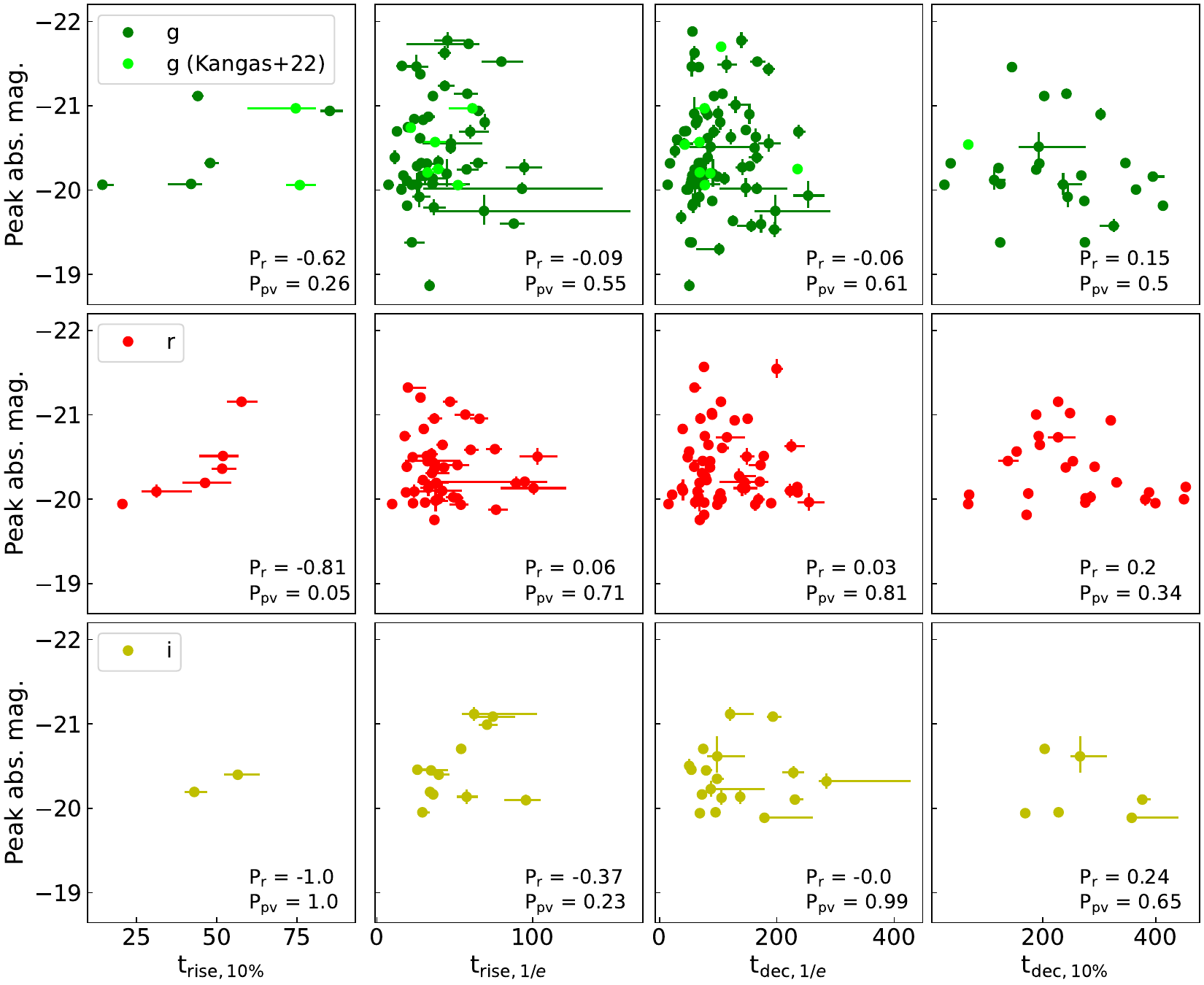}
      \caption{From top to bottom: $gri$ band (respectively) rest frame peak absolute magnitude compared to t$_{\mathrm{rise,10\%}}$ (first columns), t$_{\mathrm{rise,1/e}}$ (second columns), t$_{\mathrm{dec,1/e}}$ (third column) and t$_{\mathrm{dec,10\%}}$ (last column). The corresponding Pearson's r parameter (P$_{\mathrm{r}}$) and associated p-value (P$_{\mathrm{pv}}$) for each distribution is annotated in the bottom right of the corresponding panels, no significant correlation is found for these parameters. Lime dots in the first row show the corresponding parameter distribution of the ZTF SLSN~II sample presented by \cite{2022MNRAS.516.1193K}, see Sect.~\ref{sec:kangas} for a discussion. These additional events are not included in the correlation analysis.
      }
         \label{fig:timesVSpeak}
   \end{figure*}

We investigate correlations between the rest frame rise and decline times and rest frame peak absolute magnitude  in $gri$ bands. Fig.~\ref{fig:timesVSpeak} shows the respective scatter plots. The first column of Fig.~\ref{fig:timesVSpeak} shows t$_{\mathrm{rise,10\%}}$ against peak absolute magnitude, the scatter plot seems to point towards a correlation between these parameters, where longer rising events show brighter peaks. However, such a conclusion would suffer from low number statistics. In addition, we do not see the same trends in the second column of Fig.~\ref{fig:timesVSpeak} that shows t$_{\mathrm{rise,1/e}}$ against peak absolute magnitude  and considers a larger number of events. No correlations are seen for the decline times versus rest frame peak absolute magnitude either. To confirm the lack of correlations we use the \texttt{scipy.stats.pearsonr} package to calculate the Pearson's r parameter (P$_{\mathrm{r}}$) and its associated p-value  (P$_{\mathrm{pv}}$) for all the distributions. As anticipated, no significant correlation is found. The Pearson's parameters are annotated in the corresponding panels of Fig.~\ref{fig:timesVSpeak}. The time parameters for each event in each observed photometric band are detailed in Table~\ref{table:params}.

\begin{figure*}
   \centering
   \includegraphics[width=19cm]{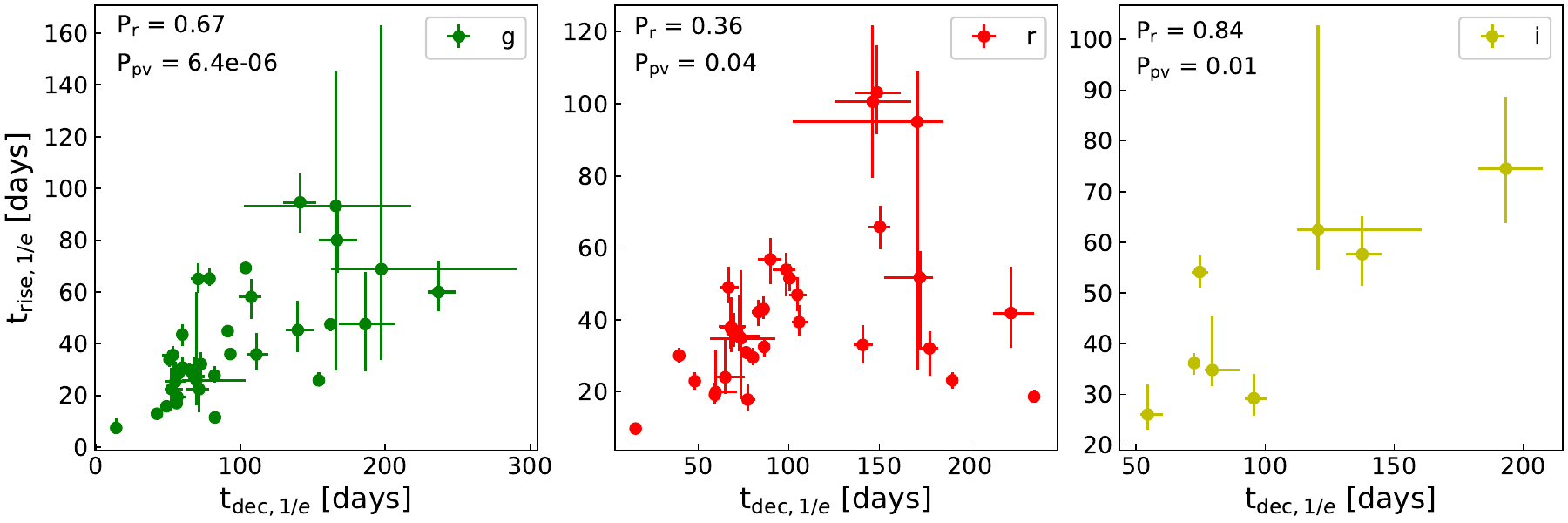}
      \caption{$gri$ (left to right panels respectively)  t$_{\mathrm{dec,1/e}}$ versus t$_{\mathrm{rise,1/e}}$. The corresponding Pearson's r parameter (P$_{\mathrm{r}}$) and associated p-value (P$_{\mathrm{pv}}$) for each distribution is annotated in the top left of the corresponding panels. Strong correlations are found for the $g$ and $i$ bands.
            }
         \label{fig:riseVSdecay}
   \end{figure*}

Since t$_{\mathrm{dec,1/e}}$ and t$_{\mathrm{rise,1/e}}$ are the best sampled timescales, we investigate possible correlations among them. We present the respective scatter plots in Fig.~\ref{fig:riseVSdecay}. We see clear trends that show that fast risers decline faster and viceversa. the Pearson's parameters show a strong correlation in the $g$ and $i$ bands, however this is not the case in the $r$ band. This lack of correlation in the $r$ band is driven by six events (SN~2018bwr, SN~2019npx, SN~2019aafk, SN~2020jgv, SN~2020yrn and SN~2021yyy) that decline slow compared to their rise.

\begin{table*}
\caption{Median rest frame peak absolute magnitude of the SLSN~II sample, dimmest and brightest rest frame peak absolute magnitudes, and median rest frame rise and decline times per ZTF filter. The errors associated to the median values correspond to the 15.9\%ile and 84.1\%ile respectively.}              
\label{table:peakAbsmagANDrisedecay}      
\centering                          
\begin{tabular}{l c c c c c c c}        
\hline\hline                
\noalign{\vskip 1mm}
Band &  $\widetilde{\mathrm{Peak}}$  & Faintest        & Brightest & $\widetilde{\mathrm{t}_{\mathrm{rise,10\%}}}$ & $\widetilde{\mathrm{t}_{\mathrm{rise,1/e}}}$ &  $\widetilde{\mathrm{t}_{\mathrm{dec,1/e}}}$ & $\widetilde{\mathrm{t}_{\mathrm{dec,10\%}}}$  \\    
     &  [mag.]             & [mag.]         &  [mag.]   & [days]               & [days]              &  [days]                & [days]               \\    			
\noalign{\vskip 1mm}
\hline 	
\noalign{\vskip 1mm}    
   g  & -20.3$_{-0.5}^{+0.6}$ & -18.9 $\pm$ 0.1 &-21.9 $\pm$ 0.1 & 44.1$_{-12.2}^{+17.4}$ & 33.8$_{-14.1}^{+28.0}$ & 75.0$_{-21.3}^{+81.0}$  & 235.8$_{-113.1}^{+100.1}$ \\		
   \noalign{\vskip 1mm}
   \hline 
   \noalign{\vskip 1mm}
   r  & -20.4$_{-0.4}^{+0.5}$ & -19.8 $\pm$ 0.1 &-21.6 $\pm$ 0.1 & 49.0$_{-20.0}^{+4.1}$  & 37.0$_{-13.3}^{+26.0}$ & 89.3$_{-28.8}^{+82.2}$  & 248.1$_{-80.4}^{+134.7}$  \\     
   \noalign{\vskip 1mm}
   \hline 
   \noalign{\vskip 1mm}
   i  & -20.3$_{-0.2}^{+0.3}$ & -19.9 $\pm$ 0.1 &-21.1 $\pm$ 0.1  & 49.8$_{-4.6}^{+4.6}$   & 46.9$_{-14.0}^{+24.7}$ & 98.3$_{-26.9}^{+105.3}$ & 246.8$_{-50.8}^{+114.6}$  \\
   \noalign{\vskip 1mm}
\hline 
\end{tabular}
\end{table*}

\subsection{Colors}
\label{sec:color}

\begin{figure*}
   \includegraphics[width=19cm]{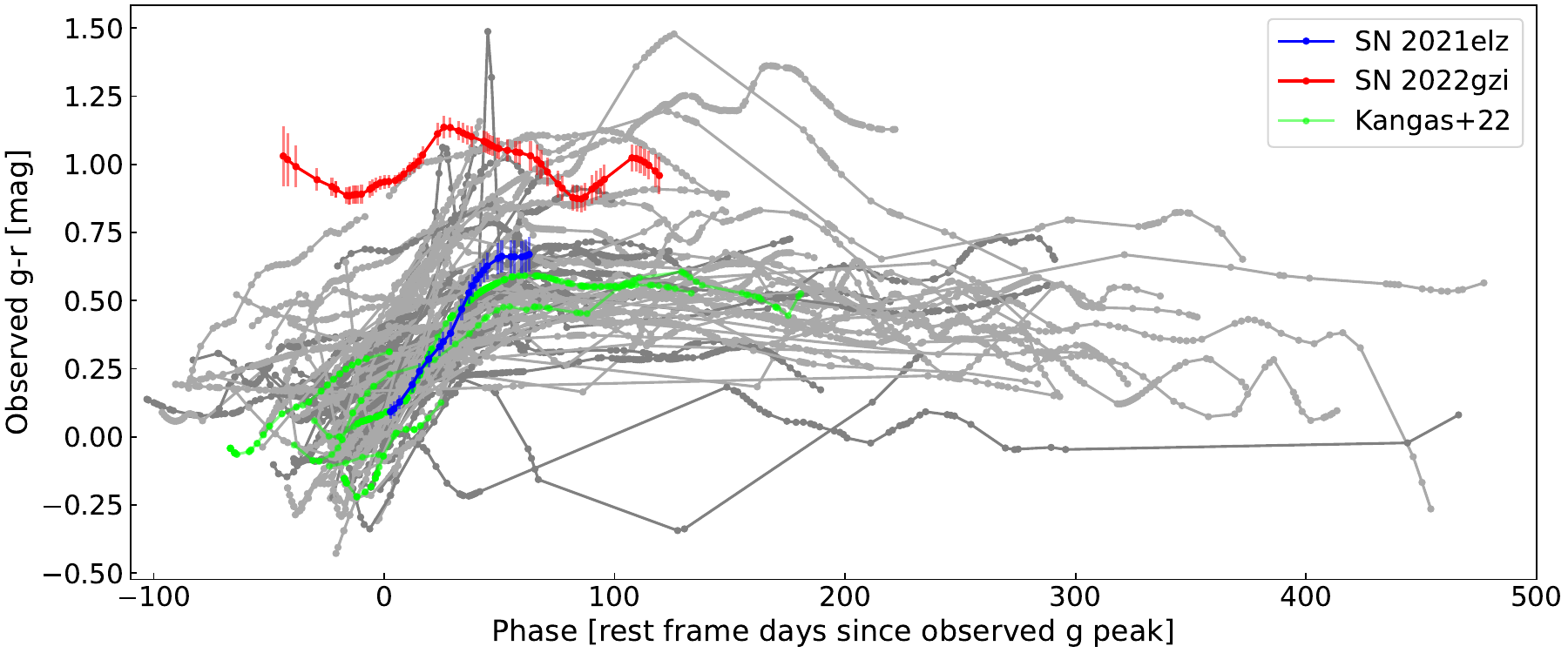}
      \caption{Observed $g-r$ colors with respect to rest frame days since $g$ band peak. In dark grey we show events at $z < 0.17$ and in light grey events at $z \geq 0.17$. In red we highlight the event with the faintest rest frame peak absolute  magnitude in $g$ band. SN~2022gzi is the reddest event at peak, indicating that it may be suffering from considerable host extinction. In blue we highlight the event with the brightest rest frame peak absolute  magnitude in $g$ band, the color evolution of this SLSN~II is consistent with the general trend of the sample. In lime we show the observed $g-r$ colors of the events with $z \leq 0.17$ in the ZTF SLSN~II sample presented by \cite{2022MNRAS.516.1193K}, see Sect.~\ref{sec:kangas} for a discussion. 
      }
         \label{fig:grcol}
   \end{figure*}

In Fig.~\ref{fig:grcol} we show observed $g-r$ colors with respect to rest frame days since observed $g$ band peak, calculated considering the $g$ and $r$ band LOESS interpolations (see Sect.~\ref{sec:analysis}). 
We see a similar behaviour for the whole sample, this is somewhat blue colors at early times that become red as the event evolves. Although these represent observed colors, the observed behaviour is similar regardless of the object's $z$.
After $\sim$ 100 rest frame days the color evolution of most SLSNe~II stalls, becoming almost constant. We note that these late phases are generally poorly sampled and better data is needed to accurately make any claim about the color behavior at these times. Fig.~\ref{fig:grcol} shows SN~2022gzi highlighted in red. If only the $g$ band light curve is considered, this event would not be classified as a SLSNe~II but as a SN~IIn however, it reaches SLSN~II peak luminosities in the redder bands. Thus, we conclude that this event is suffering from considerable host extinction. Still, given the shallower color evolution of this event compared to the rest of the sample, we consider the possibility of SN~2022gzi being a contaminant in Sect.~\ref{sec:contaminants}. SN~2021elz is also highlighted in blue in Fig.~\ref{fig:grcol}, this is the brightest event of the sample (see Sect.~\ref{sec:brightdimm}), we see that the overall color evolution of this events follows the general trend of the whole sample.

\begin{figure}
   \includegraphics[width=8.5cm]{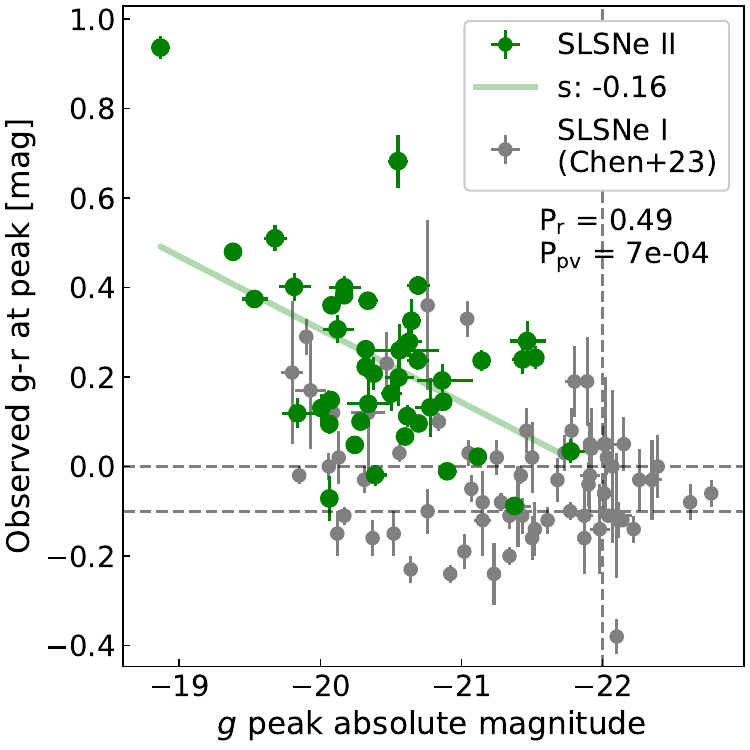}
      \caption{Observed $g-r$ colors at around $g$ band peak versus rest frame $g$ band peak absolute magnitude of our SLSN~II (green) and the sample of SLSN~I (grey) presented by \cite{2023ApJ...943...41C}. In green we include a straight line fit to the scatter plot, the slope (s) is indicated in the label of the figure. We also annotate the distribution's Pearson's r parameter and associated p-value. Grey dashed horizontal lines show the interval between the bluest color of the SLSN~II and color $= 0.0$ mag. A grey dashed vertical line indicates peak absolute magnitude $= -22$~mag.
      }
         \label{fig:grcol_peak}
   \end{figure}

In Fig.~\ref{fig:grcol_peak} we show the observed $g-r$ color at $g$ band peak against rest frame $g$ band peak absolute magnitude, SN~2022gzi is the reddest event ($g-r = 0.9$~mag) at the top left corner of the plot. We can see that fainter events seem to be redder than brighter ones. We calculate the Pearson's r parameter and associated p-value for the distribution using the \texttt{scipy.stats.pearsonr} package, and find a correlation of P$_{\mathrm{r}} = 0.49$ with an associated p-value P$_{\mathrm{pv}} = 7 \times 10^{-04}$, that we illustrate by fitting a straight line to the points (shown in light green in  Fig.~\ref{fig:grcol_peak}). The plot also shows that SLSNe~II are redder than SLSNe~I at similar phases. This is consistent with the findings of \cite{2023ApJ...943...41C} that suggest that brighter events show bluer colors at peak. Further analysis to investigate whether SLSN~II are intrinsically redder or this effect is linked to intrinsic host extinction, or is associated to dust production, or produced by CSM, is out of the scope of this work but .

\subsection{SLSN~II energetics}
\label{sec:energy}

To constrain the total radiated energy of the SLSNe~II in our sample we construct pseudo-bolometric light curves. We only consider events with an available peak date and ten or more photometric observations in all three $gri$ bands, these include 39 SLSNe~II. Two approaches are considered following the procedures presented by \cite{2024A&A...685A..20G} who considers the methods presented by \cite{2014MNRAS.437.3848L}. The first approach is to simply integrate the observed spectral energy distribution (SED), to do this we interpolate the $gri$ bands with respect to the phase of $r$ band observations using \texttt{ALR} (see Sect.~\ref{sec:analysis}) and integrate over the resulting interpolated curves. This approach gives a lower limit for the bolometric luminosity completely ignoring both the UV and NIR contributions to the SED. Given that at early times, before photons start diffusing, SNe present high temperatures, a large fraction of the early energy is emitted in the UV \citep[see for example][]{2017A&A...605A..83D}.
As SNe evolve and cool down, the contribution at longer wavelengths becomes more important and so, it is crucial to account for the NIR emission. Therefore, our second approach is to also consider extrapolations to both the UV and NIR by fitting a black body to consider the missing flux. To account for the UV flux, we fit a
black body to the available optical data and integrate it from 0~\AA\ to our $g$ band. In a similar manner, to account for the NIR flux, we integrate the black body fit to the available optical data from our $i$ band to 25000~\AA.\ Black body approximation is inaccurate when line emission becomes dominant over the SLSN radiation, this effect will become more important at later times and so we restrict our approximation to phases $< 400$ days post peak, although further multi-wavelength analysis is needed to fully understand the limits of our approach. Given the limits of our data, such analysis is out of the scope of this work.

\begin{figure*}
   \includegraphics[width=19cm]{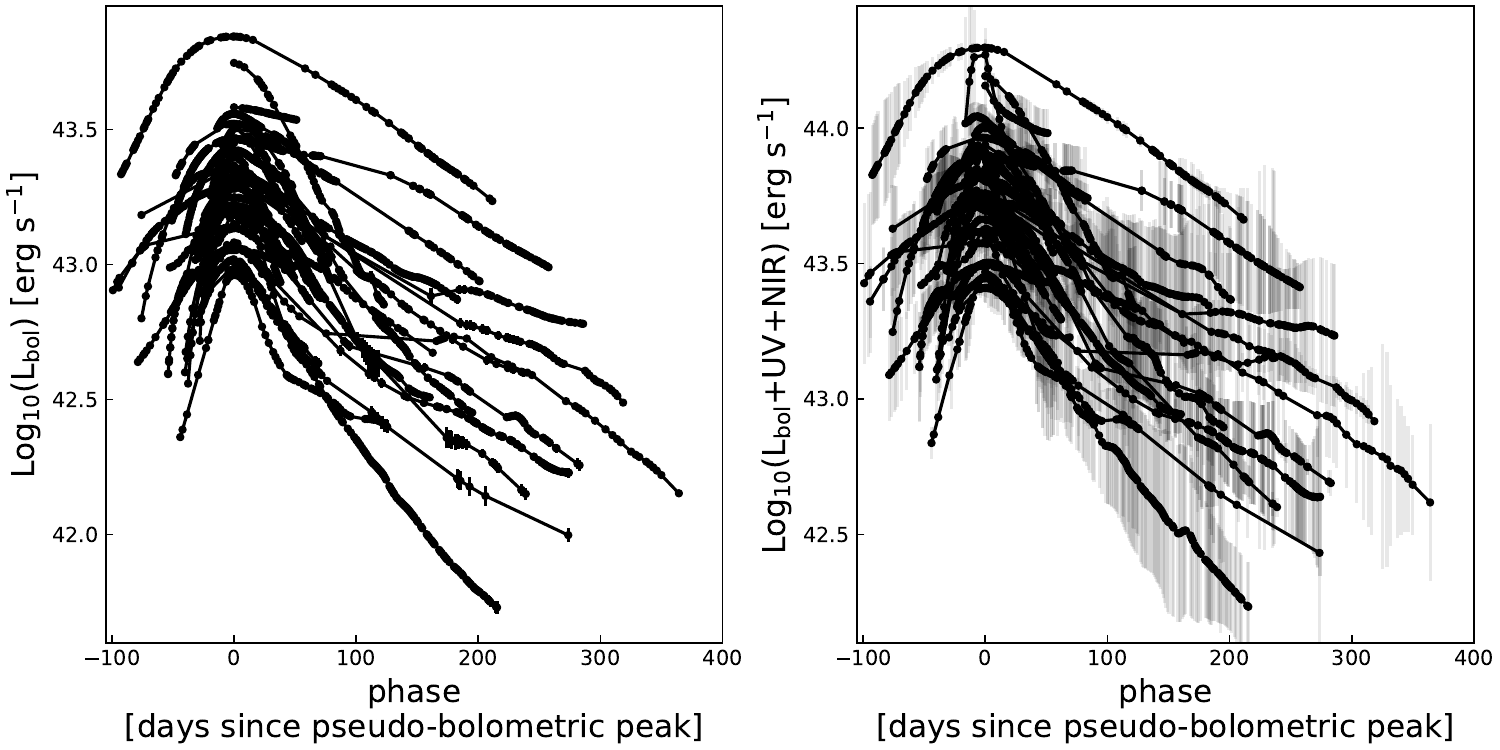}
      \caption{Pseudo-bolometric light curves of the SLSN~II sample. Left panel shows the results of integrating the observed SED considering the epochs for which we have all three $gri$ photometric observations. Right panel shows the results of adding extrapolations to both the UV and NIR to the SED obtained from the $gri$ photometric observations.
              }
         \label{fig:bols}
   \end{figure*}

In Fig.~\ref{fig:bols} we show the obtained pseudo-bolometric light curves from our first and second approach in the left and right panels respectively. The median difference between the peak luminosities of the pseudo-bolometric light curves obtained through the two methods is $\sim$ 0.5~dex and can be as high as $\sim$ 0.75~dex. We note that the error bars associated to our second approach (presented in grey color in the right panel of Fig.~\ref{fig:bols}) are quite large due to the addition of artificial UV and NIR flux. At later phases, line blanketing becomes important and the black body approximation is no longer adequate. We highlight the importance of obtaining UV photometry to better estimate SLSN~II energetics. \cite{2022A&A...660A..40M} showed that in the case of SNe~II, NIR observations are essential when line blanketing starts to affect bluer bands, since the black body approximation starts peaking at redder bands. Unfortunately, we have very few photometric bands to make a comparison with their method. Hence, we also highlight the importance of obtaining NIR photometry to estimate SLSNe~II energetics.

\begin{figure*}
   \includegraphics[width=19cm]{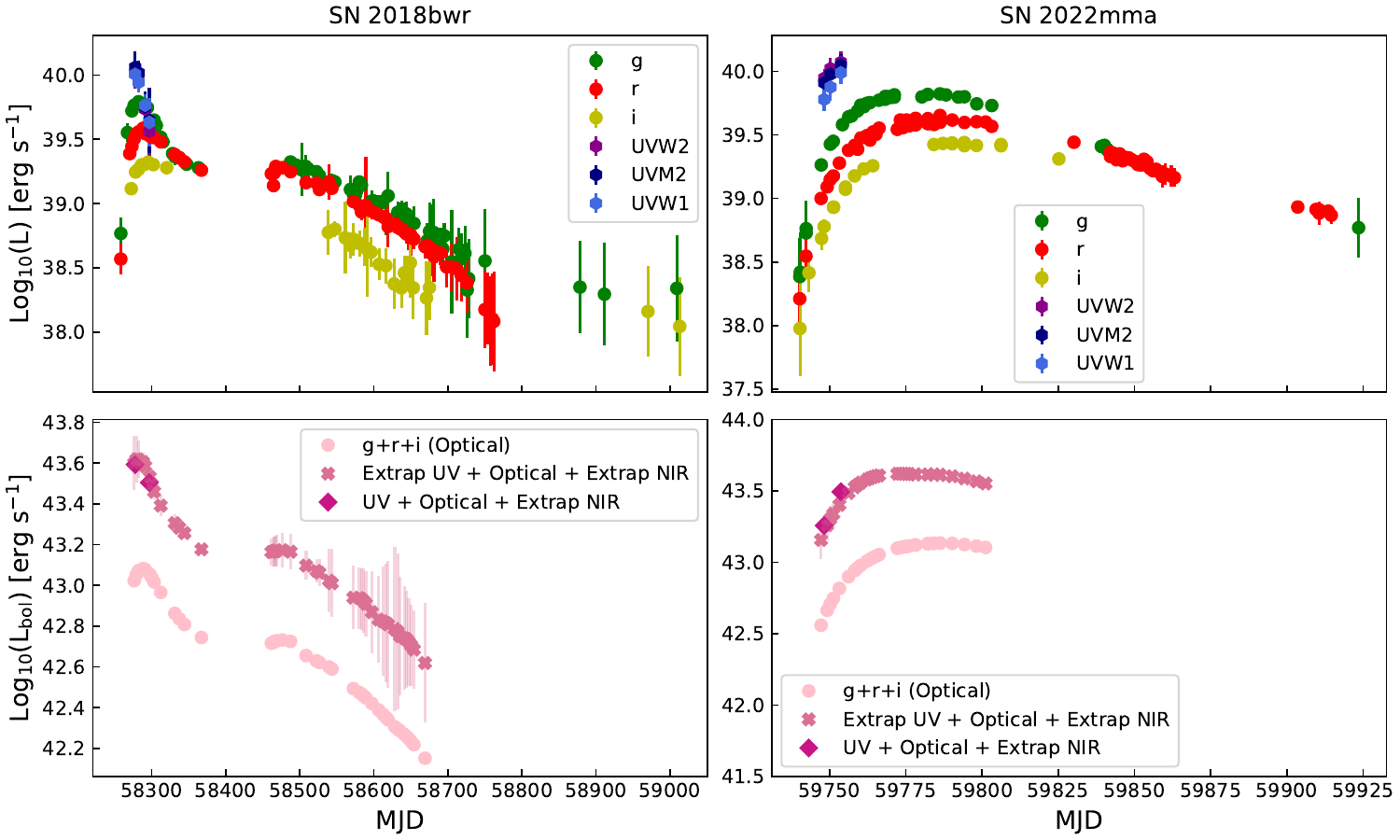}
      \caption{Pseudo-bolometric light curves including UVOT observations. Top panels show the observed $gri +$UVOT light curves and bottom panels the calculated pseudo-bolometric light curves for SN~2018bwr (left) and SN~2022mma (right).}
         \label{fig:bolsUVOT}
   \end{figure*}

Out of the fourteen events in our sample with available UV data (see Sect.~\ref{sec:swift}), only two also have observations in all three optical $gri$ bands. These two events are shown in Fig.~\ref{fig:bolsUVOT}. The top panels show the observations in all the considered bands, while the bottom panels show three calculated pseudo-bolometric light curves. Two correspond to the methods described above, and the third pseudo-bolometric light curve is calculated by direct integration of the $UV-gri$ bands in the overlapping observed phases plus an extrapolation to the NIR in the same way as described above. We can see that although both of our initial approaches to calculate pseudo-bolometric light curves only provide lower limits for the observed total luminosity, including extrapolations to the UV and NIR provides results that are closer to the real emitted total luminosity. 

\begin{figure}
   \includegraphics[width=8.5cm]{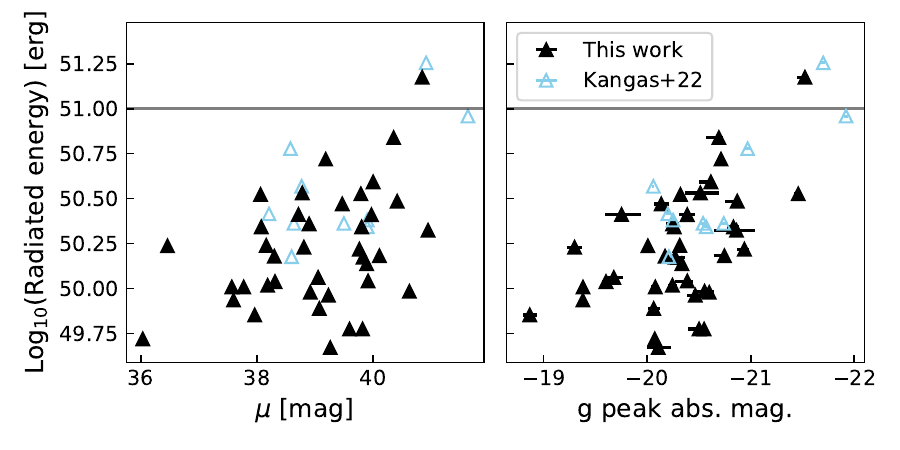}
      \caption{Total radiated energy of the full sample of SLSNe~II. Left panel compares the total radiated energy with the distance modulus and right panel with rest $g$ band peak absolute magnitude, shown as black filled markers. Empty skyblue markers show the corresponding parameters for the ZTF SLSN~II sample presented by \cite{2022MNRAS.516.1193K}, see Sect.~\ref{sec:kangas} for a discussion. 
              }
         \label{fig:TandE}
   \end{figure}

Once we have calculated pseudo-bolometric light curves, we can estimate a lower limit for the total radiated energy of each event. These are presented in Fig.~\ref{fig:TandE}. 
We see that brighter events are typically found at larger distance moduli and radiate more energy. SN~2018lzi shows a radiated energy $> 10^{51}$~erg, which is considered to be typical given the kinetic energy of CCSNe. To achieve such high energies, additional powering mechanisms are needed. \cite{Pruzhinskaya_2022} have proposed that SN~2018lzi is a PISNe, although their analysis is approximate as they did not count with spectroscopic $z$ at the time and other power sources have not been discarded. \cite{2022MNRAS.516.1193K} show that the SLSN~II, SN~2019uba, also radiates more than $10^{51}$~erg, they suggest that this may be produced by CSM interaction plus a central engine.

\subsection{Luminosity distribution}

 \begin{figure*}
   \centering
   \includegraphics[scale=0.6]{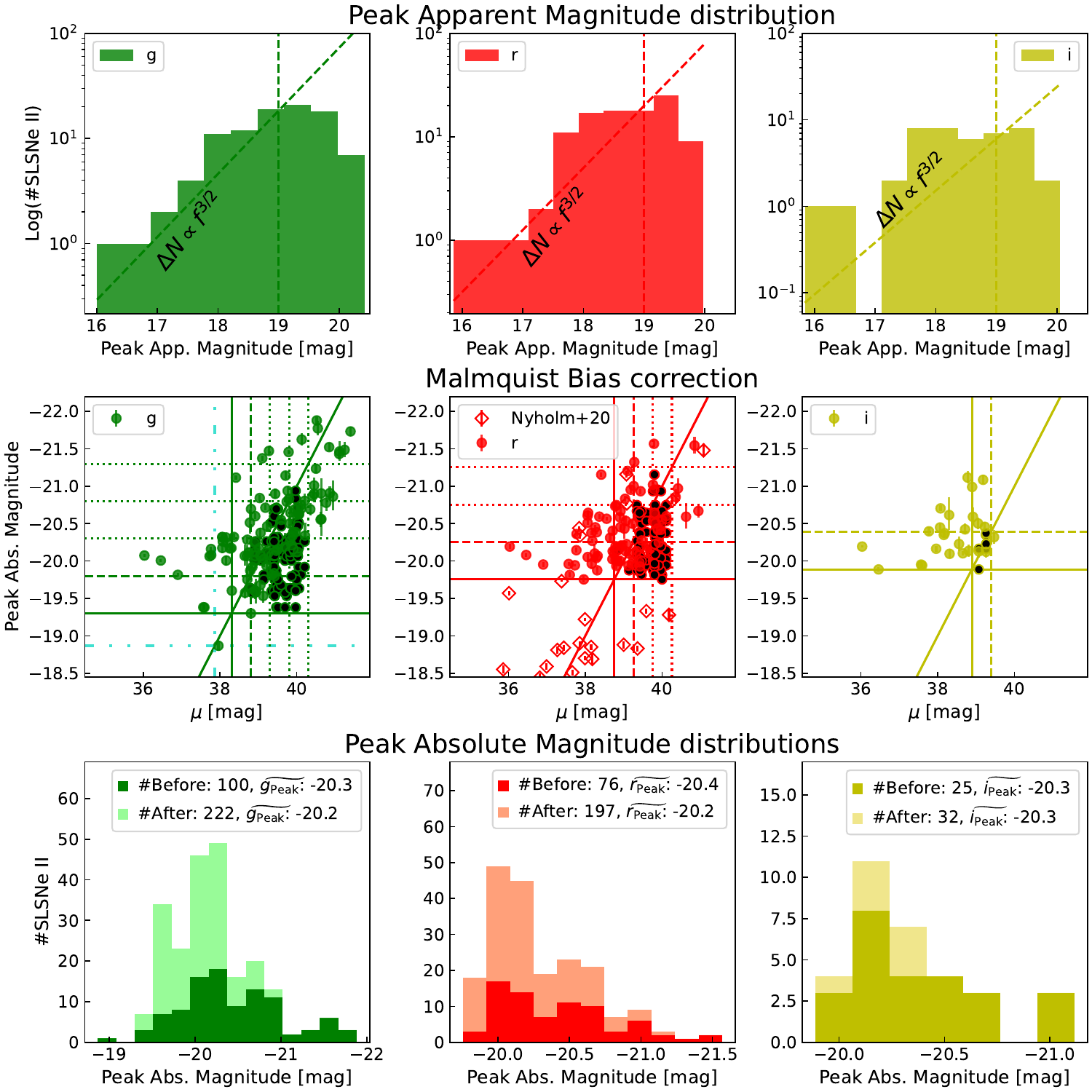}
      \caption{Malmquist bias correction. Top row shows the peak apparent magnitude distribution of the observed sample in $gri$ bands (left, middle and right panels respectively), the inclined dashed line shows the $\Delta N \propto f^{3/2}$ relation, the vertical dashed line shows the integer magnitude closest to the last position at which the slope of this relation intersects the distribution.  
      Second row shows the original rest frame peak absolute magnitude distribution with respect to the distance moduli in $gri$ bands (left, middle and right panels respectively), together with the estimated distribution of missed events, presented as black dots. The solid horizontal lines indicate the faintest event of the distribution while the solid vertical lines indicate the corresponding distance modulus. The subsequent horizontal and vertical lines indicate 0.5~mag bins from the solid lines respectively. The turquoise double dotted dashed lines in the first panel shows an event fainter than the one we consider for the analysis, we consider this event suffers from extinction and it is not considered. Third row shows the distribution of rest frame peak absolute magnitude in $gri$ bands (left, middle and right panels respectively) before (dark colors) and after (light colors) applying the Malmquist bias correction.
              }
         \label{fig:malmquist}
   \end{figure*}

To correctly analyze the luminosity distribution of our SLSN~II sample it is necessary to account for Malmquist bias \citep{1922MeLuF.100....1M}. This is a selection effect produced by the limiting magnitude associated to the telescope used to obtain the observations. The fact that telescopes have an associated limiting magnitude means that we are not detecting the intrinsically faintest events. 
To mitigate this effect, we follow the procedure described by \cite{2020A&A...637A..73N} to statistically estimate the number and distance distribution of the missing SLSNe~II. This procedure consists of defining a magnitude completeness limit based on the limiting magnitude of the telescope and the brightness of the faintest event in the sample. However, defining our sample's completeness in this way is challenging as we do not have estimates on how many SLSN remain unclassified due to reasons other than their brightness, such as failure to identify their light curve as potential SLSN candidates or lack of spectroscopic follow up resources. Although we are not considering all the effects that may be impacting our selection, we roughly estimate the magnitude completeness limit by inspecting the distribution of peak apparent magnitudes (top pannel of Fig.~\ref{fig:malmquist}). In an homogeneous, Euclidian universe, the distribution of a flux-limited survey should satisfy the relation $\Delta N \propto f^{3/2}$, where $\Delta N$ corresponds to the number of events per considered bin and $f$ is the observed flux at peak \citep[e.g.][]{2020ApJ...904...35P}. We consider that our sample is complete up to the integer magnitude closest to the position at which the slope of the $f^{3/2}$ relation intersects the peak magnitude distribution, for our sample this is 19~mag for all three $gri$ bands. The proportionality factor for each band was inferred by fitting the function to the number of events in each histogram bin, for all bins with an increasing number of events. To do this, we used the \texttt{curve\_fit} module on the \texttt{SciPy} optimization package. 
The volume defined by events with brighter rest frame peak absolute magnitude  than these limits, left to the diagonal lines and above the solid lines in the second row panels of Fig.~\ref{fig:malmquist}, is considered to be complete. The volume defined by fainter events to the right of the diagonal lines and above the solid lines, is considered to be incomplete. Note that in the case of the $g$ band, there is one event that is fainter (marked with a turquoise double dotted dashed line) than the one considered to define the complete volume, we assume that this event suffers from considerable host extinction (see Sect.~\ref{sec:color}) and remove it from the analysis. 

Once the complete volume has been defined, we consider bins of 0.5~mag both in brightness and distance moduli. These bins are marked with different line styles in Fig.~\ref{fig:malmquist}. The solid horizontal lines indicate the faintest event of the distribution, while the solid vertical lines indicate the corresponding distance modulus. Then, the first 0.5~mag absolute magnitude bin is encompassed between the solid and dashed horizontal lines and the first 0.5~mag distance moduli bin of is encompassed between the solid and dashed vertical lines. We consider that events left to the solid vertical line represent the first complete volume. Below the dashed horizontal lines are the dim events and above it the bright events. Between the solid and dashed lines is the first considered incomplete volume with a lack of dim events below the dashed horizontal line. This dim region is populated by randomly selecting peak absolute magnitude values from the dim part of the complete volume left to the vertical solid line, at randomly selected distance moduli. 
The ratio between bright and dim events in the incomplete volume should be the same as the ratio observed in the complete volume. We then consider the randomly generated points as true observations, move to the next peak absolute magnitude and distance modulus bin and repeat the procedure. When no point is found in the new bin above the horizontal line and to the right of the vertical line, the procedure stops. The true observed points are marked with colored circles and the randomly generated points are marked with black circles in the second row panels of Fig.~\ref{fig:malmquist}. 

The bottom panel of Fig.~\ref{fig:malmquist} shows the distributions of rest frame peak absolute magnitudes before and after Malmquist bias corrections. The median rest frame peak absolute magnitude of the distribution without Malmquist bias correction is $\sim$ 0.1~mag and $\sim$ 0.2~mag brighter than that of the distribution with the correction in the $g$ and $r$ bands respectively. However, it remains the same in the $i$ band. This could be a result of the lower number of events in this band. We note that this is just a rough estimate as we ignore possible systematic bias that can be introduced in a sample selected based on classified events.

\section{Extreme events}
\label{sec:extev}

Although we treat our sample of SLSNe~II as a whole, we can see that some events clearly stand out from the rest. We discuss such events in the following.

\subsection{Multi-peaked SLSNe~II}
\label{sec:multipeak}

\begin{figure*}
   \includegraphics[width=19cm]{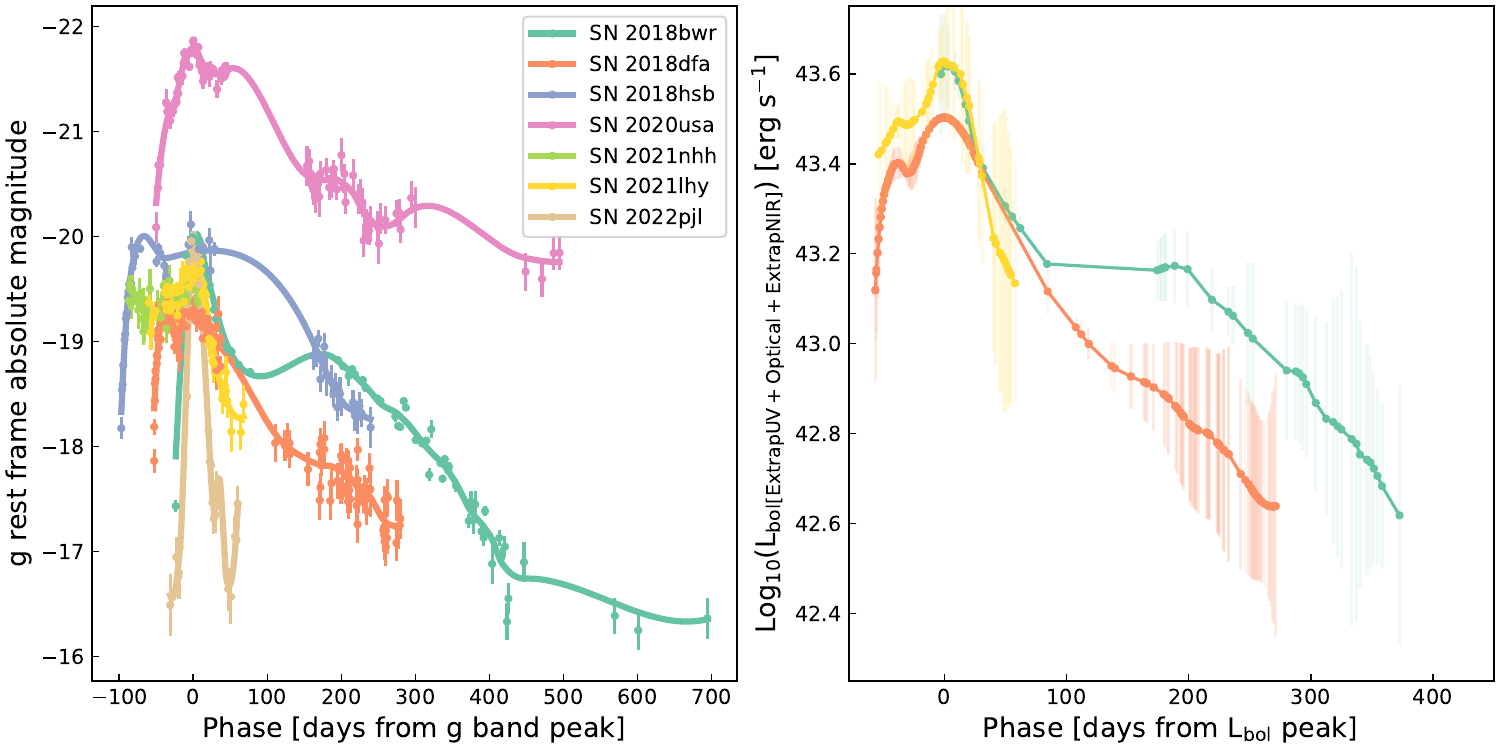}
      \caption{SLSNe~II with multiple peaks detected visually. Left panel: rest frame $g$ band absolute magnitude with respect to rest frame days since $g$ band peak. Right panel: pseudo-bolometric light curves of multipeaked events with $gri$ band observations.
              }
         \label{fig:doublepeak}
   \end{figure*}

A subgroup of SLSNe~II stand out for their visually inferred multi-peaked behavior. This subgroup is shown in Fig.~\ref{fig:doublepeak}. SN~2018dfa, SN~2018hsb and SN~2021nhh show a narrow peak before the main peak; SN~2021lhy shows a broad rise with a subtle peak before the main peak, similar to what has previously been considered to be a precursor event \citep[e.g.][]{2014ApJ...789..104O,2021ApJ...907...99S,2022ApJ...936..114M}, although brighter than usual; SN~2018bwr shows a somewhat wide secondary peak after a the primary peak; SN~2022pjl shows a secondary peak much fainter than the primary peak and; SN~2020usa shows more than two peaks, these could also be considered to be bumps \citep[e.g.][]{2017A&A...605A...6N}. SN~2020usa is also the brightest of the group, although not the brightest of our whole SLSN~II sample. Although an environment analysis is out of the scope of this work, it is worth mentioning that most of these events appear to be off-center from their host galaxies, except from SN~2020usa and SN~2021lhy. There is no additional evidence to indicate that these events are not SLSN however, other types of nuclear transients are a major source of contamination when defining SLSNe~II samples (see Sect.~\ref{sec:contaminants}) and a thorough dedicated analysis would be needed in order to further evaluate the classification of these two events.

In general, the presence of multiple peaks in the light curves is explained by invoking the presence of CSM. Differences in the mass loss mechanisms of the progenitor stars can produce different CSM structures, as the ejecta meets a different portion of the CSM, different peaks can be seen in the light curves \citep[e.g.][]{2023A&A...677A.105D,2023arXiv230403360K}. We note that these multi-peaked events represent only a $\sim$ 6.5\% of our sample of SLSNe~II. Thus, we could argue that intricate CSM configurations may be the exception rather than the rule. However, we can not rule out that we are missing some fraction of multi-peaked events due to differences in the observing cadence and duration of the sample. It is not trivial to define an optimal cadence to observe all possible light curve peaks as they appear with a variety of widths, duration and brightness. 

\subsection{Slow risers}
\label{sec:longrise}

\begin{figure*}
   \includegraphics[width=19cm]{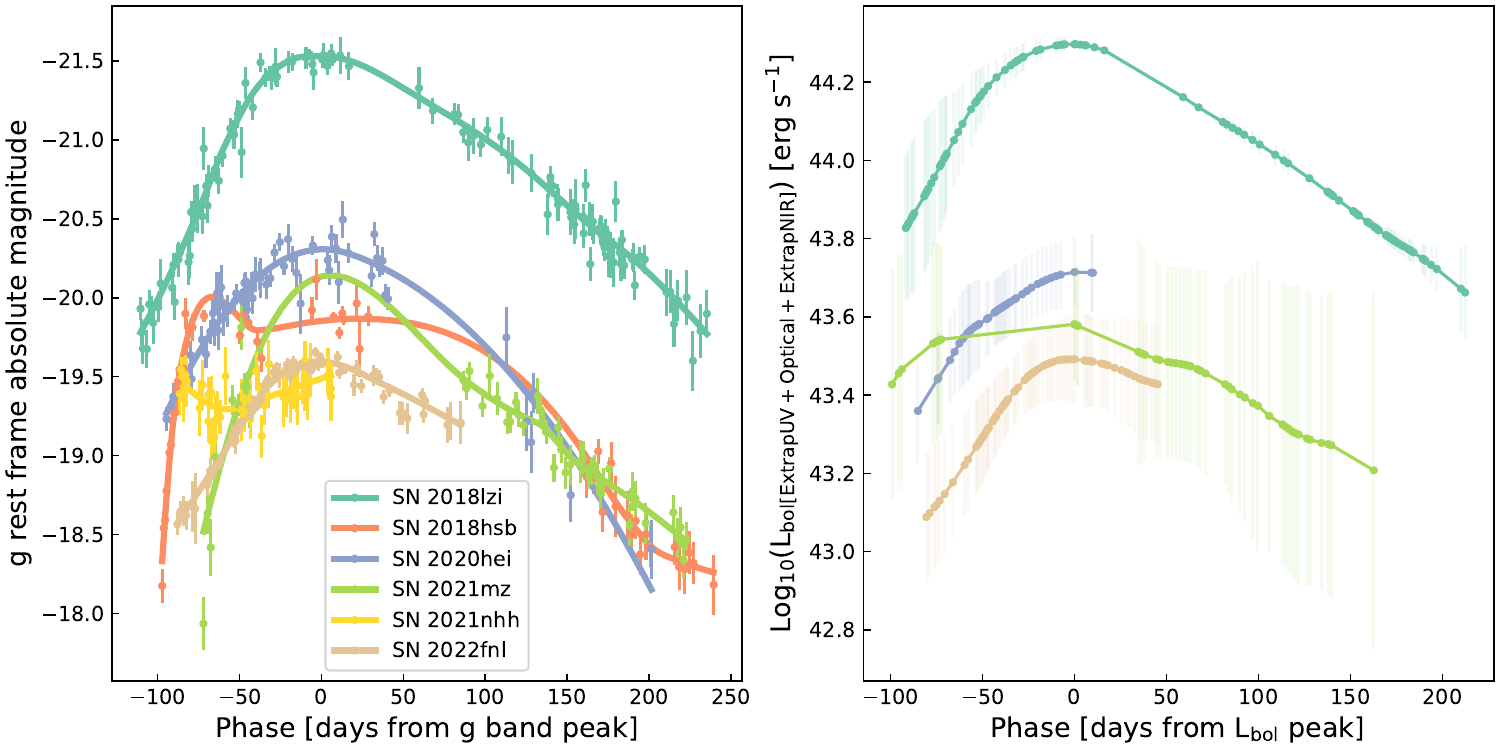}
      \caption{SLSNe~II with t$_{\mathrm{rise,1/e}} \geq 80$ rest frame days in any of the $gri$ photometric bands. Left panel: rest frame $g$ band absolute magnitude with respect to rest frame days since $g$ band peak. Right panel: pseudo-bolometric light curves of events with $gri$ band observations.
              }
         \label{fig:longrise}
   \end{figure*}

The rise time of SN is directly connected to the characteristics and environment of their progenitor, and also to their explosion mechanism \citep[e.g.][]{2022A&A...660A..42M}. 
In Fig.~\ref{fig:longrise} we show five SLSNe~II that stand out for having rise times t$_{\mathrm{rise,1/e}} \geq$ 80 rest frame days in any of the $gri$ photometric bands. This limit was chosen arbitrarily considering that the rise time of the canonical long rising SN, SN~1987A, is $\sim$ 80 days \citep[e.g.][and references therein]{1988Ap&SS.150..291M,1988PASA....7..401M,1989ARA&A..27..629A,2023ApJ...959..142S}. Note that the rise time in this work is measured differently. Nevertheless, if we would considered days elapsed from explosion, our rise times would be larger than the considered limit. The behaviour of the light curve of SN~1987A (and 87A~like events) has been explained as the product of the explosion of a compact progenitor in a formerly binary system \citep{2012A&A...537A.141P,2016A&A...588A...5T,2023ApJ...959..142S}. But 87A-like events show much lower peak luminosities than those presented in this work, so an alternative scenario is needed to explain the behaviour of this subgroup.
The slow risers in our sample are:  SN~2018hsb, SN~2018lzi, SN~2020hei, SN~2021mz, SN~2021nhh, and SN~2022fnl. In paticular, SN~2018hsb does not only have a long rise but shows a narrow peak before the main peak (see Sect.~\ref{sec:multipeak}). SN~2018lzi is the most luminous slow riser. The slowest riser is SN~2020hei, with t$_{\mathrm{rise,1/e}} = 103.2$ rest frame days in $r$ band. 

\subsection{Fast risers}
\label{sec:shotrise}

\begin{figure*}
   \includegraphics[width=19cm]{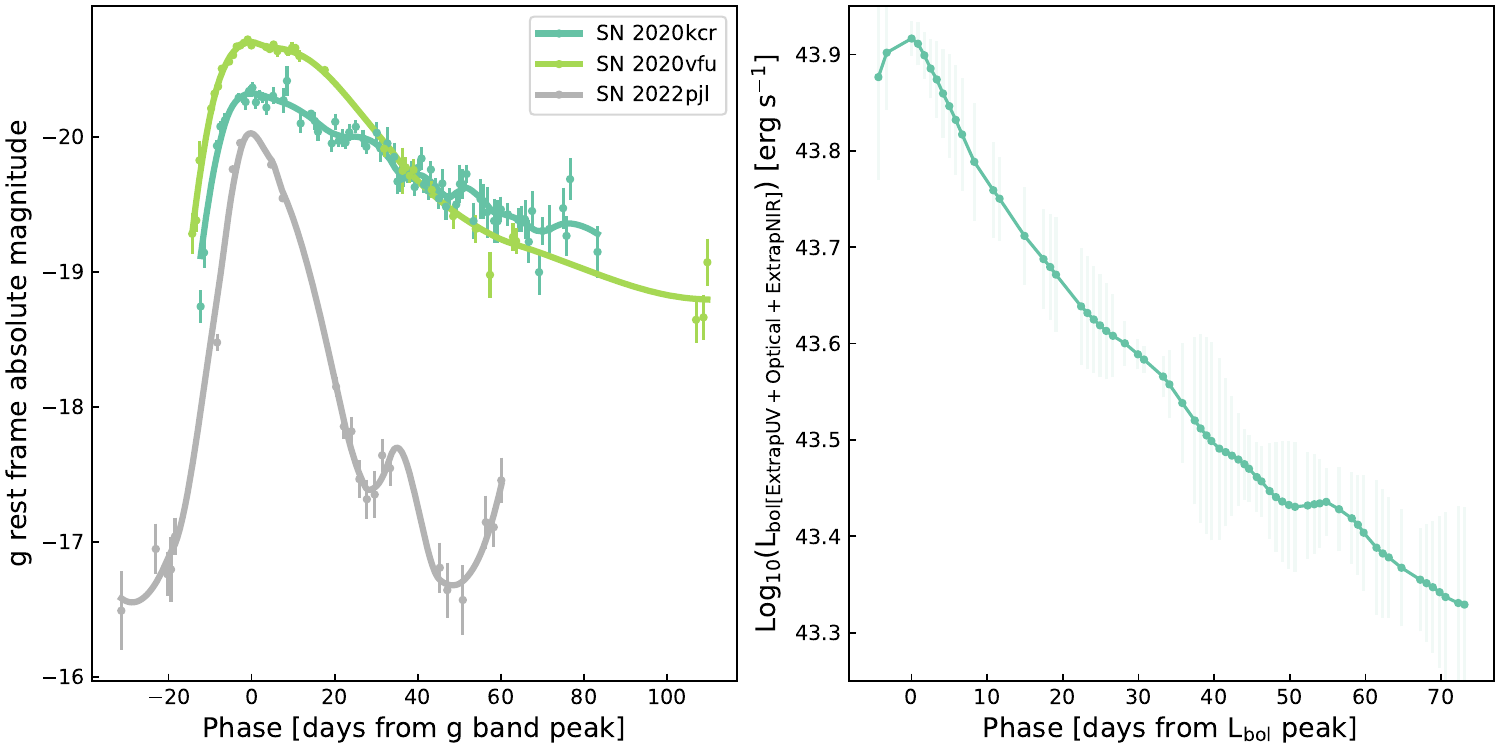}
      \caption{SLSNe~II with t$_{\mathrm{rise,1/e}} \leq 15$ rest frame days in any of the $gri$ photometric bands. Left panel: rest frame $g$ band absolute magnitude with respect to rest frame days since $g$ band peak. Right panel: pseudo-bolometric light curves of events with $gri$ band observations.
              }
         \label{fig:shortrise}
   \end{figure*}

As mentioned in the section above, the rise time of SN is directly connected to the characteristics and environment of their progenitors thus, we also highlight events with the shortest rise times. We arbitrarily considered any events with t$_{\mathrm{rise,1/e}} \leq 15$ rest frame days in any of the $gri$ photometric bands displays a fast rise time, as this limit is roughly half of the median t$_{\mathrm{rise,1/e}}$ value (see Table~\ref{table:peakAbsmagANDrisedecay}). Only three SLSNe~II present such short rise times, although other fast rising events could exist that were not classified due to initial selection effects that tend to favor slow rising objects as potential SLSN~II candidates. The fast risers in this sampler are shown in Fig.~\ref{fig:shortrise}. In the context of regular SNe~II, fast rises can be explained by a prolonged shock breakout moving through either a progenitor's extended atmosphere or surrounding CSM \citep[e.g.][]{2015MNRAS.451.2212G}. In the context of interaction powered SNe~IIn, fast risers also decline fast \citep{2020A&A...637A..73N}. Out of the three SLSNe~II that show a short rise time, only SN~2020kcr show a slow decline, whereas SN~2020vfu and SN~2022pjl show a linear fast decline. 

\subsubsection{The fastest evolving SLSN~II}

SN~2022pjl stands out as the fastest riser in the sample, with t$_{\mathrm{rise,10\%}} = 9.8$ rest frame days and t$_{\mathrm{rise,1/e}} = 7.5$ rest frame days in $g$ band. This is $\sim$ three rest frame days shorter than the next fastest event. SN~2022pjl also shows multiple peaks (see Sect.~\ref{sec:multipeak}), possibly indicating the presence of an unconventional CSM configuration.  

SN~2022pjl also stands out as the fastest decliner in the sample, with t$_{\mathrm{dec,10\%}} = 24.1$ rest frame days in $g$ band. For regular SNe~II, fast declining light curves are associated with higher explosion energies \cite[e.g:][and references therein]{2022A&A...660A..42M}. Although it has been argued that considering CSM relatively close to the progenitor star could naturally account for faster declining SNe~II \citep{2017ApJ...838...28M}, it is unclear whether a parallelism can be considered to SLSNe~II.

\subsection{Slow decliners}
\label{sec:longdecline}

\begin{figure*}
   \includegraphics[width=19cm]{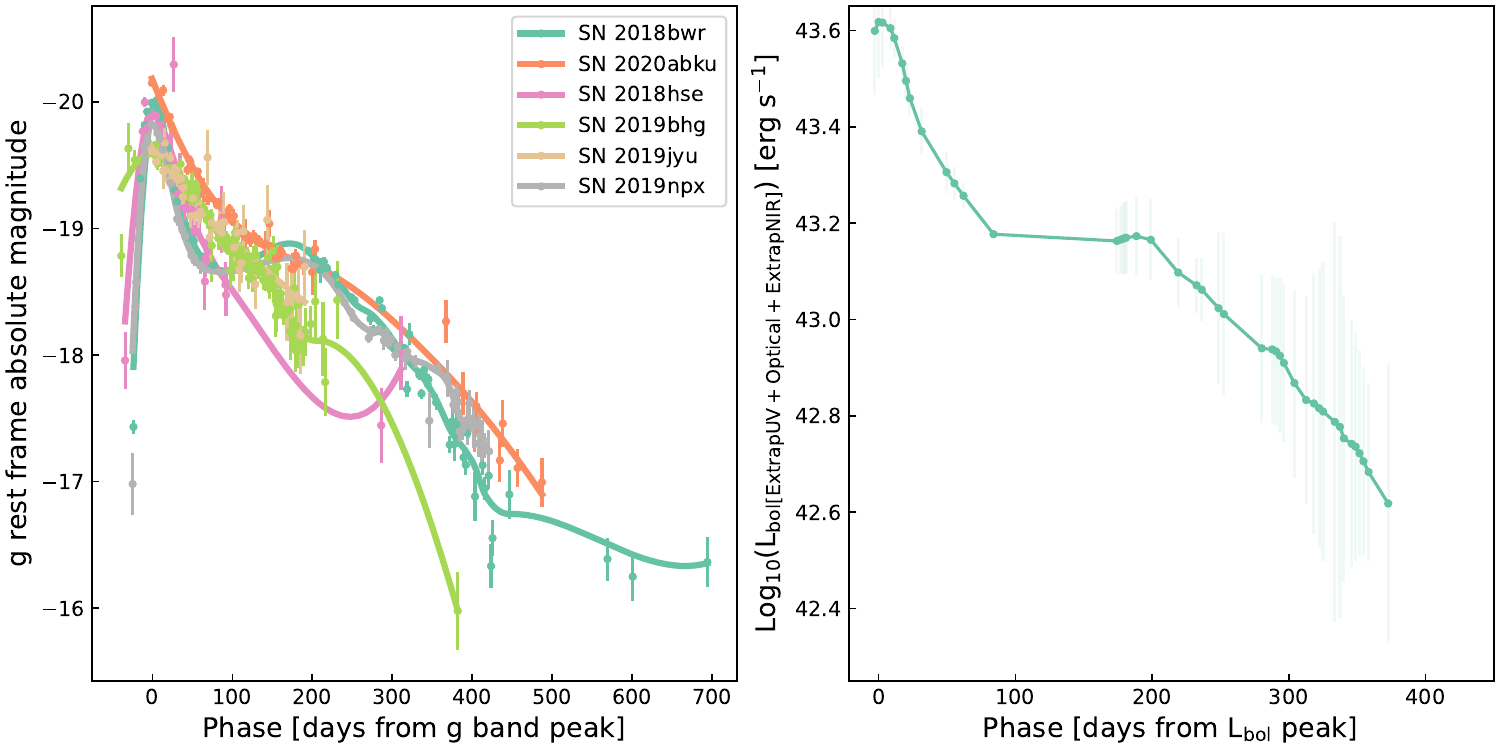}
      \caption{SLSNe~II with t$_{\mathrm{dec,10\%}} \geq 365$ rest frame days in any of the $gri$ photometric bands. Left panel: rest frame $g$ band absolute magnitude with respect to rest frame days since $g$ band peak. Right panel: pseudo-bolometric light curves of events with $gri$ band observations.
              }
         \label{fig:longdecay}
   \end{figure*}

Six SLSNe~II in our sample stand out for their long duration, showing  t$_{\mathrm{dec,10\%}} > $ 1 year in any of the ZTF $gri$ bands. These events are shown in Fig.~\ref{fig:longdecay}. Out of all the slow decliners, SN~2018bwr is also multi-peaked (see Sect.~\ref{sec:multipeak}). The other events: SN~2018hse, SN~2019bhg, SN~2019jyu, SN~2019npx and SN~2020abku, decline rather linearly after peak. Other slow decliners could have been missed due to lack of observations or because of our selection criteria (see Sect.~\ref{sec:sample}).

\subsection{The most and least luminous SLSNe~II}
\label{sec:brightdimm}

The mean peak $g$ band luminosity for our SLSN~II sample is $-20.3$~mag (see Table~\ref{table:peakAbsmagANDrisedecay}). The luminosity range of SLSNe~II could be affected by host extinction (see Sect.~\ref{sec:color}), and thus it is not trivial to define a lower luminosity threshold. Below we present the most and least luminous events of the sample.

\begin{figure*}
   \includegraphics[width=19cm]{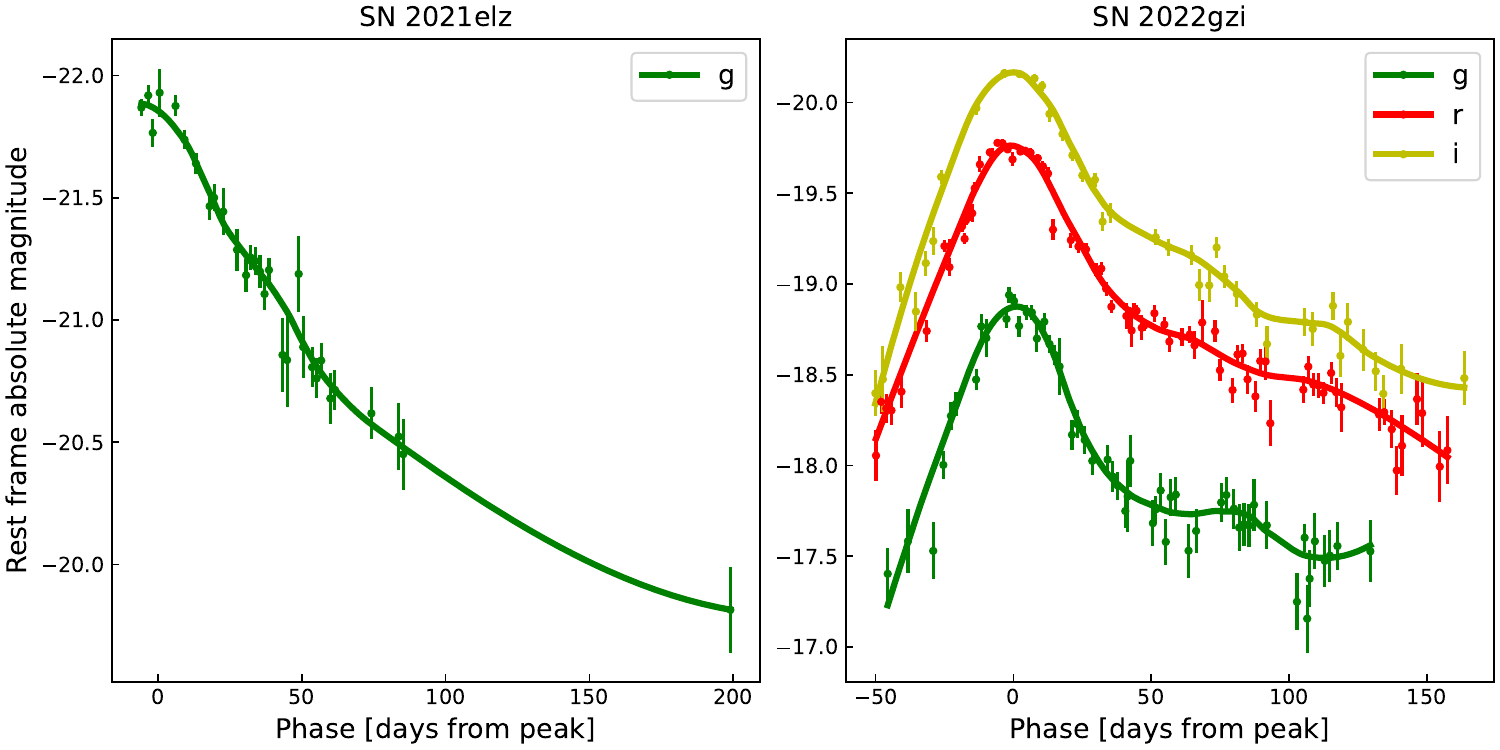}
      \caption{Left panel: Light curve of the brightest event in the SLSN II sample. Right panel: Light curves of the faintest event in the SLSN II sample.
              }
         \label{fig:brightdimm}
   \end{figure*}

The most luminous SLSN~II in our sample is SN~2021elz. Correcting by its high $z = 0.26$  we obtain a $g$ band rest frame peak absolute magnitude of $g_{\mathrm{peak}} \sim -21.9^{+0.1}_{-0.0}$~mag. The light curve of SN~2021elz is shown in the left panel of Fig.~\ref{fig:brightdimm}, it evolves smoothly, showing a linear decline after peak. 

The least luminous SLSN~II in our sample is SN~2022gzi. As expected, this event occurred at a lower $z = 0.089$ than the most luminous one. The rest frame peak absolute magnitude  are $g_{\mathrm{peak}} \sim -18.9 \pm 0.07$~mag, $r_{\mathrm{peak}} \sim -19.8 \pm 0.04$~mag, and $i_{\mathrm{peak}} \sim -20.2 \pm 0.04$~mag. This SLSN~II is the reddest one of the sample (see Sect.~\ref{sec:color}). We considered it to be a SLSN~II based on the $i$ band peak luminosity. If this band would not have been considered, this event would not have been classified as a SLSN. However, given that we consider the peak luminosity in all $gri$ bands, we can confirm this event to be superluminous. We highlight the benefit of considering redder bands when looking for SLSNe. 

\section{Comparison to other events}
\label{sec:comparisons}

When discussing SLSNe, people usually refer to hydrogen poor SLSNe~I. This is because there is much more data available on SLSNe~I than on SLSNe~II \citep[e.g:][]{2019ARA&A..57..305G}. 
This work presents the first large sample of SLSNe~II light curves. 
Some SLSNe~II exist in the literature that show narrow spectral lines and were classified as SN~IIn \citep[e.g.][]{2007ApJ...659L..13O,2008ApJ...686..467S}. The distinction between SNe~IIn and SLSNe~II is not always clear given that SNe~IIn tend to be luminous and the threshold to classify an event as SLSNe~II is somewhat arbitrary. Thus, it is often considered that SLSNe~II that show narrow lines are a mere extension in luminosity of SNe~IIn \citep[][]{2016ApJ...830...13P,2023AAS...24120716D}. In this section we compare our sample to SLSNe~I and SNe~IIn as well as to the SLSNe~II sample presented by \cite{2022MNRAS.516.1193K}.

   \begin{figure*}
   \centering
   \includegraphics[width=19cm]{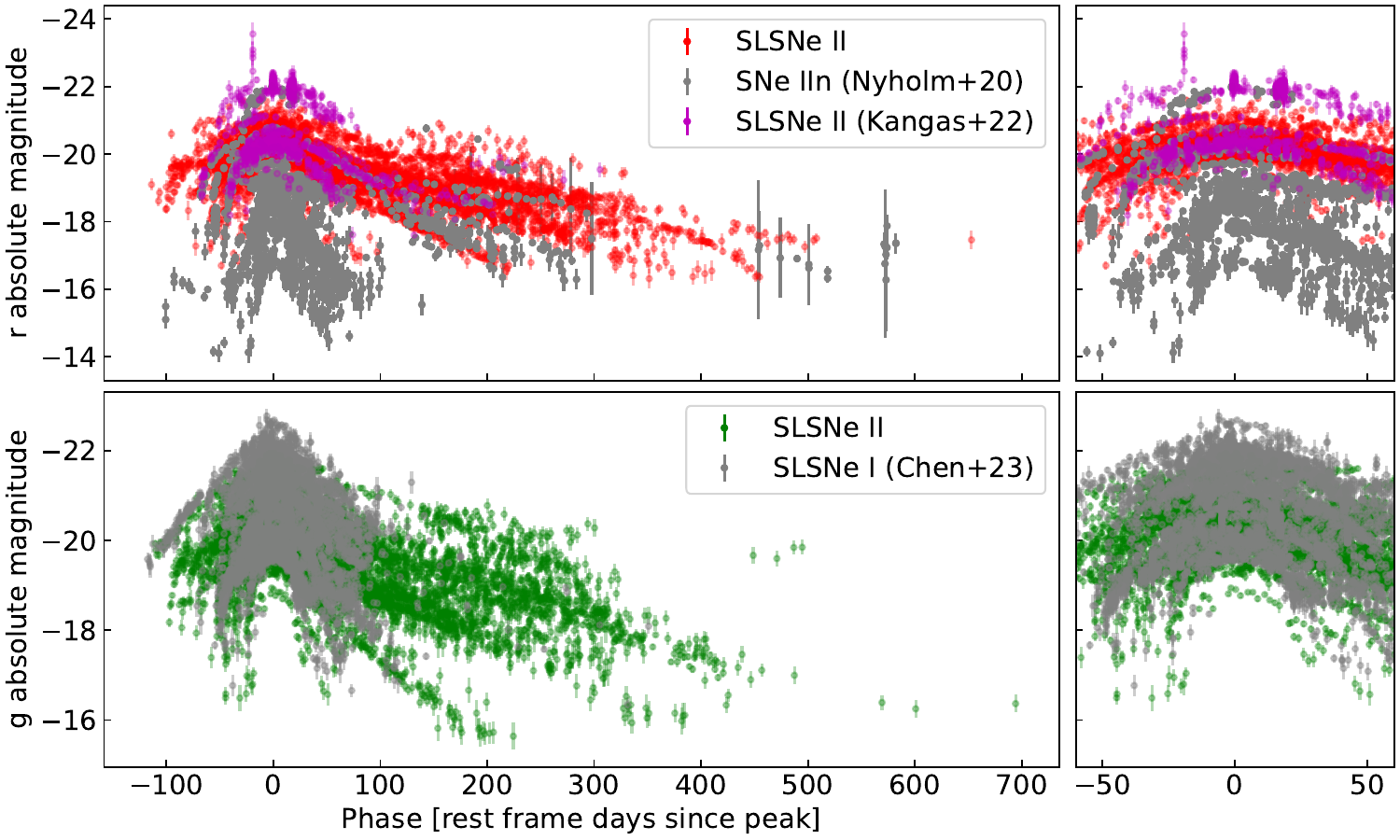}
      \caption{Top left panel: comparison of the rest frame $r$ band light curves of our sample of SLSNe~II (red dots), the sample of SLSNe~II (magenta dots) of \cite{2022MNRAS.516.1193K} and the sample of SNe~IIn (grey dots) of \cite{2020A&A...637A..73N}. Bottom left panel: comparison of the rest frame $g$ band light curves of our samples of SLSNe~II (green dots) and the SLSNe~I (grey dots) sample of \cite{2023ApJ...943...41C}. Top and bottom right panels show the same as the respective left panels but constrained to $\pm$ 60 rest frame days from peak.
              }
         \label{fig:comptoothers}
   \end{figure*}

\subsection{Other SLSNe~II}
\label{sec:kangas}

The SLSNe~II sample presented by \cite{2022MNRAS.516.1193K} was selected based on their lack of narrow spectral lines. We did not include these events in our sample but present comparisons of the corresponding distributions of $z$, rest frame $g$ band peak absolute magnitude, and total radiated energy in Figs.~\ref{fig:z}, ~\ref{fig:peakabsmag} and ~\ref{fig:TandE}. We also calculated the $g$ band rise and decline time parameters of the sample presented by \cite{2022MNRAS.516.1193K} as described in Sect.~\ref{sec:riseanddecline}, and present comparisons to our sample in Fig.~\ref{fig:risedecaydist} and in the top panel of Fig.~\ref{fig:timesVSpeak}. Fig.~\ref{fig:grcol} shows comparisons of the $g-r$ colors with respect to $g$ band peak of our sample and the events in the sample of \cite{2022MNRAS.516.1193K} with $z \leq 0.17$, as this is the $z$ limit for which we consider that K-corrections are negligible we can utilize the reported peak epochs directly.

We see that the overall $z$ distribution of both samples overlap. The same is true for the $g$ band rest frame peak absolute magnitude, and for the $g$ band rise and decline time distributions. In addition, both samples show similar observed $g-r$ colors and estimated total radiated energies. Moreover, in the top panels of Fig~\ref{fig:comptoothers} we see that the overall shape of the light curves presented by \cite{2022MNRAS.516.1193K} is similar to the light curve of the events in our sample. 
The premise behind the selection criteria of \cite{2022MNRAS.516.1193K} is that narrow H lines are irrefutable evidence of CSM interaction (see Sect.~\ref{sec:intro}) thus, excluding such events provides further insight on the powering mechanism of SLSNe~II. Nevertheless, after inspecting different models, \cite{2022MNRAS.516.1193K} conclude that only pure $^{56}$Ni can be discarded, while both the presence of a magnetar and CSM interaction reproduce the observed light curve with similar success. Still, the UV excess detected in most of their SLSN~II where good enough UV data exist, indicates the presence of interaction.
We do not find any bimodality or visual clustering in the distributions of the studied light curve parameters, this indicates that there is no photometric distinction between SLSNe~II that do not show persistent narrow lines and those that would be classified as SLSNe~IIn. It has been argued before that a continuum in observed light curve parameters point towards a common powering mechanism \citep[e.g.][]{2014ApJ...786...67A}. If this is the case for all H-rich SLSN, then further subclassification does not provide additional insight in the physics involved in powering the observed light curves. However, we can not rule out that the studied parameters are not adequate to capture differences between spectroscopic subclasses. 
   
\subsection{Type IIn SNe}

In order to compare the characteristics of SLSNe~II to those of SNe~IIn we chose the sample of SNe~IIn presented by \cite{2020A&A...637A..73N}. They present an analysis of SNe~IIn without really excluding possible SLSNe~II thus, we expect some overlap between their peak brightness measurements and ours. Most of their analysis considers the $R/r$ bands, so we use the $r$ band for light curve comparison, presented in the top panel of Fig.~\ref{fig:comptoothers}. Indeed, there seems to be a continuum of peak magnitudes between SNe~IIn and SLSNe~II, with a small gap of $\sim 0.2$~mag between both SN types, that can also be seen in the second middle panel of Fig.~\ref{fig:malmquist}. In addition, (less luminous) SNe~IIn tend to decline faster than SLSNe~II. This could indicate that if CSM interaction is the sole responsible mechansism for the additional luminosity seen in SLSNe~II, the CSM configuration around the progenitors of SNe~IIn and SLSNe~II should be different in radius, density or other morphological characteristics in order to make the light curves of the latter last longer. However, both our sample and the one presented by \cite{2020A&A...637A..73N} may be suffering from unknown selection effects, and a more detailed comparative analysis is needed to further assess the presence or absence of a continuum between SNe~IIn and SLSNe~II.

\subsection{SLSNe~I}

\begin{figure}
   \includegraphics[width=8.5cm]{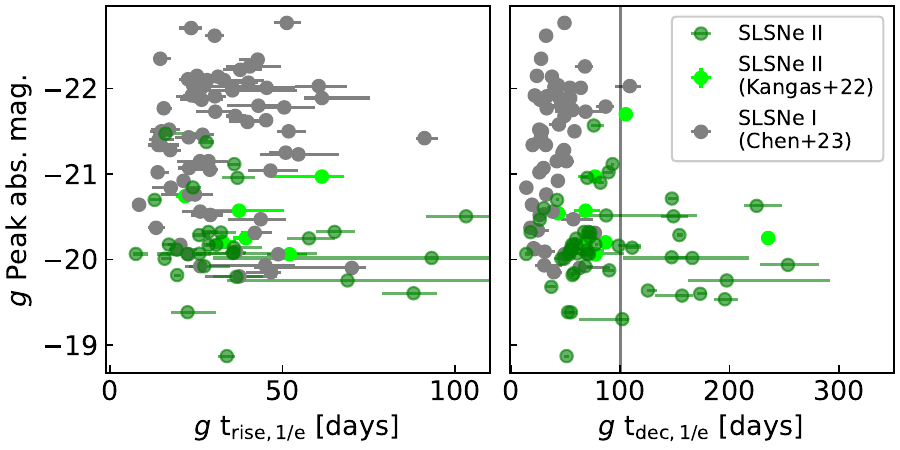}
      \caption{Left panel: $g$ band t$_{\mathrm{rise,1/e}}$ versus rest frame peak absolute magnitude for our sample of SLSNe~II (green circles) and for the SLSN~I (grey circles) sample of \cite{2023ApJ...943...41C}. Right panel: $g$ band t$_{\mathrm{dec,1/e}}$ versus rest frame peak absolute magnitude for our sample of SLSNe~II (green circles) and the SLSN~I (grey circles) sample of \cite{2023ApJ...943...41C}.
              }
         \label{fig:timecompSLSNeI}
   \end{figure}

We consider the SLSN~I sample presented by \cite{2023ApJ...943...41C} to compare to our SLSNe~II. This comparison is ideal as we both consider the same survey and thus, the same filters and reduction methods. \cite{2023ApJ...943...41C} consider the $g$ band as their primary photometric band, so we compare our $g$-band light curves to theirs in the bottom panels of Fig.~\ref{fig:comptoothers}. At earlier times the light curves are very difficult to distinguish as they present similar rise times, this can be better seen in the left panel of Fig.~\ref{fig:timecompSLSNeI}. In Fig.~\ref{fig:grcol_peak} we compare the observed $g-r$ colors at peak against the rest frame $g$-band peak absolute magnitude of both samples and see that SLSNe~I are overall bluer and brighter than SLSNe~II, without any SLSN~II being bluer than $g-r < -0.1$~mag (below the lower horizontal grey dashed line) or brighter than $g_{\mathrm{Peak}} < -22$~mag (right of the vertical grey dashed line). Another difference is the SLSNe~II present much longer decline times than SLSNe~I (see right panel of Fig.~\ref{fig:timecompSLSNeI}).

\section{Contaminants}
\label{sec:contaminants}

\begin{figure*}
   \centering
   \includegraphics[scale=0.7]{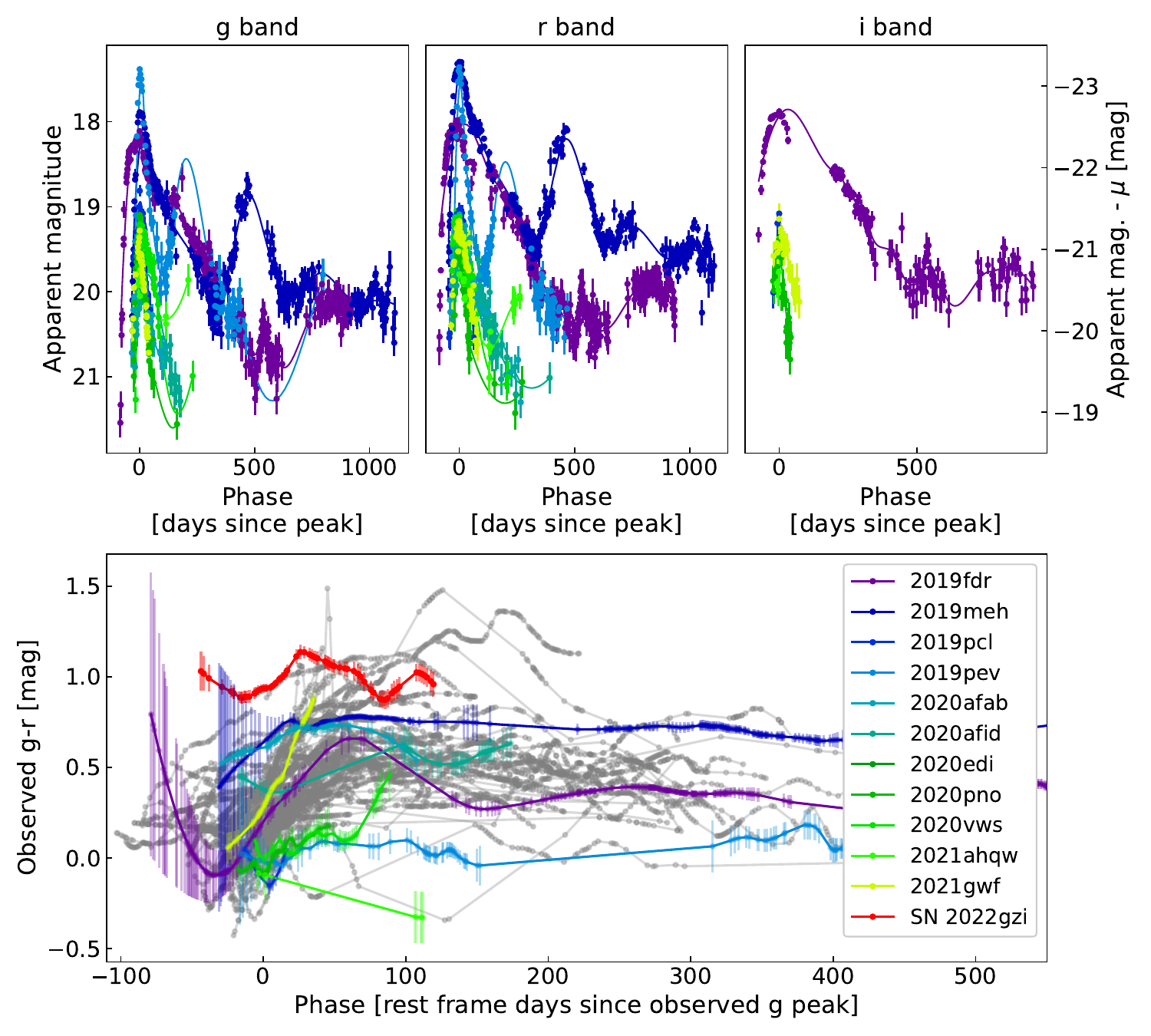}
      \caption{Top panels: $gri$ (left, center and right panel respectively) band light curves of events with ambiguous classification. The left vertical axis displays the apparent magnitude while the right vertical axis corrects for $\mu$. Bottom panel: observed $g-r$ colors at around $g$ band peak versus rest frame $g$ band peak absolute magnitude of the ambiguous transients (green to blue colors) and our SLSN~II (grey). In red we highlight SN~2022gzi as the reddest event in the sample.
              }
         \label{fig:contaminants}
   \end{figure*}

At times SLSNe~II can resemble AGNs, this is because the light curves of some SLSNe~II can reach comparable peak luminosities to those of AGNs. In addition, SLSNe~II can be long-lived and show bumps and wiggles in their light curves, as some AGNs do. Moreover, both AGNs and SLSNe~II can display narrow H emission lines \citep[e.g:][]{1989AJ.....97..726F}. When the classification is uncertain, one method to rule out AGNs is to assess whether the transient is located at the center of the host galaxy; if it is not, then it is likely that the transient is a SLSN~II. However, there is evidence that a small number of AGNs may be offset from the center of their host \citep{2021ApJ...913..102W}. In addition, poor spatial resolution or line-of-sight effects could make it difficult to study the exact position of the SNe with respect to the center of the host. All of this make AGNs the primary contaminant in SLSN~II samples. Also TDEs can contaminate SLSN~II samples, as illustrated by the case of SN~2020yue considered to be a SLSN~II by \cite{2022MNRAS.516.1193K} but later reclassified by \cite{2023ApJ...955L...6Y}.

When searching for SLSNe~II in the ZTF survey we found eleven events with ambiguous classifications\footnote{Some of the events have a ``SN'' prefix due to reported classifications; the remainder were flagged as potential SLSNe~II internally but not reported so have an ``AT'' prefix. However, as discussed, we consider these identifications ambiguous.}: SN~2019fdr, SN~2019meh, AT~2019pcl, AT~2019pev, SN~2020edi, AT~2020pno, SN~2020vws, AT~2020afab, AT~2020afid, AT~2021gwf and AT~2021ahqw. All of these events seem to be located at the center of their respective hosts and show evidence pointing towards an AGN classification. SN~2019fdr is a very interesting case that highlights the difficulties in differentiating SLSNe~II from AGN, and also from TDEs. SN~2019fdr was discovered on May 3rd 2019 by \cite{2019TNSTR.771....1N}, the report includes a last non-detection on the 27th of April 2019, hinting towards a one week constraint on the possible explosion epoch. A spectrum was obtained on the 15th of June 2019 by \cite{2019TNSCR1016....1C}. However, a classification was not possible as the nuclear location of the event and the lack of metal lines in the spectra could not rule out the possibility of AGN, TDE or SLSN. A second spectrum motivated a SLSN~II classification by \cite{2019TNSAN..45....1Y}, in this case the authors argue in favor of such classification based on UV colors and lack of previous variability. Further analysis of the characteristics of the event led \cite{2020ApJ...904...35P} and \cite{2021ApJ...920...56F} to consider SN~2019fdr to be an AGN. To add complication to the interpretation of this event, \cite{2020GCN.27872....1R} reported a possible association with neutrino emission. This association motivated a TDE classification by \cite{2022PhRvL.128v1101R} and \cite{2021ApJ...920...50A}. However, \cite{2022ApJ...929..163P} argue that a neutrino association actually favors a SLSN~II classification. 
Recently, \cite{2024arXiv240611552W} analyzed a sample of a few ambiguous nuclear transients (ANTs), that include SN~2019fdr, and conclude that such events may be obscured TDEs based on occurrence rate arguments. Further analysis is needed in order to better understand these events..

In the top panels of Fig.~\ref{fig:contaminants} we show the $gri$ light curves of the events with ambiguous classification. SN~2019fdr, SN~2019meh and AT~2019pev have very long lived, bumpy light curves indicating that these events may not be SNe. In the bottom panel of Fig.~\ref{fig:contaminants} we show the observed $g-r$ colors of the ambiguous events in comparison to those of the SLSN~II sample. The main difference seems to be that SLSN~II are blue at early times and then become redder with time, while the ambiguous events are red at early times, then become blue and then turn to redder colors again. We mention in Sect.~\ref{sec:color} that the color evolution of SN~2022gzi, the reddest event in the sample, is shallower than the one seen for other SLSN~II. We see in Fig.~\ref{fig:contaminants} that it is also shallower than that of the ambiguous events. But it does show initial bluer colors than then become bluer and redder again later on. In this work we consider SN~2022gzi to be a SLSN~II based on the available data, although further analysis in needed to confirm the true nature of this event.

Based on nebular spectral modelling, \cite{2020Sci...367..415J}  proposed that a thermonuclear Type Ia SN enshrouded in a H-rich CSM could explain the observed characteristics of the prototypical SLSN~II SN~2006gy. In this scenario, the CSM would ``hide'' the SN~Ia features during the photospheric phase and these would only become visible at late times, once the shock has crossed the whole CSM. Such events are classified as Type Ia-CSM \citep[e.g.][]{2013ApJS..207....3S}. SN~2006gy, with a peak absolute magnitude of $V_{\mathrm{Peak}} \sim -22.$~mag \citep{2007ApJ...659L..13O}, is brighter than any of the events in our SLSN~II sample. Additional nebular spectroscopy and dedicated models would be necessary to study whether this mechanism could explain any objects in our sample. Two events in our sample: SN~2018dfa and SN~2019vpk are considered to be ambiguous by \cite{2023ApJ...948...52S}, but they do not find robust spectroscopic indications that they are indeed SNe Ia-CSM. We include them here as the available spectra match other H-rich SNe. 

\section{Discussion}
\label{sec:discussion}

We have shown that there is a great diversity in our SLSN~II sample, with some events showing extreme characteristics (see Sect.~\ref{sec:extev}). 
This sample of SLSNe~II was selected without considering the morphology of the H$\alpha$ emission line profile; however, there is evidence of narrow lines in most events.
Assuming that all the observed narrow lines are intrinsic to the SLSNe~II, we could conclude that the main powering mechanism producing the observed light curves is CSM interaction. \cite{2022MNRAS.516.1193K} consider only events without narrow lines in their spectra and also find that the main powering mechanism contributing to their light curves could be CSM interaction, although additional mechanisms may be needed in the cases of their most extreme events. The most popular alternative powering mechanism for SLSNe is the presence of a magnetar. \cite{2018MNRAS.475.1046I} argues that a magnetar is better at explaining their sample of SLSNe~II, although they acknowledge that even though they selected their sample because the events do not show narrow lines, some of them do show evidence of CSM interaction during the photospheric phase. 

\cite{2019ApJ...878...56K} suggest that the relation between rise time and peak luminosity can help determine the underlying powering mechanism of the most luminous transients. When analyzing rise time versus peak absolute magnitude of our SLSNe~II, we find no significant correlation (see Sect.~\ref{sec:riseanddecline}) and no noticeable groups. We also inspected the rise times of the calculated pseudo-bolometric light curves (see Sect.~\ref{sec:energy}) against their peak luminosities finding no clear trend or correlation, although this could be due to the low number of events for which we can estimate bolometric light curves ($\sim$ 42\% of the sample).

Recently, \cite{2023arXiv230403360K} suggested that, if a light curve is powered by CSM interaction, the diversity of the observed morphologies can be explained by considering different CSM configurations. In this context, they propose that SN~IIn light curves may arise from light interior interaction whereas events with radiated energies higher than 10$^{51}$~erg may result from events with heavy interior interaction. This last scenario could explain the total radiated energy of SN~2018lzi (see Sect.~\ref{sec:energy}). 
If we assume that the light curves of all the SLSNe~II in this sample are powered by CSM interaction, we can use the relation presented by \cite{1994MNRAS.268..173C} to approximate the mass-loss rate of the progenitor star as:

\begin{equation}
    \dot{M} = \frac{2L}{\epsilon} \frac{v_{w}}{v_{\mathrm{SN}}^{3}}
\end{equation}

\begin{figure}
   \includegraphics[width=8.5cm]{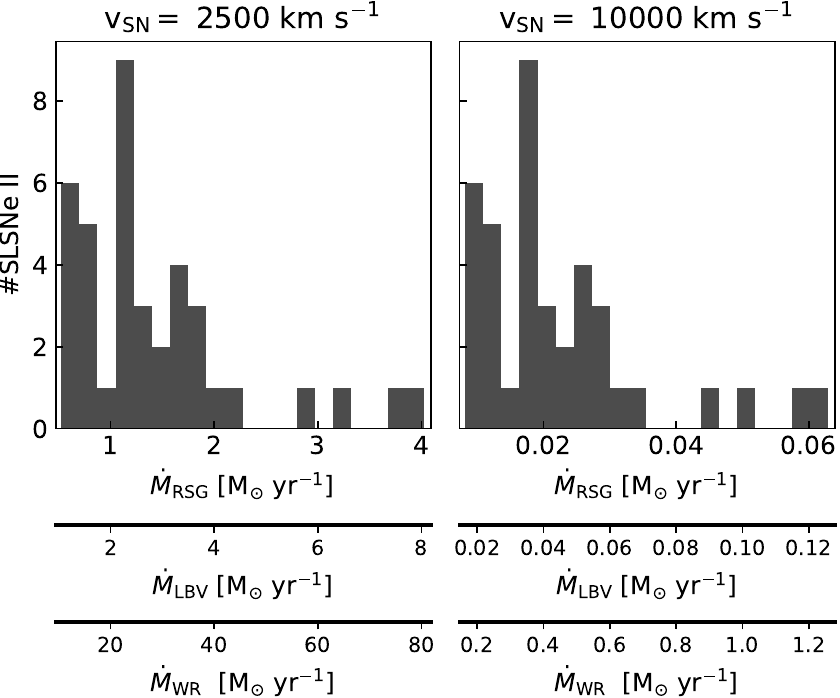}
      \caption{Mass loss rate ($\dot{M}$) of the SLSNe~II for which we can approximate a pseudo-bolometric light curve (see Sect.~\ref{sec:energy}) considering a RSG, LBV and WR progenitor. The different mass loss rate ranges are shown in consecutive horizontal axis. The left panel considers $v_{\mathrm{SN}} = 2500$~km s$^{-1}$ and the right panel $v_{\mathrm{SN}} = 10000$~km s$^{-1}$.
              }
         \label{fig:mdot}
   \end{figure}

Where $L$ is the bolometric luminosity, $\epsilon (< 1)$ is the efficiency of shock kinetic energy to optical radiation conversion, $v_{w}$ is the stellar wind velocity pre-explosion and $v_{\mathrm{SN}}$ is the post shock shell velocity. Accurately measuring all these parameters present different challenges, starting with the bolometric luminosity. $\epsilon$ is also an uncertain parameter and heavily relies on the adopted model. In addition, although $v_{\mathrm{SN}}$ could be estimated from the width of the spectral lines, narrow lines are usually produced by electron scattering and a good spectral sequence coverage (which we do not have for most events) is needed to assess the moment at which expansion takes over. Given all these uncertainties any estimation of $\dot{M}$ is rather speculative. However, it is still useful to have a sense of the possible progenitor characteristics. In Fig.~\ref{fig:mdot} we present rough approximations of $\dot{M}$ using the pseudo-bolometric luminosity at the epoch of light curve peak. We adopt $\epsilon = 0.5$, and consider two different $v_{\mathrm{SN}}$; first for interacting SNe we adopt $v_{\mathrm{SN}} = 2500$~km s$^{-1}$ \citep{2017hsn..book..403S}, second for regular SNe~II we adopt  $v_{\mathrm{SN}} = 10000$~km s$^{-1}$. We also consider three different $v_{w}$ to account for the most popular proposed interacting SN progenitors, 50~km s$^{-1}$ typical for red supergiants \citep[RSG;][]{2005A&A...438..273V}, 100~km s$^{-1}$ typical for luminous blue variables \citep[LBV;][]{2017hsn..book..403S,2020MNRAS.499..129G}, and 1000~km s$^{-1}$ typical for Wolf-Rayet stars \citep[WR;][]{2007A&G....48a..35C,2022ApJ...930..127G}. 
Considering $v_{\mathrm{SN}} = 2500$~km s$^{-1}$ returns mass-loss rates that range from over half a solar mass per year for RSG winds to tens of solar masses per year for WR winds (see left panel of Fig.~\ref{fig:mdot}). These far exceed the mass-loss rates typically inferred for regular SNe~II, although they are somewhat consistent with values deduced for other SLSNe~II \citep[e.g.][]{2024arXiv240404235D}.
While regular SNe~II show evidence for elevated mass loss in the late stages of stellar evolution, analysis of a magnitude-limited sample gives a typical value of 3 $\times$ 10$^{-3}$ M$_{\odot}$ yr$^{-1}$ for an assumed wind velocity of 10 km s$^{-1}$ \citep{hinds:inprep}, which is at least two orders of magnitude lower than the aforementioned values. 
Considering $v_{\mathrm{SN}} = 10000$~km s$^{-1}$ for our SLSNe~II, returns loss rate estimates that are more consistent with those found for regular SNe~II by \cite{hinds:inprep} (see right panel of Fig.~\ref{fig:mdot}).

While this work was being reviewed, \cite{2024arXiv241107287H} released the analysis of a large literature sample of CSM interaction powered SNe~IIn, which also includes events classified as SLSNe~II with narrow lines in their spectra. They identify a bimodal distribution in the luminosity--timescale  parameter space for their SNe~IIn, with a division near the median peak luminosity. Their analysis indicates a median peak absolute magnitude of $\sim -19.2$~mag for their sample. Based on this, they propose that SLSNe~II represent the brighter subset of SNe~IIn. In contrast, we suggest that this observed bimodality may reflect differences in the underlying powering mechanisms for events within each distribution. Consequently, it may be necessary to adjust the threshold for defining an event as a SLSN~II to include fainter peak absolute magnitudes. However, further analysis is required to determine whether this interpretation is consistent across the SN~II, LSN~II, and SLSN~II classes when spectra lack narrow emission lines. This raises the question whether it is time to reconsider the classification scheme for H-rich SNe, ensuring a clear separation between light curve characteristics and spectral features. Finally, we note that events such as those discussed in Sect.~\ref{sec:extev} are not adequately represented by the templates presented by \cite{2024arXiv241107287H}, highlighting the need for a more nuanced approach to classification.

\subsection{SLSNe II in the LSST era}

We are undoubtedly in a new era of large surveys and SNe analysis methods are gradually changing. There are so many SNe discovered every day, that it has become impossible to invest observational efforts to classify and follow up all of them. Instead, the community tries to find the most interesting events based on somewhat rare or outstanding characteristics and obtain dedicated follow up of only those objects. The follow up of the remaining SNe is doomed to rely on survey observing cadence. It seems impossible to design a better strategy as the observing resources are limited. Optimal follow up of a large number of SNe will entirely rely on coordination of those observing resources. It has occurred that interesting events have been missed and found buried in the archives long after they have disappeared, when no further follow up is possible. Interesting events could also be missed due to lower cadence in some regions of the sky that result in gaps in the data. All this is going to become more of a concern once the upcoming Vera C. Rubin Observatory Legacy Survey of Space and Time (LSST) begins operations, as the survey is expected to produce $\sim$ 10 million transient alerts per night \citep{2019ApJ...873..111I,2023PASP..135j5002H}. To try to deal with such a humongous amount of alerts, several brokers have been developed\footnote{\url{https://www.lsst.org/scientists/alert-brokers}}. These brokers currently ingest ZTF data and have been proved useful to identify rare/interesting events. 

\begin{figure*}
   \centering
   \includegraphics[scale=0.5]{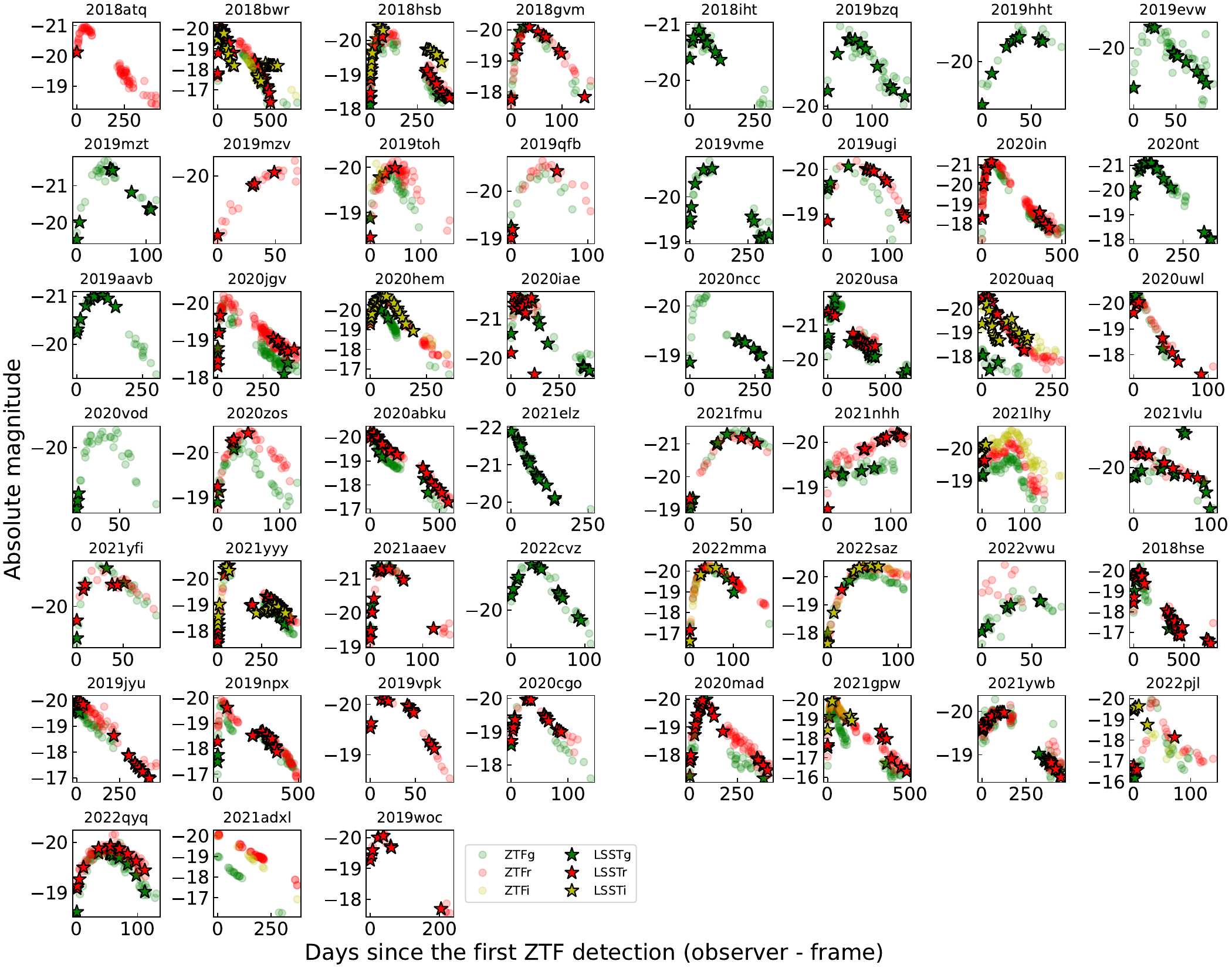}
      \caption{SLSNe~II $gri$ ZTF light curves (light green, red and yellow circles respectively) as observed with the LSST cadence in $gri$ bands (green, red and yellow stars respectively).
              }
         \label{fig:lsstlcs}
   \end{figure*}

Because SLSNe are long lived, it is often considered that they should be detected even in low cadence surveys. Thus, they should be ideal candidates to be detected and followed up with LSST. However, the SLSN class is mainly based on the light curve peak brightness. Meaning that if the peak is missed, then the event will never be classified as superluminous. To asses how many SLSNe~II could be missed due to the peak not being being observed, we used the LSST Operations Simulator\footnote{\url{https://www.lsst.org/scientists/simulations/opsim}} through the \texttt{OpSimSummary 3.0} software\footnote{\url{https://lsstdesc.org/OpSimSummary/build/html/index.html}} \citep{2022zndo...6350796R}. We identified the ZTF SLSNe~II footprint on the LSST path by considering the LSST declination and airmass limits \citep{2017arXiv170804058L}. When considering the former we found 72 of our SLSNe~II in the LSST path, 44 of which lie on the region limited by the airmass constraint. As expected, events in the airmass limited region present better light curve sampling than the rest. Overall, 50 of our SLSNe~II would have had at least one LSST detection. The obtained light curves are presented in Fig~\ref{fig:lsstlcs}. 

\begin{figure*}
   \centering
   \includegraphics[scale=0.6]{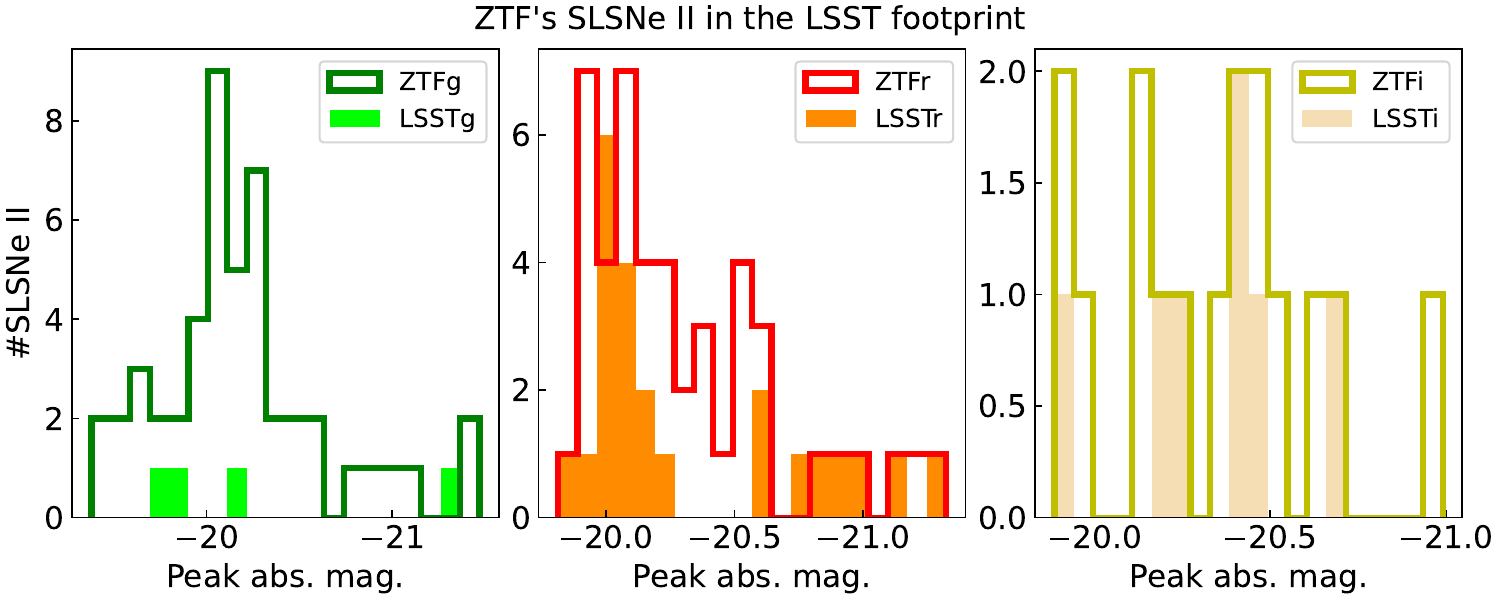}
      \caption{Distribution of $gri$ (left, center and right panels respectively) absolute peak magnitudes as observed by ZTF and LSST.
              }
         \label{fig:lsstpeak}
   \end{figure*}

Of course one detection is not enough to determine a peak absolute magnitude. We consider events with at least 5 observations in either $gri$ bands and use GP (see Sect.~\ref{sec:analysis}) to interpolate the LSST light curves and obtain an associated peak magnitude. In Fig~\ref{fig:lsstpeak} we compare the distribution of observed rest frame peak magnitudes in the ZTF and LSST overlap. We see that LSST will see the peaks of $\sim$ 20\% to 70\% of the events, with the most detections in the $r$ band. Meaning that $\geq$ 30\% of future superluminous events could be missed, at least in the $gri$ bands. LSST will have more bands that ZTF, they will become crucial in detecting SLSNe. To better determine the possible number of missed events we would need an accurate estimation of SLSN rates and we would need to fully understand their color distribution. Such analysis is out of the scope of this work. Here we limit ourselves to assess how many known events within the observable LSST sky could potentially be missed. 

Our analysis is limited to nearby events (see $z$ distribution in Fig.~\ref{fig:z}). LSST will be $\sim$ 4 times deeper than ZTF and thus, it will have the capability to detect fainter and more distant SLSNe~II. Due to time dilation, the light curves of these distant events will also be more finely sampled in the rest frame. While LSST will excel at identifying distant SLSNe~II, our findings indicate that it may be less effective at recovering the nearby population of these events, particularly in the bands analyzed here. Consequently, we caution that future LSST SLSNe~II samples may be biased towards more distant events. To mitigate this potential bias, dedicated observational efforts will be essential to address possible gaps in LSST light curves for nearby SLSNe~II. Such efforts are critical for ensuring a comprehensive and accurate understanding of the SLSNe~II population.

Based on Figures~\ref{fig:comptoothers} and~\ref{fig:timecompSLSNeI}, it seems like a photometric distinction between Type I and Type II SLSNe will be impossible at early times unless the SLSN~I shows a peak color bluer than $g-r < -0.1$~mag or a $g$ band peak absolute magnitude $g_{\mathrm{Peak}} < -22.$~mag (see Fig.~\ref{fig:grcol_peak}). Otherwise, we should wait until very late times to attempt such a separation. This implies that selecting SLSNe~II during the LSST era will be rather challenging.

\section{Conclusions}
\label{sec:conclusions}

We have presented the ZTF sample of SLSN-II, comprising 107 objects, and analyzed their light curve properties. This is the first study of its kind. Our main conclusions can be summarized as follows:

\begin{itemize}
    \item[\textbf{\Huge{.}}] Our SLSN~II sample shows a median peak magnitude of $\sim -$20.3 in the considered optical bands; a median t$_{\mathrm{rise,1/e}}$ of $\sim$ 34, 37 and 45 days in $gri$ bands respectively; a median t$_{\mathrm{rise,10\%}}$ of $\sim$ 44, 49 and 50 days in $gri$ bands respectively; a median t$_{\mathrm{dec,1/e}}$ of $\sim$ 75, 89 and 98 days in $gri$ bands respectively; and a median t$_{\mathrm{dec,10\%}}$ of $\sim$ 236, 248 and 247 days in $gri$ bands respectively. However, SLSNe-II are a heterogeneous population. The dispersion of the considered time parameters is large, with some objects rising in less than two weeks and some in over two months. Similarly, some events decline fast and some take over a year to fade. We do not find any significant correlations between peak brightness and timescale.
    \item[\textbf{\Huge{.}}] Only $\sim$ 14\%, 6\% and 2\% of objects in the $gri$ bands respectively are brighter than $-21$~mag.
    \item[\textbf{\Huge{.}}] Events that rise faster also decline faster with a few exceptions.
    \item[\textbf{\Huge{.}}] Accounting for Malmquist bias, we find that the $g$ band peaks at $M_g = -20.2$~mag. Although, there may be biasing effects related to target selection and classification, especially at the fainter end of discovered events, that are not being accounted for.
    \item[\textbf{\Huge{.}}] We find lower limits on the radiated energy typically in the range of a few times $10^{50}$~erg. Only one object in our sample has a radiated energy exceeding $10^{51}$~erg. One object exceeding this limit can also be found in the sample presented by \cite{2022MNRAS.516.1193K}. In principle, only these two events would require additional powering mechanisms, as the energetics of every other SLSN~II could be explained solely by CSM interaction.
    \item[\textbf{\Huge{.}}] Compared to the sample of SLSNe~II without narrow spectral lines presented by \cite{ 2022MNRAS.516.1193K}, we find no obvious differences in the light curve parameter distributions of our sample that includes events with narrow lines in their spectra.
    \item[\textbf{\Huge{.}}] Compared to regular SNe~IIn, in addition to being more luminous (by definition), SLSNe~II generally also have longer timescales. \cite{2024arXiv241107287H} analyzed a large sample of SNe~IIn including SLSN~II(n) finding a bimodal distribution, although at lower peak absolute magnitudes. 
    \item[\textbf{\Huge{.}}] Compared to SLSN-I, the SLSN-II sample on average shows fainter peak luminosities, redder observed peak colors, and longer-duration light curves. These properties could be used to distinguish between the populations in photometric-only surveys.
    \item[\textbf{\Huge{.}}] If we assume that the light curves are powered by circumstellar interaction, our measured luminosities imply mass loss rates in the range of less than one to tens of solar masses, depending on assumptions on the nature of the progenitor and velocity of the expanding material. Further spectroscopic analysis is needed to constrain these values.
\end{itemize}

This diversity in light curve properties could reflect a similar diversity in the CSM configuration and extent at the end of a massive star's life, even if the required mass loss rates in general need to be high. Alternatively, the diversity could be produced by different powering mechanisms. The best way to probe this would be to supplement the light curve information with both multi-epoch spectroscopy and multiwavelength observations (such as radio and X-ray), although this is in general only possible for the most nearby interacting SNe. 

In the upcoming decade, the LSST at Rubin Observatory is predicted to detect thousands of SLSNe, for which only a small fraction can be spectroscopically classified. This work represents a starting point in distinguishing SLSN-II from their stripped counterparts, as well as from other rare transients in the luminous, long-duration part of parameter space (such as pair-instability supernovae). We caution that even for such slow transients, LSST cadence may miss crucial light curve information. 

\begin{acknowledgements}

PJP thanks Laureano Martinez, Stephen Thorp and Takashi Nagao for useful discussions. 
We thank the referee for the valuable revision.
PJP acknowledges support from the European Research Council (ERC) under the European Union's Horizon Europe research and innovation program (grant agreement No. 10104229 – TransPIre). 
AGY's research is supported by the ISF GW excellence center, as well as the André Deloro Institute for Space and Optics Research, the Center for Experimental Physics, the Norman E Alexander Family M Foundation ULTRASAT Data Center Fund, and Yeda-Sela;  AGY is the incumbent of the The Arlyn Imberman Professorial Chair. 
T.-W.C. acknowledges the Yushan Fellow Program by the Ministry of Education, Taiwan for the financial support (MOE-111-YSFMS-0008-001-P1).
MMK acknowledges generous support from the David and Lucille Packard Foundation. This work was supported by the GROWTH project \citep{2019PASP..131c8003K} funded by the National Science Foundation under Grant No 1545949.
This work is based on observations obtained with the Samuel Oschin Telescope 48-inch and the 60-inch Telescope at the Palomar Observatory as part of the Zwicky Transient Facility project. ZTF is supported by the National Science Foundation under Grants No. AST-1440341 and AST-2034437 and a collaboration including current partners Caltech, IPAC, the Weizmann Institute of Science, the Oskar Klein Center at Stockholm University, the University of Maryland, Deutsches Elektronen-Synchrotron and Humboldt University, the TANGO Consortium of Taiwan, the University of Wisconsin at Milwaukee, Trinity College Dublin, Lawrence Livermore National Laboratories, IN2P3, University of Warwick, Ruhr University Bochum, Northwestern University and former partners the University of Washington, Los Alamos National Laboratories, and Lawrence Berkeley National Laboratories. Operations are conducted by COO, IPAC, and UW.
The ZTF forced-photometry service was funded under the Heising-Simons Foundation grant No. 12540303 (PI: Graham).
The SED Machine is based upon work supported by the National Science Foundation under Grant No. 1106171
This work has made use of data from the Asteroid Terrestrial-impact Last Alert System (ATLAS) project. The Asteroid Terrestrial-impact Last Alert System (ATLAS) project is primarily funded to search for near earth asteroids through NASA grants NN12AR55G, 80NSSC18K0284, and 80NSSC18K1575; byproducts of the NEO search include images and catalogs from the survey area. This work was partially funded by Kepler/K2 grant J1944/80NSSC19K0112 and HST GO-15889, and STFC grants ST/T000198/1 and ST/S006109/1. The ATLAS science products have been made possible through the contributions of the University of Hawaii Institute for Astronomy, the Queen’s University Belfast, the Space Telescope Science Institute, the South African Astronomical Observatory, and The Millennium Institute of Astrophysics (MAS), Chile.
This research has made use of the NASA/IPAC Extragalactic Database, which is funded by the National Aeronautics and Space Administration and operated by the California Institute of Technology.
\end{acknowledgements}

\bibliography{SLSNII} 
\onecolumn
\begin{appendix} 

\section{Sample information, observing logs and measured parameters}

\begin{ThreePartTable}
\begin{TableNotes}
      \small  
      \item The first and second columns show the internal ZTF name and the corresponding IAU name for each SLSN~II. The third and fourth column show right ascension and declination respectively. Column five shows the heliocentric redshift. The sixt colum shows the Landolt $V$ band absorption value obtained from NED. The last two columns present the citations for the corresponding discovery and classification report respectively.
\end{TableNotes}
\onecolumn
\setlength{\tabcolsep}{6pt} 
\begin{longtable}{llccccll}
    \caption{ZTF SLSNe~II sample.}
    \label{tab:sample}
    \endfirsthead
    \hline\hline          
ZTFName        & IAUName     & R.A.        &  Dec.        &   z  & Av     &  Discov.              & Class. 	           \\
               &             & [J2000]     &  [J2000]     &      & [mag]  &  Report               &  Report                \\
\hline  
ZTF18aahmhxu & SN~2018atq   & 11:47:04.08   & +19:33:02.90  & 0.167  & 0.083 & \scalebox{.6}{\cite{2018TNSTR.483....1X}}  & \scalebox{.6}{\cite{2018TNSCR.529....1M}} \\
ZTF18aasyjhd & SN~2022lvm   & 14:49:17.64  & +27:16:47.96  & 0.143  & 0.082 & \scalebox{.6}{\cite{2022TNSTR1534....1M}}  & \scalebox{.6}{\cite{2023TNSCR.207....1G}} \\
ZTF18aautopz & SN~2018lzi   & 13:43:53.35  & +61:33:16.96  & 0.296  & 0.050 & \scalebox{.6}{\cite{2021TNSTR3839....1P}}  & \scalebox{.6}{Pessi~et~al.~2024}          \\
ZTF18aavjcpf & SN~2018ddq   & 11:23:14.91  & +61:37:44.01  & 0.169  & 0.037 & \scalebox{.6}{\cite{2018TNSTR.943....1F}}  & \scalebox{.6}{\cite{2018TNSCR1038....1F}} \\
ZTF18aavskep & SN~2018bwr   & 15:28:26.16  & +08:48:22.17  & 0.046  & 0.101 & \scalebox{.6}{\cite{2018TNSTR.695....1T}}  & \scalebox{.6}{\cite{2018TNSCR.762....1F}} \\
ZTF18aazydub & SN~2018lxa   & 17:00:16.32  & +70:30:50.92  & 0.201  & 0.133 & \scalebox{.6}{\cite{2021TNSTR1713....1N}}  & \scalebox{.6}{Pessi~et~al.~2024}          \\
ZTF18abcfdzu & SN~2018dfa   & 15:20:52.12  & +54:12:55.99  & 0.128  & 0.032 & \scalebox{.6}{\cite{2018TNSTR.949....1T}}  & \scalebox{.6}{\cite{2018TNSCR1100....1F}} \\
ZTF18abgrlpv & SN~2018lpu   & 18:55:44.98  & +47:26:28.45  & 0.210  & 0.151 & \scalebox{.6}{\cite{2019TNSTR1135....1L}}  & \scalebox{.6}{\cite{2019TNSCR1140....1L}} \\
ZTF18ablwafp & SN~2018ffs   & 20:54:37.15  & +22:04:51.77  & 0.141  & 0.308 & \scalebox{.6}{\cite{2018TNSTR1210....1T}}  & \scalebox{.6}{\cite{2018TNSCR1396....1G}} \\
ZTF18abuicad & SN~2018hsb   & 01:18:00.21  & +02:03:27.66  & 0.135  & 0.095 & \scalebox{.6}{\cite{2018TNSTR1674....1F}}  & \scalebox{.6}{\cite{2018TNSCR1877....1F}} \\
ZTF18abvfecb & SN~2018gvm   & 03:10:21.32  & -07:41:06.13  & 0.125  & 0.189 & \scalebox{.6}{\cite{2018TNSTR1463....1F}}  & \scalebox{.6}{\cite{2018TNSCR1760....1F}} \\
ZTF18abwlupf & SN~2018hxe   & 14:44:10.23  & +62:53:42.64  & 0.134  & 0.041 & \scalebox{.6}{\cite{2018TNSTR1714....1T}}  & \scalebox{.6}{\cite{2019TNSCR.329....1F}} \\
ZTF18abxbmqh & SN~2018iaw   & 09:35:27.81  & +52:37:55.99  & 0.206  & 0.051 & \scalebox{.6}{\cite{2018TNSTR1725....1N}}  & \scalebox{.6}{\cite{2018TNSCR1886....1F}} \\
ZTF18acbvhfl & SN~2018hse   & 03:50:06.27  & -11:00:12.31  & 0.130  & 0.114 & \scalebox{.6}{\cite{2018TNSTR1674....1F}}  & \scalebox{.6}{\cite{2018TNSCR1877....1F}} \\
ZTF18acbwdxy & SN~2018ksc   & 08:08:41.54  & +38:52:50.66  & 0.213  & 0.130 & \scalebox{.6}{\cite{2018TNSTR1973....1C}}  & \scalebox{.6}{Pessi~et~al.~2024}          \\
ZTF18accjvic & SN~2018iht   & 02:57:47.77  & -03:07:42.12  & 0.253  & 0.127 & \scalebox{.6}{\cite{2018TNSTR1746....1T}}  & \scalebox{.6}{Pessi~et~al.~2024}          \\
ZTF18acnnevs & SN~2018lng   & 11:36:40.39  & +55:10:12.48  & 0.191  & 0.037 & \scalebox{.6}{\cite{2019TNSTR.326....1F}}  & \scalebox{.6}{\cite{2019TNSCR.329....1F}} \\
ZTF19aaadwfi & SN~2018lnb   & 10:38:32.75  & +48:16:31.13  & 0.222  & 0.036 & \scalebox{.6}{\cite{2019TNSTR1249....1F}}  & \scalebox{.6}{\cite{2019TNSCR.329....1F}} \\
ZTF19aaeopgw & SN~2019aje   & 12:53:45.04  & +46:38:52.50  & 0.127  & 0.045 & \scalebox{.6}{\cite{2019TNSTR.177....1T}}  & \scalebox{.6}{\cite{2019TNSCR1242....1T}} \\
ZTF19aafljiq & SN~2019afz   & 11:27:25.62  & +19:23:52.19  & 0.106  & 0.049 & \scalebox{.6}{\cite{2019TNSTR.157....1N}}  & \scalebox{.6}{\cite{2019TNSCR.329....1F}} \\
ZTF19aafnend & SN~2019bhg   & 16:59:26.20  & +46:47:56.90  & 0.120  & 0.102 & \scalebox{.6}{\cite{2019TNSTR.299....1F}}  & \scalebox{.6}{\cite{2019TNSCR.652....1F}} \\
ZTF19aailptb & SN~2019bzq   & 14:01:38.39  & +14:24:55.00  & 0.287  & 0.045 & \scalebox{.6}{\cite{2019TNSTR.410....1N}}  & \scalebox{.6}{\cite{2019TNSCR.417....1B}} \\
ZTF19aailsyx & SN~2019avv   & 15:44:41.52  & +51:08:58.69  & 0.221  & 0.038 & \scalebox{.6}{\cite{2019TNSTR.236....1N}}  & \scalebox{.6}{\cite{2019TNSCR1186....1P}} \\
ZTF19aalbrgu & SN~2019cmv   & 18:57:53.00  & +45:35:23.99  & 0.097  & 0.160 & \scalebox{.6}{\cite{2019TNSTR.464....1N}}  & \scalebox{.6}{\cite{2019TNSCR.590....1F}} \\
ZTF19aariuyd & SN~2019hht   & 12:20:26.07  & +17:19:30.58  & 0.230  & 0.080 & \scalebox{.6}{\cite{2019TNSTR.984....1T}}  & \scalebox{.6}{\cite{2019TNSCR1186....1P}} \\
ZTF19aarixve & SN~2019evw   & 13:53:12.91  & +08:09:54.37  & 0.199  & 0.073 & \scalebox{.6}{\cite{2019TNSTR.735....1T}}  & \scalebox{.6}{Pessi~et~al.~2024}          \\
ZTF19aaserwb & SN~2019kwv   & 15:00:39.08  & +20:16:45.37  & 0.328  & 0.112 & \scalebox{.6}{\cite{2019TNSTR1265....1P}}  & \scalebox{.6}{\cite{2019TNSCR1186....1P}} \\
ZTF19aaynqaj & SN~2019kww   & 22:57:48.66  & +42:56:58.98  & 0.272  & 0.569 & \scalebox{.6}{\cite{2019TNSTR1266....1P}}  & \scalebox{.6}{\cite{2019TNSCR1186....1P}} \\
ZTF19aayvycv & SN~2019kkj   & 02:26:43.48  & +37:24:59.96  & 0.216  & 0.125 & \scalebox{.6}{\cite{2019TNSTR1147....1N}}  & \scalebox{.6}{Pessi~et~al.~2024}          \\
ZTF19abcejsg & SN~2019jyu   & 00:22:38.69  & -00:09:28.80  & 0.134  & 0.079 & \scalebox{.6}{\cite{2019TNSTR1116....1N}}  & \scalebox{.6}{\cite{2019TNSCR1372....1D}} \\
ZTF19abdkgwo & SN~2019khb   & 02:00:30.9   & +37:03:03.14  & 0.098  & 0.156 & \scalebox{.6}{\cite{2019TNSTR1134....1N}}  & \scalebox{.6}{\cite{2019TNSCR1193....1F}} \\
ZTF19abiagjr & SN~2019mzt   & 20:15:26.44  & -00:34:30.07  & 0.329  & 0.339 & \scalebox{.6}{\cite{2019TNSTR1460....1P}}  & \scalebox{.6}{\cite{2019TNSCR1556....1W}} \\
ZTF19abiubpd & SN~2019qhz   & 16:25:10.89  & +42:47:32.01  & 0.368  & 0.028 & \scalebox{.6}{\cite{2019TNSTR1849....1T}}  & \scalebox{.6}{Pessi~et~al.~2024}          \\
ZTF19ablzfzd & SN~2019mzv   & 01:40:31.18  & +22:02:24.55  & 0.103  & 0.222 & \scalebox{.6}{\cite{2019TNSTR1454....1T}}  & \scalebox{.6}{Pessi~et~al.~2024}          \\
ZTF19abpvbzf & SN~2019npx   & 20:32:52.07  & -12:51:00.34  & 0.056  & 0.105 & \scalebox{.6}{\cite{2019TNSTR1514....1N}}  & \scalebox{.6}{\cite{2019TNSCR1595....1B}} \\
ZTF19abucgpn & SN~2019toh   & 01:37:18.96  & +15:33:07.09  & 0.165  & 0.176 & \scalebox{.6}{\cite{2019TNSTR2205....1C}}  & \scalebox{.6}{Pessi~et~al.~2024}          \\
ZTF19abxekxi & SN~2019qfb   & 20:45:12.03  & -12:00:39.89  & 0.142  & 0.118 & \scalebox{.6}{\cite{2019TNSTR1844....1N}}  & \scalebox{.6}{\cite{2019TNSCR1991....1D}} \\
ZTF19abyhhkb & SN~2019ynt   & 23:22:05.71  & +10:38:47.84  & 0.273  & 0.190 & \scalebox{.6}{\cite{2019TNSTR2722....1C}}  & \scalebox{.6}{Pessi~et~al.~2024}          \\
ZTF19abzvbra & SN~2019aafk  & 22:56:20.93  & +45:12:45.20  & 0.160  & 0.528 & \scalebox{.6}{\cite{2020TNSTR2673....1Y}}  & \scalebox{.6}{\cite{2020TNSCR2676....1D}} \\
ZTF19acmezdo & SN~2019vme   & 23:35:22.69  & -11:28:53.68  & 0.215  & 0.085 & \scalebox{.6}{\cite{2019TNSTR2447....1N}}  & \scalebox{.6}{Pessi~et~al.~2024}          \\
ZTF19acmwszk & SN~2019ugi   & 10:44:58.30  & -12:36:46.05  & 0.120  & 0.160 & \scalebox{.6}{\cite{2019TNSTR2282....1N}}  & \scalebox{.6}{\cite{2020TNSCR.601....1D}} \\
ZTF19actabny & SN~2019vas   & 17:13:55.47  & +21:40:04.48  & 0.155  & 0.136 & \scalebox{.6}{\cite{2019TNSTR2386....1H}}  & \scalebox{.6}{\cite{2020TNSCR1933....1S}} \\
ZTF19acvkibv & SN~2019vpk   & 04:16:58.47  & -16:06:43.01  & 0.101  & 0.098 & \scalebox{.6}{\cite{2019TNSTR2440....1F}}  & \scalebox{.6}{\cite{2020TNSCR..25....1D}} \\
ZTF19acxyugk & SN~2019wky   & 14:18:23.25  & +52:10:53.85  & 0.208  & 0.026 & \scalebox{.6}{\cite{2019TNSTR2580....1D}}  & \scalebox{.6}{Pessi~et~al.~2024}          \\
ZTF19acyjqzd & SN~2019aava  & 12:29:25.96  & +76:13:33.66  & 0.308  & 0.108 & \scalebox{.6}{Pessi~et~al.~2024}           & \scalebox{.6}{Pessi~et~al.~2024}          \\
ZTF19acykbce & SN~2019woc   & 23:36:20.73  & -04:42:39.80  & 0.104  & 0.103 & \scalebox{.6}{\cite{2019TNSTR2597....1T}}  & \scalebox{.6}{Pessi~et~al.~2024}          \\
ZTF20aaaweke & SN~2020in    & 09:53:1.85   & +20:14:03.50  & 0.108  & 0.077 & \scalebox{.6}{\cite{2020TNSTR..62....1N}}  & \scalebox{.6}{\cite{2020TNSCR.205....1G}} \\
ZTF20aadcbvz & SN~2020nt    & 15:16:12.87  & +02:19:33.07  & 0.187  & 0.119 & \scalebox{.6}{\cite{2020TNSTR..71....1N}}  & \scalebox{.6}{\cite{2020TNSCR1735....1Y}} \\
ZTF20aafduse & SN~2019aavb  & 02:56:08.27  & +01:17:57.98  & 0.253  & 0.230 & \scalebox{.6}{Pessi~et~al.~2024}           & \scalebox{.6}{Pessi~et~al.~2024}          \\
ZTF20aammdfk & SN~2020cgo   & 12:56:02.68  & -14:23:17.89  & 0.097  & 0.130 & \scalebox{.6}{\cite{2020TNSTR.448....1C}}  & \scalebox{.6}{\cite{2020TNSCR.688....1D}} \\
ZTF20aampiit & SN~2020jgv   & 07:28:58.12  & +30:44:08.33  & 0.147  & 0.180 & \scalebox{.6}{\cite{2020TNSTR1267....1P}}  & \scalebox{.6}{Pessi~et~al.~2024}          \\
ZTF20aatrkif & SN~2020hei   & 13:39:20.51  & +38:04:55.17  & 0.196  & 0.018 & \scalebox{.6}{\cite{2020TNSTR1031....1C}}  & \scalebox{.6}{Pessi~et~al.~2024}          \\
ZTF20aaurfwa & SN~2020hem   & 15:02:40.15  & +09:18:13.86  & 0.094  & 0.086 & \scalebox{.6}{\cite{2020TNSTR1027....1F}}  & \scalebox{.6}{\cite{2020TNSCR1176....1D}} \\
ZTF20aavbvjr & SN~2020jvj   & 11:54:08.65  & +61:28:57.81  & 0.258  & 0.118 & \scalebox{.6}{\cite{2020TNSTR1327....1S}}  & \scalebox{.6}{Pessi~et~al.~2024}          \\
ZTF20aavrulv & SN~2020iae   & 14:03:13.22  & +12:39:57.98  & 0.347  & 0.074 & \scalebox{.6}{\cite{2020TNSTR1107....1F}}  & \scalebox{.6}{Pessi~et~al.~2024}          \\
ZTF20aazffau & SN~2020kro   & 00:40:08.16  & +38:49:54.32  & 0.127  & 0.141 & \scalebox{.6}{\cite{2020TNSTR1440....1T}}  & \scalebox{.6}{\cite{2020TNSCR2394....1B}} \\
ZTF20aazppax & SN~2020kcr   & 15:35:42.82  & +33:09:01.75  & 0.203  & 0.098 & \scalebox{.6}{\cite{2020TNSTR1374....1P}}  & \scalebox{.6}{Pessi~et~al.~2024}          \\
ZTF20abawhtd & SN~2020mad   & 14:15:52.75  & +05:26:35.62  & 0.123  & 0.070 & \scalebox{.6}{\cite{2020TNSTR1716....1S}}  & \scalebox{.6}{\cite{2020TNSCR3149....1G}} \\
ZTF20abbufrx & SN~2020ncc   & 16:15:46.83  & +00:39:59.56  & 0.186  & 0.293 & \scalebox{.6}{\cite{2020TNSTR1861....1T}}  & \scalebox{.6}{Pessi~et~al.~2024}          \\
ZTF20ablgmou & SN~2020afic  & 15:11:26.84  & +62:35:25.00  & 0.271  & 0.049 & \scalebox{.6}{Pessi~et~al.~2024}           & \scalebox{.6}{Pessi~et~al.~2024}          \\
ZTF20abrgzfn & SN~2020rmn   & 20:59:01.80  & +18:46:15.64  & 0.153  & 0.362 & \scalebox{.6}{\cite{2020TNSTR2520....1N}}  & \scalebox{.6}{Pessi~et~al.~2024}          \\
ZTF20acbcfaa & SN~2020usa   & 22:26:13.58  & -02:16:06.70  & 0.263  & 0.174 & \scalebox{.6}{\cite{2020TNSTR3011....1N}}  & \scalebox{.6}{\cite{2020TNSCR3641....1Y}} \\
ZTF20aceokvr & SN~2020uaq   & 15:34:49.03  & +10:58:37.33  & 0.115  & 0.099 & \scalebox{.6}{\cite{2020TNSTR2913....1J}}  & \scalebox{.6}{\cite{2021TNSCR.449....1S}} \\
ZTF20acgnenc & SN~2020uwl   & 01:29:26.60  & -15:24:45.46  & 0.093  & 0.043 & \scalebox{.6}{\cite{2020TNSTR3013....1F}}  & \scalebox{.6}{\cite{2020TNSCR3109....1A}} \\
ZTF20acgywhr & SN~2020vod   & 21:30:04.44  & -02:54:24.12  & 0.172  & 0.138 & \scalebox{.6}{\cite{2020TNSTR3064....1F}}  & \scalebox{.6}{Pessi~et~al.~2024}          \\
ZTF20achncvv & SN~2020vfu   & 07:17:05.83  & +49:29:16.65  & 0.128  & 0.239 & \scalebox{.6}{\cite{2020TNSTR3052....1F}}  & \scalebox{.6}{\cite{2020TNSCR3298....1Y}} \\
ZTF20aclnqle & SN~2020ysh   & 10:53:10.09  & +38:27:28.81  & 0.146  & 0.039 & \scalebox{.6}{\cite{2020TNSTR3332....1N}}  & \scalebox{.6}{\cite{2020TNSCR3638....1Y}} \\
ZTF20acoawtj & SN~2020yrn   & 22:12:56.07  & +12:48:30.95  & 0.123  & 0.189 & \scalebox{.6}{\cite{2020TNSTR3332....1N}}  & \scalebox{.6}{\cite{2020TNSCR3533....1I}} \\
ZTF20acqgklx & SN~2020zos   & 05:06:32.09  & +07:45:36.38  & 0.140  & 0.422 & \scalebox{.6}{\cite{2020TNSTR3437....1F}}  & \scalebox{.6}{\cite{2020TNSCR3730....1B}} \\
ZTF20acusylb & SN~2020abku  & 14:00:51.22  & -00:11:08.10  & 0.082  & 0.119 & \scalebox{.6}{\cite{2020TNSTR3629....1M}}  & \scalebox{.6}{\cite{2021TNSCR.223....1D}} \\
ZTF20acveyyv & SN~2020vci   & 15:04:22.64  & +51:04:56.77  & 0.193  & 0.040 & \scalebox{.6}{\cite{2020TNSTR3053....1T}}  & \scalebox{.6}{\cite{2021TNSCR.223....1D}} \\
ZTF20acxmxtu & SN~2021mz    & 23:01:51.22  & +13:57:23.83  & 0.123  & 0.289 & \scalebox{.6}{\cite{2021TNSTR..69....1T}}  & \scalebox{.6}{\cite{2021TNSCR1607....1D}} \\
ZTF21aaabxbd & SN~2021bn    & 16:51:32.97  & +38:00:39.71  & 0.076  & 0.050 & \scalebox{.6}{\cite{2021TNSTR..24....1F}}  & \scalebox{.6}{\cite{2021TNSCR.272....1P}} \\
ZTF21aahfjrr & SN~2021bwf   & 16:35:02.95  & +56:28:38.40  & 0.194  & 0.025 & \scalebox{.6}{\cite{2021TNSTR.342....1M}}  & \scalebox{.6}{\cite{2021TNSCR.903....1P}} \\
ZTF21aamwsud & SN~2021jmn   & 17:10:25.09  & +64:09:41.69  & 0.298  & 0.069 & \scalebox{.6}{\cite{2021TNSTR1193....1S}}  & \scalebox{.6}{Pessi~et~al.~2024}          \\
ZTF21aanefkx & SN~2021elz   & 14:00:10.83  & -12:18:58.77  & 0.260  & 0.226 & \scalebox{.6}{\cite{2021TNSTR.695....1F}}  & \scalebox{.6}{\cite{2021TNSCR1030....1P}} \\
ZTF21aaokvio & SN~2021fmu   & 10:31:56.24  & -15:57:19.64  & 0.145  & 0.193 & \scalebox{.6}{\cite{2021TNSTR.763....1T}}  & \scalebox{.6}{\cite{2021TNSCR.914....1P}} \\
ZTF21aaqhqke & SN~2021gpw   & 13:21:54.85  & +16:44:39.73  & 0.075  & 0.063 & \scalebox{.6}{\cite{2021TNSTR.836....1T}}  & \scalebox{.6}{\cite{2021TNSCR1071....1P}} \\
ZTF21aavgnld & SN~2021nhh   & 15:24:33.27  & -13:32:48.16  & 0.135  & 0.340 & \scalebox{.6}{\cite{2021TNSTR1780....1C}}  & \scalebox{.6}{Pessi~et~al.~2024}          \\
ZTF21aavuqzr & SN~2021kat   & 19:50:31.66  & +57:59:28.09  & 0.101  & 0.384 & \scalebox{.6}{\cite{2021TNSTR1237....1F}}  & \scalebox{.6}{\cite{2021TNSCR1593....1S}} \\
ZTF21aazgkjf & SN~2021lhy   & 21:32:14.95  & -00:30:05.12  & 0.142  & 0.141 & \scalebox{.6}{\cite{2021TNSTR1465....1M}}  & \scalebox{.6}{\cite{2021TNSCR1767....1P}} \\
ZTF21abeoaio & SN~2021sto   & 17:29:54.65  & +65:48:48.45  & 0.243  & 0.123 & \scalebox{.6}{\cite{2021TNSTR2378....1P}}  & \scalebox{.6}{Pessi~et~al.~2024}          \\
ZTF21abiwpjm & SN~2021rll   & 13:45:21.99  & +26:45:00.72  & 0.103  & 0.035 & \scalebox{.6}{\cite{2021TNSTR2277....1M}}  & \scalebox{.6}{\cite{2021TNSCR3998....1C}} \\
ZTF21abrprgb & SN~2021vlu   & 03:13:14.8   & +00:20:09.90  & 0.178  & 0.222 & \scalebox{.6}{\cite{2021TNSTR2755....1F}}  & \scalebox{.6}{\cite{2021TNSCR3674....1I}} \\
ZTF21abtdvpg & SN~2021waf   & 14:00:44.84  & +49:22:22.69  & 0.126  & 0.039 & \scalebox{.6}{\cite{2021TNSTR2816....1F}}  & \scalebox{.6}{\cite{2021TNSCR3188....1G}} \\
ZTF21abzbwfo & SN~2021yfi   & 00:57:20.83  & +17:52:41.27  & 0.220  & 0.133 & \scalebox{.6}{\cite{2021TNSTR3075....1M}}  & \scalebox{.6}{\cite{2021TNSCR3843....1G}} \\
ZTF21acaranm & SN~2021ywb   & 05:33:01.70  & -14:03:51.91  & 0.152  & 0.399 & \scalebox{.6}{\cite{2021TNSTR3175....1T}}  & \scalebox{.6}{\cite{2021TNSCR3210....1H}} \\
ZTF21accgsbf & SN~2021yyy   & 22:40:07.56  & -05:00:12.56  & 0.093  & 0.134 & \scalebox{.6}{\cite{2021TNSTR3190....1M}}  & \scalebox{.6}{\cite{2021TNSCR3519....1C}} \\
ZTF21aceqrju & SN~2021aaev  & 01:23:07.82  & -03:11:13.16  & 0.156  & 0.100 & \scalebox{.6}{\cite{2021TNSTR3364....1M}}  & \scalebox{.6}{\cite{2021TNSCR3546....1C}} \\
ZTF21achdvyu & SN~2021aadc  & 00:04:32.67  & +19:45:41.18  & 0.195  & 0.094 & \scalebox{.6}{\cite{2021TNSTR3371....1J}}  & \scalebox{.6}{\cite{2021TNSCR3822....1G}} \\
ZTF21ackxdos & SN~2021adxl  & 11:48:06.94  & -12:38:41.79  & 0.018  & 0.081 & \scalebox{.6}{\cite{2021TNSTR3820....1F}}  & \scalebox{.6}{Pessi~et~al.~2024}          \\
ZTF21acqvuyb & SN~2022akb   & 07:29:00.31  & +48:02:59.34  & 0.229  & 0.267 & \scalebox{.6}{\cite{2022TNSTR.195....1T}}  & \scalebox{.6}{\cite{2022TNSCR.675....1G}} \\
ZTF22aaacvyy & SN~2022pr    & 09:53:49.55  & +43:09:39.17  & 0.150  & 0.045 & \scalebox{.6}{\cite{2022TNSTR.109....1C}}  & \scalebox{.6}{\cite{2022TNSCR.649....1P}} \\
ZTF22aabomyi & SN~2022cvz   & 11:45:57.03  & +05:51:41.38  & 0.189  & 0.059 & \scalebox{.6}{\cite{2022TNSTR.465....1F}}  & \scalebox{.6}{\cite{2022TNSCR.583....1S}} \\
ZTF22aadesjc & SN~2022fnl   & 15:33:42.48  & +43:44:45.41  & 0.104  & 0.073 & \scalebox{.6}{\cite{2022TNSTR.828....1T}}  & \scalebox{.6}{\cite{2022TNSCR1282....1S}} \\
ZTF22aaetqzk & SN~2022gzi   & 17:46:04.85  & +42:16:34.59  & 0.089  & 0.141 & \scalebox{.6}{\cite{2022TNSTR.914....1M}}  & \scalebox{.6}{\cite{2022TNSCR1478....1P}} \\
ZTF22aajjojx & SN~2022jna   & 22:52:10.01  & +73:31:30.17  & 0.093  & 1.473 & \scalebox{.6}{\cite{2022TNSTR1215....1F}}  & \scalebox{.6}{\cite{2022TNSCR1726....1P}} \\
ZTF22aanwibf & SN~2022mma   & 14:39:01.50  & +15:59:11.77  & 0.038  & 0.073 & \scalebox{.6}{\cite{2022TNSTR1626....1P}}  & \scalebox{.6}{\cite{2022TNSCR1677....1P}} \\
ZTF22aapkbkl & SN~2022odp   & 23:45:33.33  & +29:38:25.96  & 0.133  & 0.358 & \scalebox{.6}{\cite{2022TNSTR1856....1A}}  & \scalebox{.6}{\cite{2022TNSCR2172....1P}} \\
ZTF22aasousd & SN~2022opm   & 22:11:17.17  & +29:18:01.79  & 0.249  & 0.201 & \scalebox{.6}{\cite{2022TNSTR1927....1F}}  & \scalebox{.6}{\cite{2022TNSCR2488....1G}} \\
ZTF22aaspkif & SN~2022pjl   & 21:56:40.95  & -09:30:52.35  & 0.080  & 0.094 & \scalebox{.6}{\cite{2022TNSTR2073....1A}}  & \scalebox{.6}{\cite{2022TNSCR2464....1P}} \\
ZTF22abajyqm & SN~2022rmg   & 17:02:49.55  & -21:48:08.28  & 0.057  & 0.970 & \scalebox{.6}{\cite{2022TNSTR2378....1F}}  & \scalebox{.6}{\cite{2022TNSCR2407....1A}} \\
ZTF22abbczmk & SN~2022qyq   & 02:24:32.74  & -06:38:42.42  & 0.166  & 0.086 & \scalebox{.6}{\cite{2022TNSTR2263....1J}}  & \scalebox{.6}{\cite{2022TNSCR2729....1G}} \\
ZTF22abcesfo & SN~2022saz   & 01:49:09.00  & +08:30:35.23  & 0.082  & 0.226 & \scalebox{.6}{\cite{2022TNSTR2447....1F}}  & \scalebox{.6}{\cite{2022TNSCR2743....1G}} \\
ZTF22abghrui & SN~2022vwu   & 09:11:29.20  & +16:39:46.70  & 0.197  & 0.096 & \scalebox{.6}{\cite{2022TNSTR2772....1A}}  & \scalebox{.6}{\cite{2022TNSCR3251....1S}} \\
ZTF22abjzweh & SN~2022wku   & 08:37:48.20  & +44:30:00.54  & 0.134  & 0.069 & \scalebox{.6}{\cite{2022TNSTR2852....1F}}  & \scalebox{.6}{\cite{2022TNSCR3054....1S}} \\
ZTF22abnfjsm & SN~2022rze   & 12:22:52.16  & +76:02:47.25  & 0.081  & 0.123 & \scalebox{.6}{\cite{2022TNSTR2428....1H}}  & \scalebox{.6}{\cite{2022TNSCR3514....1H}} \\
\hline
\insertTableNotes
\end{longtable}
\twocolumn
\end{ThreePartTable}

\begin{ThreePartTable}
\begin{TableNotes}
      \small  
      \item The first and second columns show the internal ZTF name and the corresponding IAU name for each ambiguous event. The third and fourth column show right ascension and declination respectively. Column five shows the heliocentric redshift. The sixt colum shows the Landolt $V$ band absorption value obtained from NED. The last two columns present the citations for the corresponding discovery and classification report respectively.
\end{TableNotes}
\onecolumn
\begin{longtable}{llccccll}
    \caption{ZTF contaminants sample.}
    \label{tab:contaminants}
    \endfirsthead
    \hline\hline          
ZTFName        & IAUName     & R.A.        &  Dec.        &   z  & Av     &  Discov.              & Class. 	           \\
               &             & [J2000]     &  [J2000]     &      & [mag]  &  Report               &  Report                \\
\hline  
ZTF19aatubsj & 2019fdr    & 17:09:06.85 & +26:51:20.50    &   0.267 & 0.145 & \scalebox{.6}{\cite{2019TNSTR.771....1N}}  & \scalebox{.6}{\cite{2019TNSCR1016....1C}} \\
ZTF19abclykm & 2019meh    & 21:27:17.45 & +64:24:59.07    &   0.093 & 2.072 & \scalebox{.6}{\cite{2019TNSTR1378....1N}}  & \scalebox{.6}{\cite{2019TNSCR1586....1N}} \\
ZTF19abrbskk & 2019pcl    & 22:54:21.00 & -00:36:27.81    &   0.504 & 0.219 & \scalebox{.6}{\cite{2019TNSTR1676....1T}}  & \scalebox{.6}{Pessi~et~al.~2024} \\
ZTF19abvgxrq & 2019pev    & 04:29:22.74 & +00:37:07.45    &   0.097 & 0.225 & \scalebox{.6}{\cite{2019TNSTR1684....1F}}  & \scalebox{.6}{Pessi~et~al.~2024} \\
ZTF19abkdlkl & 2020afab   & 02:53:20.19 & +44:18:05.4     &   0.288 & 0.382 & \scalebox{.6}{\cite{2021TNSTR3927....1N}}  & \scalebox{.6}{Pessi~et~al.~2024} \\
ZTF20aafquqw & 2020afid   & 03:24:54.13 & +23:53:19.1     &   0.394 & 0.471 & \scalebox{.6}{Pessi~et~al.~2024}           & \scalebox{.6}{Pessi~et~al.~2024} \\
ZTF20aasuiks & 2020edi    & 13:16:13.68 & -22:33:30.30    &   0.160 & 0.294 & \scalebox{.6}{\cite{2020TNSTR.732....1F}}  & \scalebox{.6}{\cite{2021TNSCR.433....1T}} \\
ZTF20abjwqqq & 2020pno    & 16:46:37.06 & +55:36:26.6     &   0.279 & 0.035 & \scalebox{.6}{\cite{2020TNSTR2167....1N}}  & \scalebox{.6}{Pessi~et~al.~2024} \\
ZTF20achupkw & 2020vws    & 02:41:39.83 & +19:25:34.22    &   0.384 & 0.226 & \scalebox{.6}{\cite{2020TNSTR3095....1N}}  & \scalebox{.6}{\cite{2021TNSCR.374....1Y}} \\
ZTF21aalyubu & 2021gwf    & 15:06:13.89 & +20:04:14.42    &   0.242 & 0.097 & \scalebox{.6}{\cite{2021TNSTR.879....1Y}}  & \scalebox{.6}{Pessi~et~al.~2024} \\
ZTF21acqgkkq & 2021ahqw   & 09:51:46.69 & +54:11:55.63    & 0.230   & 0.019 & \scalebox{.6}{\cite{2022TNSTR.644....1F}}  & \scalebox{.6}{Pessi~et~al.~2024}          \\
\hline
\insertTableNotes
\end{longtable}
\twocolumn
\end{ThreePartTable}

\begin{table}
 \begin{threeparttable}
\caption{ZTF forced photometry}              
\label{table:ZTFforcedphot}      
\centering                          
\begin{tabular}{l c c c c c}        
\hline\hline                
\noalign{\vskip 1mm}
Name &  MJD   & Filter  & ABMag & err\_ABMag & Flag \\    
     &  [days]          & [mag.]  &  [mag.]      &      \\    			
\noalign{\vskip 1mm}
\hline 	
\noalign{\vskip 1mm}    
2018atq &  58215.26 &  g &  19.08 &  0.19 &  F \\  
2018atq &  58230.29 &  g &  18.81 &  0.18 &  F \\  
2018atq &  58230.33 &  g &  18.56 &  0.03 &  F \\  
2018atq &  58233.21 &  g &  18.49 &  0.05 &  F \\  
2018atq &  58235.32 &  g &  18.43 &  0.09 &  F \\  
\hline 
\end{tabular}
\begin{tablenotes}
      \small
      \item The F flag in the last column means real detection and the T flag means upper limit. The full table is available in supplementary material.
    \end{tablenotes}
  \end{threeparttable}
\end{table}

\begin{table}
 \begin{threeparttable}
\caption{ATLAS forced photometry}              
\label{table:ATLASforcedphot}      
\centering                          
\begin{tabular}{l c c c c c}        
\hline\hline                
\noalign{\vskip 1mm}
Name &  MJD   & Filter  & ABMag & err\_ABMag & Flag \\    
     &  [days]          & [mag.]  &  [mag.]      &      \\    			
\noalign{\vskip 1mm}
\hline 	
\noalign{\vskip 1mm}    
2018atq & 58205.45 & o & 18.34 & $\cdots$ & T \\ 
2018atq & 58213.47 & o & 19.69 & 0.19 & F \\ 
2018atq & 58221.47 & o & 18.96 & 0.07 & F \\ 
2018atq & 58231.41 & o & 18.69 & 0.06 & F \\ 
2018atq & 58243.41 & o & 18.36 & 0.09 & F \\ 
\hline 
\end{tabular}
\begin{tablenotes}
      \small
      \item The F flag in the last column means real detection and the T flag means upper limit. The full table is available in supplementary
material.
    \end{tablenotes}
  \end{threeparttable}
\end{table}

\begin{table}
 \begin{threeparttable}
\caption{Swift forced photometry}              
\label{table:ATLASforcedphot}      
\centering                          
\begin{tabular}{l c c c c}        
\hline\hline                
\noalign{\vskip 1mm}
Name &  MJD   & Filter  & AB Mag. & AB Mag. err. \\    
     &  (days) &        & [mag.]  &  [mag.]       \\    			
\noalign{\vskip 1mm}
\hline 	
\noalign{\vskip 1mm}    
2018atq  & 58254.35 & UVW2 & 20.26 & 0.15 \\
2018atq  & 58270.67 & UVW2 & 21.31 & 0.20 \\
2018atq  & 58278.73 & UVW2 & 21.22 & 0.24 \\
2018atq  & 58282.18 & UVW2 & 21.36 & 0.22 \\
2018atq  & 58286.83 & UVW2 & 21.78 & 0.29 \\
2018atq  & 58254.36 & UVM2 & 19.85 & 0.12 \\
\hline 
\end{tabular}
  \end{threeparttable}
\end{table}

\begin{table*}
 \begin{threeparttable}
\caption{Light curve parameters}              
\label{table:params}      
\centering                          
\begin{tabular}{l c c c c c c c c c c c}        
\hline\hline                
\noalign{\vskip 1mm}
Name &  Filter  & PeakMJD &  PeakAbsMag  & t$_{\mathrm{rise,10\%}}$ & t$_{\mathrm{rise,1/e}}$ & t$_{\mathrm{dec,1/e}}$ & t$_{\mathrm{dec,10\%}}$ \\    
     &          & [days]  &  [mag]       & [days]              &  [days]              & [days]              &  [days]            \\    			
\noalign{\vskip 1mm}
\hline 	
\noalign{\vskip 1mm}
2018atq  & g &  58241.64 $_{-2.42}^{3.55}$    & -20.90 $_{-0.07}^{0.07}$ &    $\cdots$ &     $\cdots$ &   81.95      &  302.27 \\
2018atq  & r &  58260.54 $_{-3.48}^{4.94}$    & -20.93 $_{-0.04}^{0.04}$ &    $\cdots$ &     $\cdots$ &  128.18      &  320.36 \\
2018atq  & i &       $\cdots$                &    $\cdots$             &    $\cdots$ &     $\cdots$ &     $\cdots$ &     $\cdots$ \\
2018bwr  & g &  58283.10 $_{-1.11}^{1.79}$    & -20.01 $_{-0.02}^{0.02}$ &    $\cdots$ &   15.81      &   48.87      &  365.08 \\
2018bwr  & r &  58286.40 $_{-1.44}^{1.84}$    & -20.08 $_{-0.03}^{0.03}$ &    $\cdots$ &   18.71      &  235.40      &  388.25 \\
2018bwr  & i &  58298.81 $_{-5.39}^{83.04}$   & -19.89 $_{-0.03}^{0.03}$ &    $\cdots$ &     $\cdots$ &  178.90      &  357.58 \\
\hline 
\end{tabular}
\begin{tablenotes}
      \small
      \item The first colums shows the corresponding IAU name for each SLSN~II. Second column indicated the observed filter. Third column shows the peak epoch and associated errorbars. The peak absolute magnitude associated errorbars are shown in the fourth column. The remaining columns indicate the different timescales. We adopt the error bars of the peak epoch as associted errors for the different timescales. The full table is available in supplementary material.
    \end{tablenotes}
  \end{threeparttable}
\end{table*}

\begin{table}
\begin{threeparttable}
\caption{pseudo-Bolometric light curves}              
\label{table:pbolLCs}      
\centering                          
\begin{tabular}{l c c c }        
\hline\hline                
\noalign{\vskip 1mm}
Name &  MJD    & Log$_{10}$(L$_{\mathrm{bol}}$)  & Log$_{10}$(L$_{\mathrm{bol}}+UV+NIR$) \\ 
     &  [days] & [erg s$^{-1}$] &    [erg s$^{-1}$]       \\    			
\noalign{\vskip 1mm}
\hline 	
\noalign{\vskip 1mm}     
2018bwr & 58276.28 & 43.02 $\pm$ 0.00 & 43.60 $\pm$ 0.13 \\ 
2018bwr & 58279.25 & 43.05 $\pm$ 0.00 & 43.62 $\pm$ 0.11 \\ 
2018bwr & 58282.21 & 43.07 $\pm$ 0.00 & 43.62 $\pm$ 0.09 \\ 
2018bwr & 58288.25 & 43.08 $\pm$ 0.00 & 43.60 $\pm$ 0.04 \\ 
2018bwr & 58291.23 & 43.08 $\pm$ 0.00 & 43.58 $\pm$ 0.04 \\ 
\hline 
\end{tabular}
\begin{tablenotes}
      \small
      \item Pseuso-bolometric light curves calculated as described in Sect.~\ref{sec:energy}. The full table is available in supplementary material.
    \end{tablenotes}
  \end{threeparttable}
\end{table}

\begin{table}
\caption{Estimated total radiated energy for events with available $gri$ photometry. See Sect.~\ref{sec:energy}}              
\label{table:radiatedEnergy}      
\centering                          
\begin{tabular}{l c}        
\hline\hline                
\noalign{\vskip 1mm}
Name &  Log$_{10}$(Radiated energy) \\    
     &   [erg]     \\    			
\noalign{\vskip 1mm}
\hline 	
\noalign{\vskip 1mm}
SN~2018bwr  & $>$ 50.24 \\
SN~2018ddq  & $>$ 50.47 \\
SN~2018dfa  & $>$ 50.23 \\
SN~2018lng  & $>$ 50.22 \\
SN~2018lpu  & $>$ 50.59 \\
SN~2018lxa  & $>$ 50.14 \\
SN~2018lzi  & $>$ 51.17 \\
SN~2019aava & $>$ 50.32 \\
SN~2019cmv  & $>$ 50.24 \\
SN~2019khb  & $>$ 50.02 \\
SN~2019vas  & $>$ 49.67 \\
SN~2019wky  & $>$ 50.41 \\
SN~2020afic & $>$ 49.98 \\
SN~2020hei  & $>$ 50.17 \\
SN~2020hem  & $>$ 50.34 \\
SN~2020kcr  & $>$ 50.04 \\
SN~2020rmn  & $>$ 49.96 \\
SN~2020vci  & $>$ 50.53 \\
SN~2021aadc & $>$ 49.77 \\
SN~2021bn   & $>$ 49.93 \\
SN~2021bwf  & $>$ 50.34 \\
SN~2021gpw  & $>$ 50.01 \\
SN~2021lhy  & $>$ 50.06 \\
SN~2021mz   & $>$ 50.41 \\
SN~2021rll  & $>$ 50.18 \\
SN~2021sto  & $>$ 50.84 \\
SN~2021vlu  & $>$ 49.77 \\
SN~2021waf  & $>$ 50.53 \\
SN~2021yfi  & $>$ 50.18 \\
SN~2021yyy  & $>$ 50.52 \\
SN~2022fnl  & $>$ 50.04 \\
SN~2022gzi  & $>$ 49.85 \\
SN~2022lvm  & $>$ 49.89 \\
SN~2022mma  & $>$ 49.72 \\
SN~2022odp  & $>$ 50.36 \\
SN~2022opm  & $>$ 50.48 \\
SN~2022pr   & $>$ 50.72 \\
SN~2022saz  & $>$ 50.01 \\
SN~2022wku  & $>$ 49.98 \\
\hline 
\end{tabular}
\end{table}

  \begin{figure*}
   \includegraphics[scale=0.35]{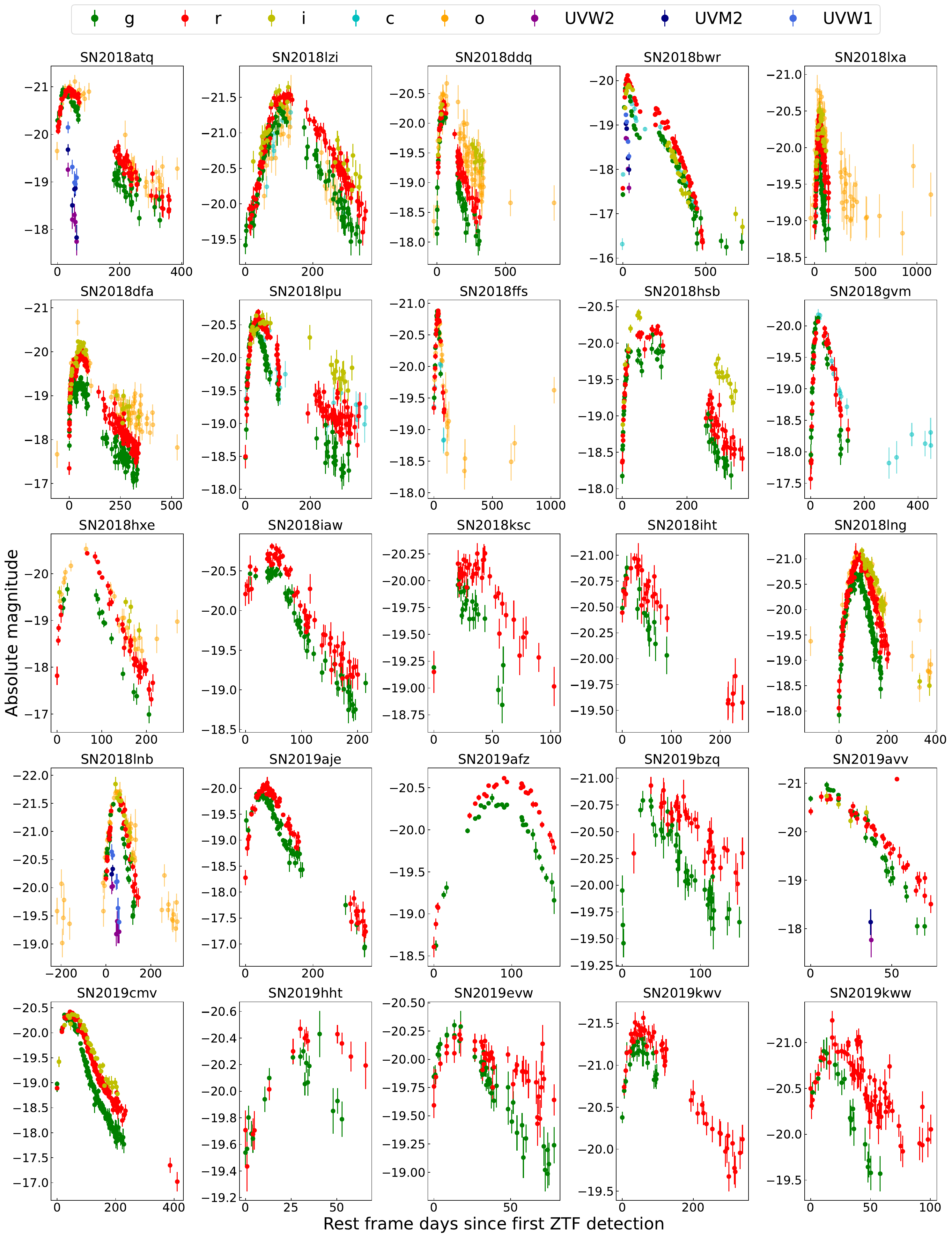}
      \caption{SLSNe~II light curves. The vertical axis shows the absolute magnitudes calculated as decribed in Sect.~\ref{sec:abs_mag}. The colors indicate observed photometric bands. Some ATLAS light curves seem to show additional pre or post-peak detections. We caution that these could not be significant and may have survived our re-processing due to the lack of further quality constrains (see Sect.~\ref{sec:ATLAS}). The horizontal axis shows rest frame days from ZTF's first detection.
              }
         \label{fig:alllcs1}
   \end{figure*}

     \begin{figure*}
   \includegraphics[scale=0.35]{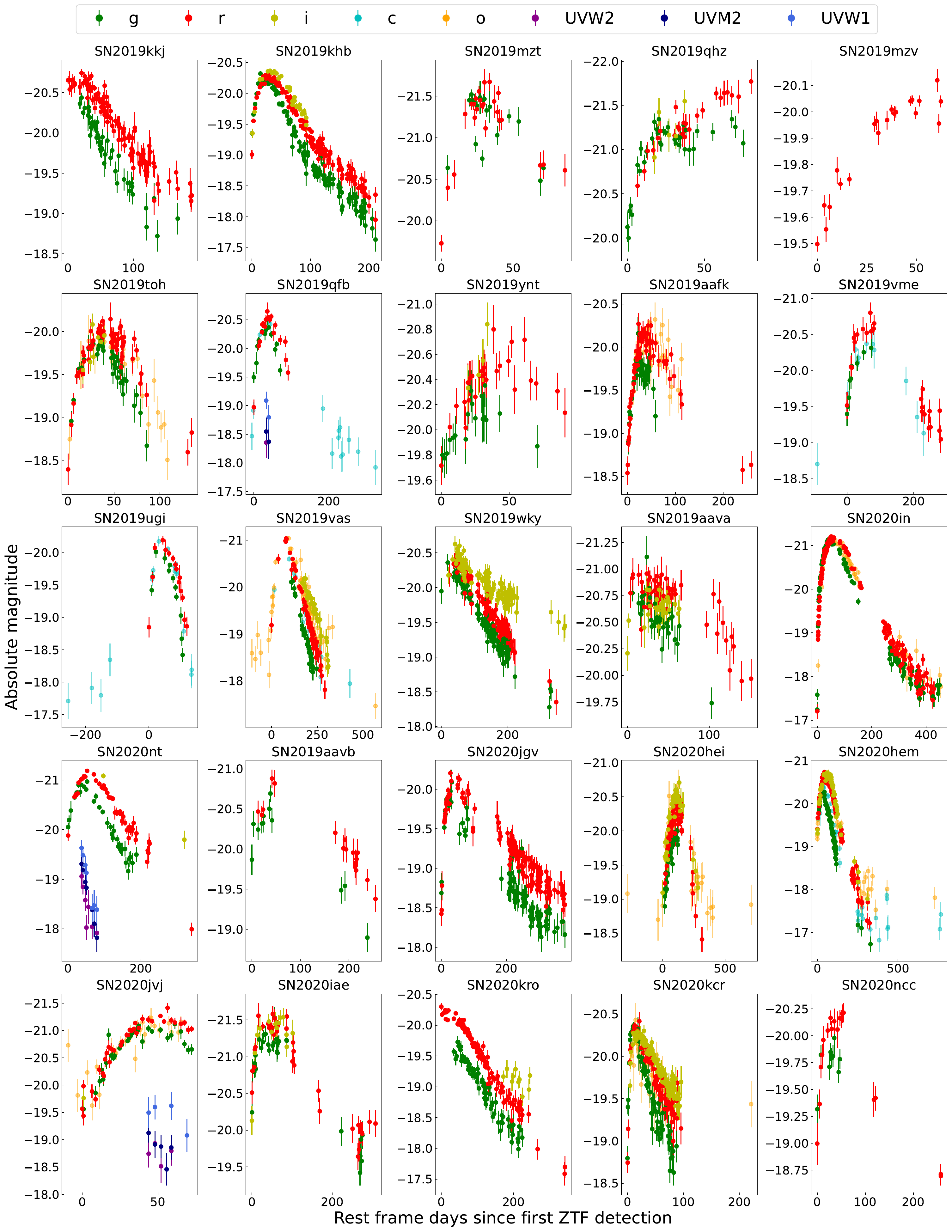}
      \caption{Similar to Fig.~\ref{fig:alllcs1}.
              }
         \label{fig:alllcs2}
   \end{figure*}

     \begin{figure*}
   \includegraphics[scale=0.35]{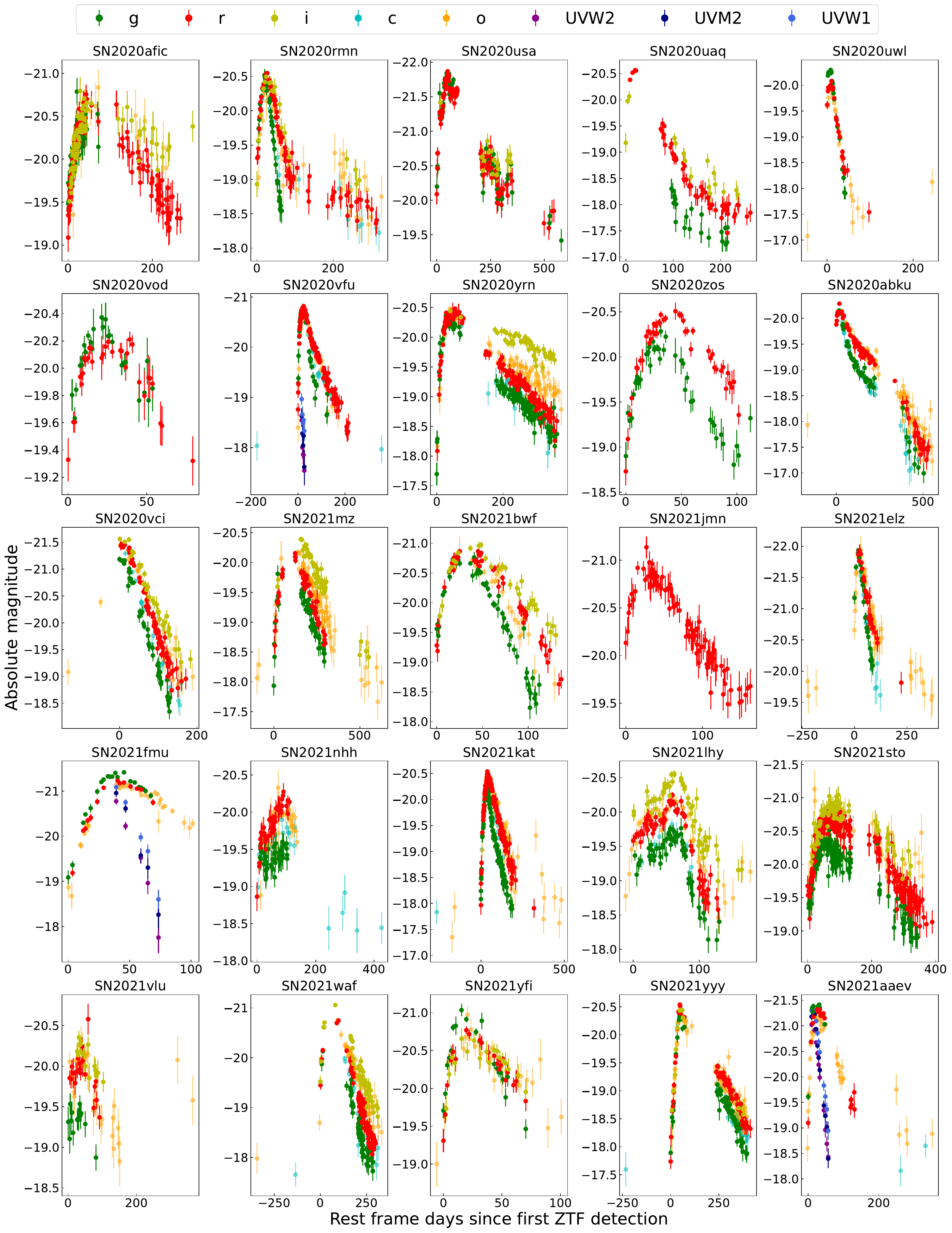}
      \caption{Similar to Fig.~\ref{fig:alllcs1}.
              }
         \label{fig:alllcs3}
   \end{figure*}

     \begin{figure*}
   \includegraphics[scale=0.35]{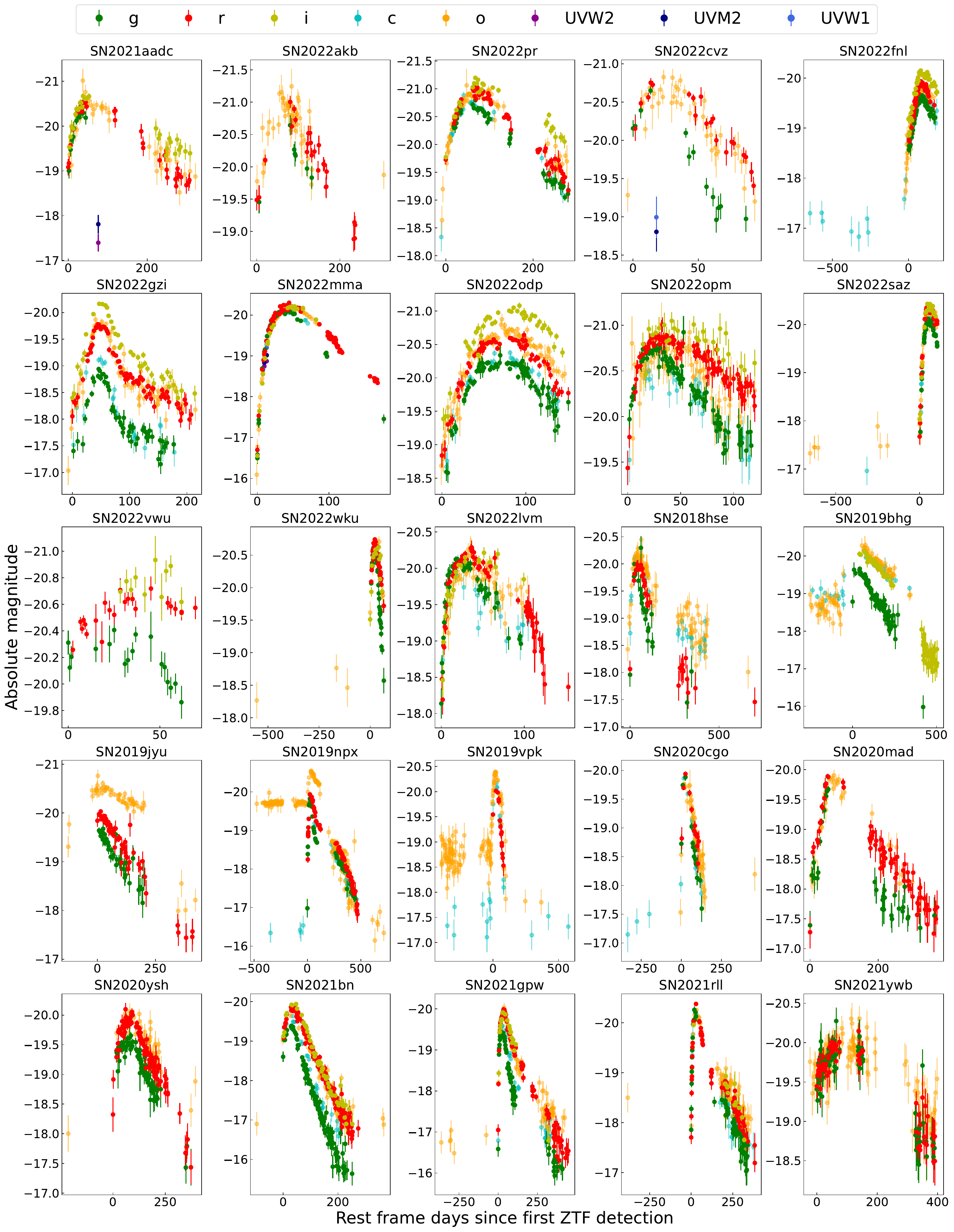}
      \caption{Similar to Fig.~\ref{fig:alllcs1}.
              }
         \label{fig:alllcs4}
   \end{figure*}

     \begin{figure*}
   \includegraphics[scale=0.5]{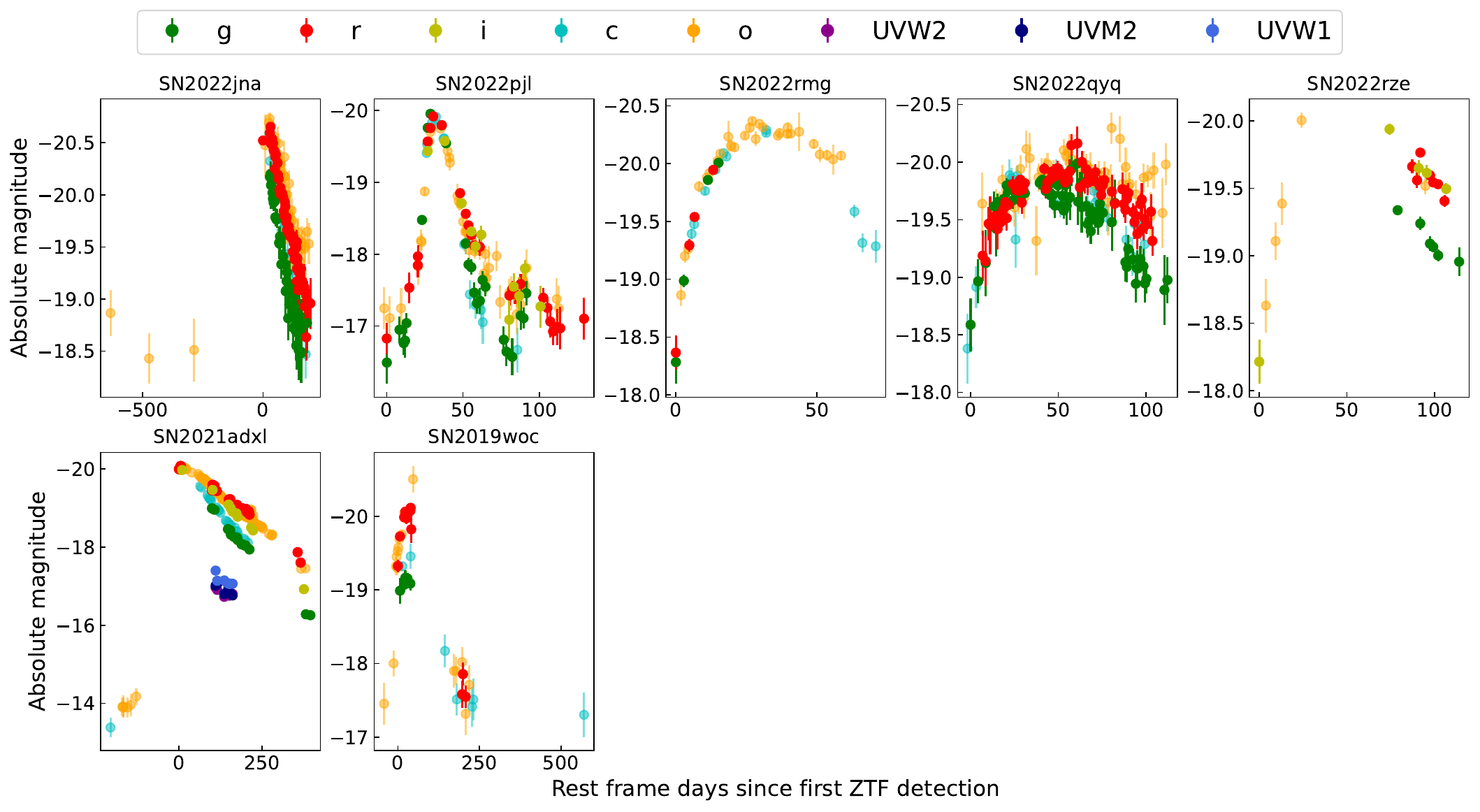}
      \caption{Similar to Fig.~\ref{fig:alllcs1}.
              }
         \label{fig:alllcs5}
   \end{figure*}
   
\end{appendix}

\end{document}